\documentstyle[10pt,epsf,epsfig,hangcaption,xspace,amssymb,amsfonts,amsmath,amsthm,cite,dp_delphititle,lscape,dcolumn,tabularx,dectab,multirow,subfig]{dp_delphi}
\makeindex
\def\DpPaperGroup{PH--EP}
\def\DpPaperRef{2010--057}
\def\DpDate{27 October 2010}
\def\DpAuthors{DELPHI Collaboration}
\def\DpSubmit{(Accepted by Eur. Phys. J. C)}
\def\DpTitle{{A study of the\\
b-quark fragmentation function \\
with the DELPHI detector at LEP I \\
and an averaged distribution \\
obtained at the $\Zz$ Pole}}
\def\DpComment{}
\def\DpEMail{}

\newcommand{\BP}   {\rm B^+} 
\newcommand{\BS}   {\rm B_{\rm s}^0}
\newcommand{\BD}   {\rm B_{\rm d}^0}
 
\newcommand{\B}    {{\rm B}}
\newcommand{\K}    {{\rm K}}
\newcommand{\Bstar}{\mbox{B}^{\ast}}
\newcommand{\Bsstar}{\mbox{B}^{\ast \ast}}

\newcommand{\Zz}{{\rm Z}} 
\newcommand{\Dp}{{\rm D}^+}

\newcommand{\Do}{{\rm D}^0}
\newcommand{\Ds}{{\rm D_s}}
\newcommand{\Ztobb} {$\rm Z \to \rm b\overline{\rm b}$} 
\newcommand{\Ztocc} {$\rm Z \to \rm c\overline{\rm c}$} 
\newcommand{\Ztoqq} {$\rm Z \to \rm q\overline{\rm q}$} 
\newcommand{\xb}{x^{\mathrm weak}_{\mathrm B}}
\newcommand{\xbrec}{x^{\mathrm weak}_{{\mathrm B}, rec}}
\newcommand{\xbgen}{x^{\mathrm weak}_{{\mathrm B}, gen}}
\newcommand{\avxb}{\langle x^{\mathrm weak}_{\mathrm B} \rangle}
\newcommand{\xp}{x^{\mathrm weak}_{p}}

\newcommand{\Eb} {E_{\mathrm B}^{\mathrm weak}}
\newcommand{\micron} {\mu {\mathrm m}}
\newcommand{\GeV}{\rm{GeV}}
\newcommand{\bunit}{\rm{GeV}^{-2}}

\begin{document}
\makeatletter
\makeatother


\begin{titlepage}
\pagenumbering{roman}

\CERNpreprint{\DpPaperGroup}{\DpPaperRef}   
\date{{\small\DpDate}}                      
\title{\DpTitle}                            
\address{\DpAuthors}                        

\begin{shortabs}                            
\noindent
\noindent

The nature of b-quark jet hadronisation has been
investigated using data taken at
the $\Zz$~peak by the DELPHI detector at LEP.
Two complementary methods 
are used to reconstruct the energy of 
weakly decaying b-hadrons, $\Eb$. 
The average value of $\xb= \Eb/E_{beam}$ is measured to be
$0.699 \pm 0.011$.
The resulting $\xb$~distribution is then
analysed in the framework of two choices for the perturbative
contribution (parton shower and Next to Leading Log QCD calculation) in order to 
extract measurements of the non-perturbative contribution to be used
in studies of b-hadron production  
in other experimental environments than LEP.
In the parton shower framework, data favour the Lund model ansatz and 
corresponding values of its parameters have been determined within PYTHIA~6.156 from DELPHI data:
\begin{equation}
a= 1.84^{+0.23}_{-0.21}~{\rm and} ~b=0.642^{+0.073}_{-0.063}\ \bunit~, \nonumber
\end{equation}
with a  correlation factor $\rho = 92.2\%$.

Combining the data on the b-quark fragmentation distributions with those obtained at the $\Zz$~peak by
ALEPH, OPAL and SLD, the average value of $\xb$ is found to be $0.7092 \pm 0.0025$ and the non-perturbative fragmentation component is extracted. 
Using the combined distribution,
a better determination of the Lund parameters is
also obtained:
\begin{equation}
a= 1.48^{+0.11}_{-0.10}~{\rm and} ~b=0.509^{+0.024}_{-0.023}\ \bunit~, \nonumber
\end{equation}
with a correlation factor $\rho = 92.6\%$.                        
\end{shortabs}

\vfill

\begin{center}
\DpSubmit \ \\          
\DpComment \ \\
\DpEMail \ \\
\end{center}

\vfill
\clearpage

\headsep 10.0pt

\addtolength{\textheight}{10mm}
\addtolength{\footskip}{-5mm}
\begingroup
%
\newcommand{\DpName}[2]{\hbox{#1$^{\ref{#2}}$},\hfill}
\newcommand{\DpNameTwo}[3]{\hbox{#1$^{\ref{#2},\ref{#3}}$},\hfill}
\newcommand{\DpNameThree}[4]{\hbox{#1$^{\ref{#2},\ref{#3},\ref{#4}}$},\hfill}
\newskip\Bigfill \Bigfill = 0pt plus 1000fill
\newcommand{\DpNameLast}[2]{\hbox{#1$^{\ref{#2}}$}\hspace{\Bigfill}}
%
\footnotesize
\noindent
\DpName{J.Abdallah}{LPNHE}
\DpName{P.Abreu}{LIP}
\DpName{W.Adam}{VIENNA}
\DpName{P.Adzic}{DEMOKRITOS}
\DpName{T.Albrecht}{KARLSRUHE}
\DpName{R.Alemany-Fernandez}{CERN}
\DpName{T.Allmendinger}{KARLSRUHE}
\DpName{P.P.Allport}{LIVERPOOL}
\DpName{U.Amaldi}{MILANO2}
\DpName{N.Amapane}{TORINO}
% ...
\normalsize
\endgroup

\newpage
\titlefoot{Department of Physics and Astronomy, Iowa State
     University, Ames IA 50011-3160, USA
    \label{AMES}}
\titlefoot{Physics Department, Universiteit Antwerpen,
     Universiteitsplein 1, B-2610 Antwerpen, Belgium
    \label{ANTWERP}}
\titlefoot{IIHE, ULB-VUB,
     Pleinlaan 2, B-1050 Brussels, Belgium
    \label{BRUSSELS}}
\titlefoot{Physics Laboratory, University of Athens, Solonos Str.
     104, GR-10680 Athens, Greece
    \label{ATHENS}}
\titlefoot{Department of Physics, University of Bergen,
     All\'egaten 55, NO-5007 Bergen, Norway
    \label{BERGEN}}
\titlefoot{Dipartimento di Fisica, Universit\`a di Bologna and INFN,
     Viale C. Berti Pichat 6/2, IT-40127 Bologna, Italy
    \label{BOLOGNA}}
\titlefoot{Centro Brasileiro de Pesquisas F\'{\i}sicas, rua Xavier Sigaud 150,
     BR-22290 Rio de Janeiro, Brazil
    \label{BRASIL-CBPF}}
\titlefoot{Inst. de F\'{\i}sica, Univ. Estadual do Rio de Janeiro,
     rua S\~{a}o Francisco Xavier 524, Rio de Janeiro, Brazil
    \label{BRASIL-IFUERJ}}
\titlefoot{Coll\`ege de France, Lab. de Physique Corpusculaire, IN2P3-CNRS,
     FR-75231 Paris Cedex 05, France
    \label{CDF}}
\titlefoot{CERN, CH-1211 Geneva 23, Switzerland
    \label{CERN}}
\titlefoot{Institut Pluridisciplinaire Hubert Curien, Universit\'e de Strasbourg,
     IN2P3-CNRS, BP28, FR-67037 Strasbourg \indent~~Cedex~2, France
    \label{CRN}}
\titlefoot{Now at DESY-Zeuthen, Platanenallee 6, D-15735 Zeuthen, Germany
    \label{DESY}}
\titlefoot{Institute of Nuclear Physics, N.C.S.R. Demokritos,
     P.O. Box 60228, GR-15310 Athens, Greece
    \label{DEMOKRITOS}}
\titlefoot{FZU, Inst. of Phys. of the C.A.S. High Energy Physics Division,
     Na Slovance 2, CZ-182 21, Praha 8, Czech Republic
    \label{FZU}}
\titlefoot{Dipartimento di Fisica, Universit\`a di Genova and INFN,
     Via Dodecaneso 33, IT-16146 Genova, Italy
    \label{GENOVA}}
\titlefoot{Institut des Sciences Nucl\'eaires, IN2P3-CNRS, Universit\'e
     de Grenoble 1, FR-38026 Grenoble Cedex, France
    \label{GRENOBLE}}
\titlefoot{Helsinki Institute of Physics and Department of Physical Sciences,
     P.O. Box 64, FIN-00014 University of Helsinki, 
     \indent~~Finland
    \label{HELSINKI}}
\titlefoot{Joint Institute for Nuclear Research, Dubna, Head Post
     Office, P.O. Box 79, RU-101 000 Moscow, Russian Federation
    \label{JINR}}
\titlefoot{Institut f\"ur Experimentelle Kernphysik,
     Universit\"at Karlsruhe, Postfach 6980, DE-76128 Karlsruhe,
     Germany
    \label{KARLSRUHE}}
\titlefoot{Institute of Nuclear Physics PAN,Ul. Radzikowskiego 152,
     PL-31142 Krakow, Poland
    \label{KRAKOW1}}
\titlefoot{Faculty of Physics and Nuclear Techniques, University of Mining
     and Metallurgy, PL-30055 Krakow, Poland
    \label{KRAKOW2}}
\titlefoot{LAL, Univ Paris-Sud, CNRS/IN2P3, Orsay, France
    \label{LAL}}
\titlefoot{School of Physics and Chemistry, University of Lancaster,
     Lancaster LA1 4YB, UK
    \label{LANCASTER}}
\titlefoot{LIP, IST, FCUL - Av. Elias Garcia, 14-$1^{o}$,
     PT-1000 Lisboa Codex, Portugal
    \label{LIP}}
\titlefoot{Department of Physics, University of Liverpool, P.O.
     Box 147, Liverpool L69 3BX, UK
    \label{LIVERPOOL}}
\titlefoot{Dept. of Physics and Astronomy, Kelvin Building,
     University of Glasgow, Glasgow G12 8QQ, UK
    \label{GLASGOW}}
\titlefoot{LPNHE, Univ.~Pierre et Marie Curie, Univ.~Paris Diderot, CNRS/IN2P3,
     4 pl. Jussieu, 75252 Paris cedex 05, France
    \label{LPNHE}}
\titlefoot{Department of Physics, University of Lund,
     S\"olvegatan 14, SE-223 63 Lund, Sweden
    \label{LUND}}
\titlefoot{Universit\'e Claude Bernard de Lyon, IPNL, IN2P3-CNRS,
     FR-69622 Villeurbanne Cedex, France
    \label{LYON}}
\titlefoot{Dipartimento di Fisica, Universit\`a di Milano and INFN-MILANO,
     Via Celoria 16, IT-20133 Milan, Italy
    \label{MILANO}}
\titlefoot{Dipartimento di Fisica, Univ. di Milano-Bicocca and
     INFN-MILANO, Piazza della Scienza 3, IT-20126 Milan, Italy
    \label{MILANO2}}
\titlefoot{IPNP of MFF, Charles Univ., Areal MFF,
     V Holesovickach 2, CZ-180 00, Praha 8, Czech Republic
    \label{NC}}
\titlefoot{NIKHEF, Postbus 41882, NL-1009 DB
     Amsterdam, The Netherlands
    \label{NIKHEF}}
\titlefoot{National Technical University, Physics Department,
     Zografou Campus, GR-15773 Athens, Greece
    \label{NTU-ATHENS}}
\titlefoot{Physics Department, University of Oslo, Blindern,
     NO-0316 Oslo, Norway
    \label{OSLO}}
\titlefoot{Dpto. Fisica, Univ. Oviedo, Avda. Calvo Sotelo
     s/n, ES-33007 Oviedo, Spain
    \label{OVIEDO}}
\titlefoot{Department of Physics, University of Oxford,
     Keble Road, Oxford OX1 3RH, UK
    \label{OXFORD}}
\titlefoot{Dipartimento di Fisica, Universit\`a di Padova and
     INFN, Via Marzolo 8, IT-35131 Padua, Italy
    \label{PADOVA}}
\titlefoot{Rutherford Appleton Laboratory, Chilton, Didcot
     OX11 OQX, UK
    \label{RAL}}
\titlefoot{Dipartimento di Fisica, Universit\`a di Roma II and
     INFN, Tor Vergata, IT-00173 Rome, Italy
    \label{ROMA2}}
\titlefoot{Dipartimento di Fisica, Universit\`a di Roma III and
     INFN, Via della Vasca Navale 84, IT-00146 Rome, Italy
    \label{ROMA3}}
\titlefoot{DAPNIA/Service de Physique des Particules,
     CEA-Saclay, FR-91191 Gif-sur-Yvette Cedex, France
    \label{SACLAY}}
\titlefoot{Instituto de Fisica de Cantabria (CSIC-UC), Avda.
     los Castros s/n, ES-39006 Santander, Spain
    \label{SANTANDER}}
\titlefoot{Inst. for High Energy Physics, Serpukov
     P.O. Box 35, Protvino, (Moscow Region), Russian Federation
    \label{SERPUKHOV}}
\titlefoot{J. Stefan Institute, Jamova 39, SI-1000 Ljubljana, Slovenia
    \label{SLOVENIJA1}}
\titlefoot{Laboratory for Astroparticle Physics,
     University of Nova Gorica, Kostanjeviska 16a, SI-5000 Nova Gorica, Slovenia
    \label{SLOVENIJA2}}
\titlefoot{Department of Physics, University of Ljubljana,
     SI-1000 Ljubljana, Slovenia
    \label{SLOVENIJA3}}
\titlefoot{Fysikum, Stockholm University,
     Box 6730, SE-113 85 Stockholm, Sweden
    \label{STOCKHOLM}}
\titlefoot{Dipartimento di Fisica Sperimentale, Universit\`a di
     Torino and INFN, Via P. Giuria 1, IT-10125 Turin, Italy
    \label{TORINO}}
\titlefoot{INFN,Sezione di Torino and Dipartimento di Fisica Teorica,
     Universit\`a di Torino, Via Giuria 1,
     IT-10125 Turin, Italy
    \label{TORINOTH}}
\titlefoot{Dipartimento di Fisica, Universit\`a di Trieste and
     INFN, Via A. Valerio 2, IT-34127 Trieste, Italy
    \label{TRIESTE}}
\titlefoot{Istituto di Fisica, Universit\`a di Udine and INFN,
     IT-33100 Udine, Italy
    \label{UDINE}}
\titlefoot{Univ. Federal do Rio de Janeiro, C.P. 68528
     Cidade Univ., Ilha do Fund\~ao
     BR-21945-970 Rio de Janeiro, Brazil
    \label{UFRJ}}
\titlefoot{Department of Radiation Sciences, University of
     Uppsala, P.O. Box 535, SE-751 21 Uppsala, Sweden
    \label{UPPSALA}}
\titlefoot{IFIC, Valencia-CSIC, and D.F.A.M.N., U. de Valencia,
     Avda. Dr. Moliner 50, ES-46100 Burjassot (Valencia), Spain
    \label{VALENCIA}}
\titlefoot{Institut f\"ur Hochenergiephysik, \"Osterr. Akad.
     d. Wissensch., Nikolsdorfergasse 18, AT-1050 Vienna, Austria
    \label{VIENNA}}
\titlefoot{Inst. Nuclear Studies and University of Warsaw, Ul.
     Hoza 69, PL-00681 Warsaw, Poland
    \label{WARSZAWA}}
\titlefoot{Now at Department of Physics, University of Warwick,
     Coventry CV4 7AL, UK
    \label{WARWICK}}
\titlefoot{Fachbereich Physik, University of Wuppertal, Postfach
     100 127, DE-42097 Wuppertal, Germany \\
\noindent
{$^\dagger$~deceased}
    \label{WUPPERTAL}}
\nopagebreak
\clearpage

\headsep 30.0pt
\end{titlepage}

%
\pagenumbering{arabic}                              
\renewcommand{\multirowsetup}{\centering}           
\setcounter{footnote}{0}                            %
\large
\newcommand{\aerr}[2]{\mbox{$^{+#1}_{-#2}$}}
\setlength{\extrarowheight}{2pt}
\section{Introduction and overview}
\label{SECT-INTRO}
The fragmentation  of a  $\mathrm{b\overline{b}}$~quark pair from $\Zz$~decay, into jets of particles including the parent b-quarks
bound inside b-hadrons,  is a process that can be viewed in two stages. 
The first stage involves the b-quarks radiating  hard gluons 
at scales of $Q^2 \gg \Lambda_{{\rm QCD}}^2$~for which the strong coupling is
small $\alpha_s \ll 1$. These gluons can themselves  split into further
gluons or quark pairs in a kind of `parton shower'. 
By virtue of the small coupling, this stage can be described by perturbative QCD
implemented  either as exact QCD matrix elements or leading-log parton shower cascade models in
event generators.
As the partons separate, the energy scale drops to $\sim
\Lambda_{{\rm QCD}}^2$ and the strong coupling becomes large,
corresponding to a
regime where perturbation theory no longer applies. Through the
self interaction of radiated gluons, the colour field
energy density between partons builds up to the point where there is
sufficient energy to create new quark pairs from the vacuum. This
process continues with the result that colourless clusters of quarks
and gluons with low internal momentum become bound up together to form
hadrons. This `hadronisation' process represents the second stage of the
b-quark fragmentation which cannot be calculated in perturbation
theory and must
be modelled in some way.
In simulation programs this is made via a `fragmentation function' which, 
in the case
of b-hadron production, parameterises how
energy/momentum is shared between the parent b-quark
and its final state b-hadron. Important steps for the understanding
of the hadronisation mechanism are given in references \cite{hadroniz1, hadroniz2, hadroniz3, hadroniz4}.

The purpose of this study is to measure the
non-perturbative contribution to b-quark fragmentation in a way that
is independent of any non-perturbative hadronisation model. Up to the choice of either
QCD matrix element or leading-log parton shower to represent the
perturbative phase, results are
obtained that are  applicable to any b-hadron production
environment in addition to the  \Ztobb~data on which the
measurements were made.

Results from two  analyses are reported which measure
the b-quark fragmentation function from the data taken
in 1994 by the DELPHI detector at LEP. 
Several definitions of the functions and variables used in the measurement
of the b-quark fragmentation distribution are given in Section 
\ref{FRAGFUN}. Section \ref{delphidet} contains a short description
of the DELPHI detector with emphasis on components which are
relevant for the present measurement.
Section~\ref{SECT-XWEAK}
describes how two different approaches 
(Regularised Unfolding and  Weighted Fitting) have
been used to extract from the data the underlying energy distribution
of weakly decaying b-hadrons. These measurements are then combined in
Section~\ref{SECT-COMBINE} and
interpreted (in Section~\ref{SECT-NONPERTB}) as the combined result of a perturbative and a 
 non-perturbative part. Corresponding fragmentation functions are
 determined  by (a) finding the best fit to the data with a full simulation
 of the hadronisation process, where the perturbative contribution is
 made by a parton shower model,
and (b) by describing the perturbative 
part with a NLL QCD calculation and using the inverse Mellin
transformation to solve for the non-perturbative part.
Present measurements are combined in Section \ref{sec:world_average} 
with previous
experimental results to obtain a world averaged  b-quark
fragmentation distribution.

\section{Fragmentation functions}
\label{FRAGFUN}

Various models of the hadronisation process have been incorporated into
simulation packages in the past with varying degrees of success in 
reproducing the data. In practice these models are implemented 
via a {\it fragmentation function} $D^{\mathrm B}_{\mathrm b}(v)$ (parameterised
in terms of some kinematical variable $v$), which can be 
interpreted as the probability density function that a hadron ${\mathrm B}$, containing 
the original quark b, is produced with a given value of $v$.
In order to reproduce the data accurately, the fragmentation function
must have an appropriate form  with parameters that are tuned to the data.

Although the definition of $v$~varies from model to model,
generally speaking it is  a quantity that reflects the 
fraction of the {\it available} energy that the b-hadron receives
from the hadronisation process.  For models
relevant to b-quark fragmentation from $\Zz$~decay, the choice of
fragmentation variable $v$~usually falls 
into one of two broad categories:
\begin{itemize}
\item $z$~is a fraction normalised to kinematical properties of the parent 
b-quark just before the  hadronisation process begins;
\item $x$~is a fraction normalised to the electron/positron beam energy 
i.e. $\sqrt{s}/2$. 
\end{itemize}
From a phenomenological point of view, $z$~is the relevant 
choice of variable for a parameterisation  implemented in
an event generator algorithm. However, because  $z$~depends explicitly  on the properties 
of the parent b-quark, it is not a quantity that can be directly
measured by experiments. For this reason all existing measurements of 
 $D^{\mathrm B}_{\mathrm b}(v)$~are based on the reconstruction of $x$.  

Throughout this paper, the Lund fragmentation
model~\cite{LUND}  definition of $z$ is employed.
In the Lund model, hadronisation is described by breaks in a 
{\it string} linking two partons which mimics the 
colour field energy density between them crossing the threshold for 
the creation of a new quark pair.
The fragmentation variable, for the case of an initial ${\mathrm b}\overline{{\mathrm b}}$~quark system in the absence
of gluon radiation, is defined as
\begin{eqnarray}
\label{eqn:defz}
  z & = & \frac{(E+p_{||})_{\mathrm B}}{(E+p)_{\mathrm b}}.
\end{eqnarray}
Here, $p_{||}$~represents the hadron momentum in the direction of the
b-quark and $(E+p)_{{\mathrm b}}$~is the sum of the energy and momentum of the
b-quark just before fragmentation begins.

When discussing $x$, it is necessary  to be clear about
exactly which b-hadron is being considered.
The {\it primary b-hadron} is the state created directly after the
hadronisation phase, whereas the {\it weakly decaying b-hadron} is the state that finally
decays somewhere in the detector volume in a flavour-changing process.
Primary b-hadrons are either mesons (about $90\%$) or baryons (about
$10\%$)\cite{PDG}. In the case of mesons, measurements suggest that
about $25\%$ of  primary b-hadrons are orbitally excited
$\Bsstar$~mesons \cite{ref:bsstar1,ref:bsstar2}, about $52\%$ are $\Bstar$~mesons and 
only about $18\%$ are weakly decaying $\BP,\,
\BD$ or $\BS$ mesons \cite{ref:otherb1,ref:otherb2,ref:otherb3}.
 $\Bsstar$~and  $\Bstar$~mesons decay via kaon, pion or photon emission into
weakly decaying ground state mesons, which then carry less energy than their
parents. For both analyses presented here, the  b-hadron under consideration
is always the weakly decaying state.
Two  choices for the  $x$~fragmentation variable in common
use are $\xb$~and  $\xp$:
\begin{eqnarray}
\xb  & = & \frac{E_{\mathrm B}^{\mathrm weak}}{E_{\mathrm b}}
\end{eqnarray}
is the fraction of the energy taken by the  b-hadron
with respect to the energy  of the b-quark directly after its
production i.e. before any gluons have been radiated. This definition
is particularly suited to $e^+e^-$~annihilation as both the
numerator and denominator are directly  observable. This follows
since, in the absence of initial state radiation, the 
quark energy is equal to the electron beam energy:
\begin{eqnarray}
\xb  & = & \frac{ 2E_{\mathrm B}^{\mathrm weak} }{\sqrt{s}}= \frac{E_{\mathrm B}^{\mathrm weak}}{E_{beam}}.
\end{eqnarray}
The variable $\xp$~is defined as the ratio of the three momenta ($p$) which, assuming
$m_{\mathrm B}=m_{\mathrm b}$, can be expressed as,
\begin{eqnarray}
\xp  & = & \frac{p_{\mathrm B}^{\mathrm weak}}{p_{{\mathrm B},max.}^{\mathrm weak}}=\frac{ \sqrt{{\xb}^2-x_{min}^2} }{\sqrt{1-x_{min}^2}}
\end{eqnarray}
where $x_{min}=\frac{2m_{\mathrm B}}{\sqrt{s}}$~is the minimum value of $\xb$
and $p_{{\mathrm B},max.}^{\mathrm weak}$ is the maximum momentum
taken by the b-hadron assuming that its energy is equal to the beam energy.

\section{The DELPHI detector and b-tagging}
\label{delphidet}
A complete overview of the  DELPHI detector and its
 performance
have been described elsewhere~\cite{delphi,delphi_perf}. What follows is a
short description of the elements most relevant to this analysis.

In the barrel region, charged particle tracking was performed by the  Vertex Detector (VD), the
Inner Detector, the Time Projection Chamber (TPC) and the Outer Detector. 
In the end-cap regions, two sets of drift chambers (FCA and FCB) were situated
at about 160 cm and 275 cm from the interacion point (IP) respectively.
They covered polar angles, $\theta$, in the range $\left [11^{\circ},~36^{\circ}\right ]$
and $\left [144^{\circ},~169^{\circ}\right ]$\footnote{The DELPHI coordinate system is right
handed with the $Z$-axis 
collinear with the incoming electron beam and the $X$-axis pointing
to the center of the LEP accelerator. The radius and azimuth in the 
$XY$ plane are denoted by $R$ and $\phi$, and $\theta$ is the polar angle 
to the $Z$-axis.}.
A highly uniform magnetic field of 1.23~T parallel to the $e^+e^-$~beam
direction, was provided by the  superconducting solenoid throughout the
tracking volume. The momentum of charged particles was measured with a
precision of  $\sigma_p/p \leq 1.5 \%$~in the $\theta$ region 
 $\left [40^\circ,~140^{\circ}\right ]$~and 
for $p<10$~GeV/$c$.
The VD  consisted
of three layers of silicon micro-strip devices with an  intrinsic resolution
of about 8 $\micron$ in the $R-\phi$~plane  transverse to the beam line. In
addition, the  inner- and outer-most layers were instrumented with
double-sided devices providing coordinates of similar precision
in the $RZ$~plane along the direction of the beams.  For charged particles with hits in all three $R\phi$ VD layers the impact parameter  resolution was $\sigma_{R\phi}^2=([61 / (p \sin ^{3/2} \theta)]^2 + 20^2) \,\, \micron^2$    and for tracks with hits in both $RZ$ layers and with $\theta \approx 90^{\circ}$, $\sigma_{RZ}^2=([67 / (p \sin ^{5/2} \theta )]^2 + 33^2) \,\, \micron^2$ ($p$ is in GeV/$c$).

Calorimeters detected photons and neutral hadrons by the total absorption
of their energy. The High-density Projection Chamber (HPC) provided
electromagnetic calorimetry coverage in the region
$46^{\circ}<\theta<134^{\circ}$~giving a  relative precision on the measured
energy $E$~of  $\sigma_E/E=0.32/\sqrt{E} \oplus 0.043$~($E$~in GeV). In
addition, each HPC module worked essentially as a small TPC charting the
spatial development of showers and so providing an improved angular resolution,
which is better than that from the detector granularity alone.
For high energy photons
the angular precisions were $\pm 1.7$~mrad in the azimuthal angle $\phi$~and
 $\pm 1.0$~mrad in $\theta$. The Forward Electromagnetic 
Calorimeter consisted of two arrays of 4532 Cherenkov lead glass blocks
with 20 radiation lengths. The front faces of the blocks were placed at 
$\pm$284 cm from the IP, covering the polar angle in the ranges
$[8^{\circ},~35^{\circ}]$ and $[145^{\circ},~172^{\circ}]$.
The relative precision on the measured energy could be parameterised
as $\sigma_E/E=0.03 \oplus  0.12/\sqrt{E} \oplus 0.11/E$~($E$~in GeV).
For neutral showers of energy larger than 2 GeV, the average precision
on the reconstructed hit position in X and Y was about 0.5 cm.
The Hadron Calorimeter was installed in the return yoke of the  DELPHI solenoid and provided a relative precision on the  measured energy of $\sigma_E/E=1.12/\sqrt{E} \oplus 0.21$~($E$~in GeV).

Powerful particle identification was made possible by the combination of
$dE/dx$~information from the TPC (and to a lesser extent from the VD) with
information from the Ring Imaging CHerenkov counters (RICH) in
both the forward and barrel regions. The RICH devices utilised both liquid
and gas radiators in order to optimise coverage across a wide momentum range:
liquid was used for the momentum range from 0.7 GeV/$c$ to 8 GeV/$c$ and the
 gas radiator for the  range 2.5 GeV/$c$ to 25 GeV/$c$. 

The impact parameters provided the main variable for b-tagging.
For all the charged particle tracks in the jet, the impact parameters and 
resolutions were 
combined into a single variable, the lifetime probability, which measured the
consistency with the hypothesis that all tracks come directly from the primary
vertex. For events without long-lived particles, this variable should be 
uniformly distributed between zero and unity. In contrast, for b-jets
it has predominantly small values. This information is used in the
weighted fitting algorithm whereas additional characteristics of 
$\mathrm b \overline{b}$-events are included in the other approach.
Other features of the event are also sensitive to the presence of b-quarks,
and some of them are used together with the impact parameters information to construct
a `combined' tag. For example, b-hadrons have a 10$\%$ probability
of decaying to electrons or muons, and these often have a transverse momentum
with respect to the b-jet axis of around 1 GeV/$c$ or larger. The 
combined tag also makes use of other variables that have significantly 
different distributions for b-quark and for other events, e.g. 
the charged particle rapidities with respect to the jet axis. Further
details on the b-tagging algorithm can be found in reference
\cite{btag}.

In the analyses described in this paper, the primary and the secondary vertices are reconstructed in 3 dimensions.

\section{Measuring $f(\xb)$}
\label{SECT-XWEAK}
This paper describes two independent methods of reconstructing 
$\xb$~from the data: one which unfolds the underlying physics
distribution from the measured quantity and one which fits for the 
physics distribution by a weighting technique.
The former is described in Section~\ref{sect:unfold} and the latter in Section~\ref{sec:Orsay}.
The two methods differ also in the way particles are classified as originating from a b-hadron decay or from fragmentation. The first method is using extensively Neural Networks whereas the second is based on different techniques. 
Both methods 
are  independent of any initial
assumption regarding  the actual shape of the underlying fragmentation
function in simulation.
Throughout this section all charged particles are assumed to be pions,
and for photons and neutral hadrons we use the candidates measured in
calorimeters as described in Section~\ref{delphidet}.

\subsection{The regularised unfolding analysis}
\label{sect:unfold}
The experimental challenge of this method is to determine from the measured
distribution in data\footnote{
Throughout the paper, the subscripts and the superscripts
rec, gen and sim designate, respectively, reconstructed quantities (in data or simulation),
generated ``true'' values and quantities from the simulation.}
$g(\xbrec)$, the underlying fragmentation function  $f(\xb)$.
In general $g(\xbrec)$~will differ from $f(\xb)$~due to:
\begin{itemize}
\item[(a)] finite detector resolution;
\item[(b)] limited measurement acceptance;
\item[(c)] variable transformation, i.e. any biases or distortions that may
be present in the measured quantity.
\end{itemize}
Mathematically, the distributions are related by:
\begin{equation}
  \label{gAfb}
  g(\xbrec)  =    \int R(\xbrec;\xb)f(\xb)d\xb+b(\xbrec)~,
\end{equation}
where $R(\xbrec;\xb)$~is the {\it response function}
which describes the
mapping  of  $\xbrec$~onto true $\xb$~and thus contains all the effects of
resolution, acceptance and variable transformation mentioned above.
 The term $b(\xbrec)$~is the background
contribution and is taken from simulation.

\subsubsection{Hadronic event selection}
Hadronic $\Zz$~decays were selected by the following requirements:
\begin{itemize}
   \item[(a)]  at least 5 reconstructed charged particles;
   \item[(b)]  the summed energy in charged particles
         with momentum greater
         than 0.2~GeV/$c$ had to be larger than 12\% of the centre-of-mass
         energy, with at least 3\% of it in each of the forward and
         backward hemispheres defined with respect to the beam axis.
\end{itemize}
These requirements resulted in the selection of  about
1.36 million events from data.
The simulated sample of \Ztoqq~events, details
of which are listed in Table~\ref{tab:ourmc},
contained approximately three times the number of data events.
The generated events were passed through a full detector
simulation\cite{delphi_perf} and the same multihadronic
selection criteria as the data.
\begin{table}[h!]
\begin{center}
\begin{tabular}{|l|l|} \hline
Event Generator            & JETSET 7.3\cite{jetset1,jetset2}         \\
Perturbative ansatz        & Parton shower ($\Lambda_{{\rm QCD}}=0.346$~GeV,$Q_0=2.25$~GeV)\cite{tuningdelphi} \\
Non-perturbative ansatz    & String fragmentation \\
Fragmentation function     & Peterson~\cite{peterson} ($\epsilon_{\mathrm b}=0.002326$) \\
Bose-Einstein correlations & Enabled \\
\hline
\end{tabular}
\caption
{\label{tab:ourmc} Details of the
event generator used together
with some of the more relevant parameter values that have been
tuned to the DELPHI data.}
\end{center}
\end{table}

\subsubsection{Event hemisphere selection}
\label{hemsel}
In each event, particles are distributed in two hemispheres
depending on their direction relative to the thrust axis.
Event hemispheres used for the analysis were accepted if
the following criteria were fulfilled:
\begin{itemize}
   \item[(a)]   $|\cos \theta_{thrust}|<0.7$, where $\theta_{thrust}$
is the polar angle of the event thrust axis relative to the beam direction;
   \item[(b)]   the hemisphere was tagged as a \Ztobb~candidate  event by the
         standard DELPHI b-tagging package\cite{btag};
   \item[(c)]   the secondary vertex fit converged successfully;
   \item[(d)]   $0.5< E_{hem}/E_{beam}<1.1$
where $E_{hem}$ is equal to the sum of the energy of particles contained
in the hemisphere.

\end{itemize}
After this selection, 227940 hemispheres remained in the data with a purity
(as calculated from the simulation)
in ${\mathrm b\overline{b}}$~events of 96\%.
\subsubsection{The reconstruction of $\Eb$}
\label{EB}

The following corrections were applied to the
simulation to
account for known discrepancies with the data which could affect modelling of
the B-energy scale:
\begin{itemize}
\item[(a)] The reconstructed energy distributions per charged or neutral
particle
were separately shifted and smeared\footnote{For charged particles the shift
in the mean was 0.01~GeV and a Gaussian smearing of 3\% (relative) applied.
For neutral clusters the corresponding numbers were 0.04~GeV and 20\%.}
in the simulation to bring them
into better agreement with the data
(based on a $\chi^2$-histogram comparison).
\item[(b)] The multiplicities of:
\begin{itemize}
\item fragmentation charged particles (identified by a selection cut on the
TrackNet\footnote{The TrackNet is a neural network trained to
distinguish between charged particles from the b-hadron decay chain and
those originating from the event primary vertex. See also Appendix A.}$<0.5$),
\item b-hadron weak decay products (identified by a selection cut on the TrackNet$>0.5$),
\item neutral particles,
\end{itemize}
were fixed separately in the simulation by a weighting function, 
to agree with the data.
\item[(c)] After applying the above two corrections, a very small 
residual difference remained between data and simulation in the total energy 
of charged particles (``charged energy'') and neutral particles (``neutral 
energy'') which was accounted for by a further weighting function.
\end{itemize}

The  energy $\Eb$ of a b-hadron undergoing weak decay within the
hemisphere of a $\Zz$~hadronic-decay
event, was reconstructed using the Neural Network (NN) package,
Neurobayes~\cite{NEUROBAYES}. The full list of variables
that the NN was trained on is presented in  Appendix A.
Since the degree of correlation of the
inputs to the network target value naturally varies from case to case,
a pre-processing stage to the network algorithm was used to
suppress
the influence of the inputs with low correlation automatically 
and so retain optimal performance.
The network was trained to return a complete  probability 
density function (p.d.f.) for the energy, on a hemisphere-by-hemisphere basis, 
and  $\Eb$~was defined to be the median of this
distribution. Full details of this approach can be found in reference~\cite{NEUROBAYES}.

\begin{figure}[ht]
\begin{center}
\includegraphics[width=8.0cm]{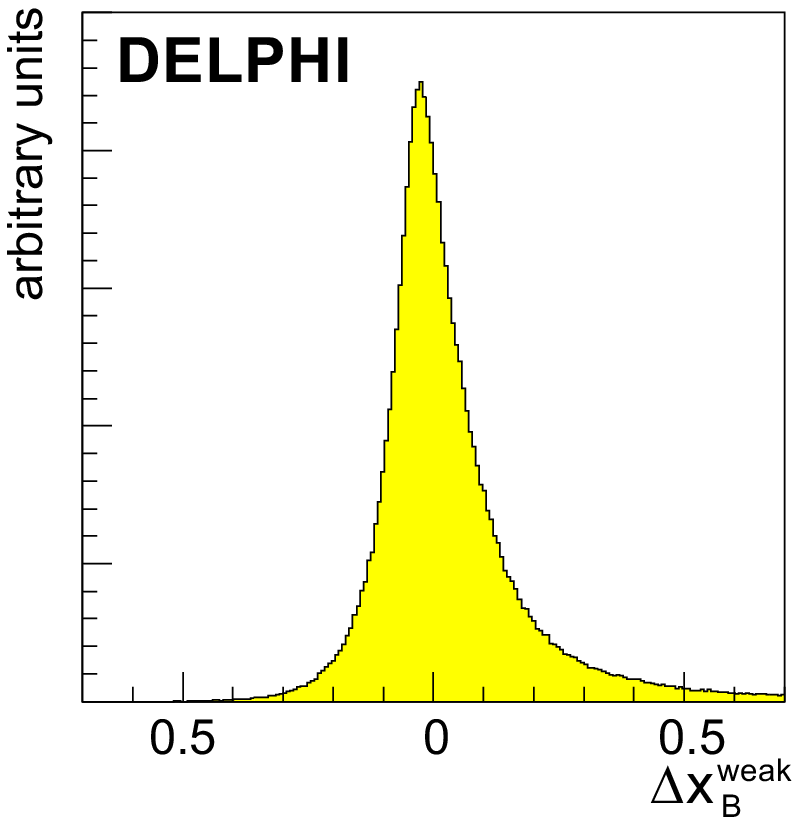}
\caption[]{Distribution of the precision of the NN estimator for  $\xb$, 
defined as $\Delta\xb=(\xbrec-\xbgen)/\xbgen$.
\label{resolution}} 
\end{center}
\end{figure}

The precision of the resulting estimator, based on a statistically
independent
simulated event sample to that used for training
and after all  analysis selection cuts have been applied,
is shown  in Figure~\ref{resolution}. The full width at half maximum is 
$14.0\%$.

\subsubsection{The unfolding method}
\label{sec:unfold}

The solution of Equation~(\ref{gAfb}) for $f(\xb)$~is a non-trivial
problem since the solution can be highly oscillatory.
A practical solution to this is provided by the 
{\sc Run} ({\bf R}egularised {\bf UN}folding) program \cite{run}
which applies regularisation techniques to impose the condition
that the solution must be smooth. 
In practice, the algorithm defines a function $W(\xb)$~used to provide a  weight to
the simulated distribution  $g_{sim}(\xbrec)$~such that it
reproduces the data distribution  $g(\xbrec)$~as well as possible, i.e.
$W(\xb)$~is determined by a {\it fit} to the data.
The result of the unfolding, up to a normalisation factor, is then given by
\begin{eqnarray}
\label{resspline}
f(\xb) & = & W(\xb)\cdot f_{sim}(\xb)
\end{eqnarray}
where $f_{sim}(\xb)$~is the fragmentation function  used to generate the 
simulated events. By summing over bins in $\xb$, unfolded binned points 
are determined together with a complete covariance matrix.

It is important to note that internally to  {\sc Run}, the 
weight factors are defined as a sum over orthogonal
polynomials taken to be basis splines $P(\xb)$,
\begin{eqnarray}
\label{weightsum}
W(\xb) & = & \sum_{j=1}^{m} a_j \cdot P_j(\xb)
\end{eqnarray}
where $a_j$~are suitable expansion coefficients.
Consequently, the difficult task of solving~(\ref{gAfb}) reduces to 
deciding at which point to cutoff the sum in~(\ref{weightsum}). 
This point, $j=m$, 
is referred to in what follows as the {\it number of degrees of freedom} of 
the unfolding procedure. 
Full details of the unfolding method can be found in reference~\cite{unfolding2}.


\begin{figure}[ht]
\begin{center}
\leavevmode
\includegraphics[width=1.00\textwidth]{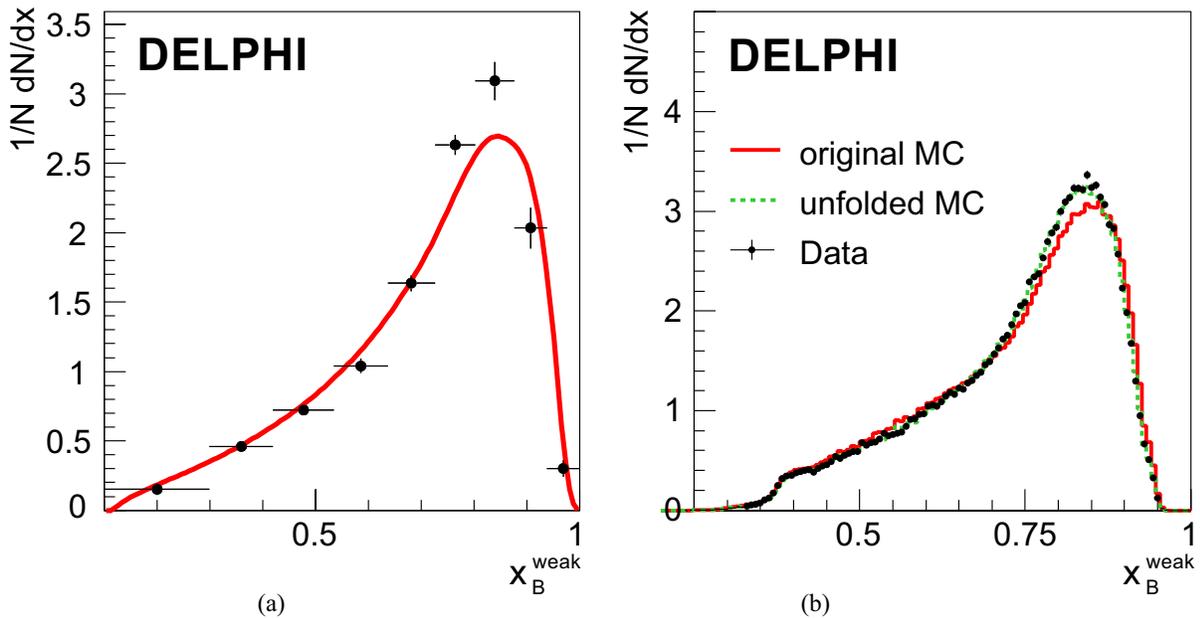}
\caption
{\label{ufoerg} {\bf a} The result of unfolding $\xb$~from
real data (points), and the generator-level $f_{sim}(\xb)$~distribution,
before applying weights (curve). {\bf b} Distribution of $\xb$~in the data,
$g(\xbrec)$, compared to both the default simulation $g_{sim}(\xbrec)$ and the 
simulation weighted for the results of the fragmentation function
unfolding result shown in {\bf a}. }
\end{center}
\end{figure}
\subsubsection{Unfolding results}
\label{results}
The result of the unfolding applied to the real data set is displayed in
Figure~\ref{ufoerg}a.
The plot shows the unfolded,  binned,
data points together with an overlay of the  `truth' or  generated
$f(\xb)$~distribution that is the input to the detector simulation.
The binning of the unfolded points was chosen to match
the observed resolution in $\xb$~according to the measurement  uncertainties
 described in Section~\ref{EB}. 
For the case of $\xb$~the median (relative) error varied from about 
$5\%$~at an $\xb$~value close to $1.0$~and degraded to about 
$65\%$~at~$\xb=0.2$. The number of degrees of freedom in the unfolding
procedure was chosen to be as low as possible (i.e. five) in order to ensure a 
smooth result. The lower limit is constrained by the need to include all
terms in the summation~(\ref{weightsum}) for which the  size of the 
expansion coefficients  $a_j$~are significant.

The results show that there is a basic disagreement in shape between the
distribution unfolded from data and the corresponding truth distribution from
the simulation before the application of weights.
Figure~\ref{ufoerg}b shows the excellent agreement that exists between 
data and 
simulation after   appropriately weighting  
the generator distribution to agree with the result  of the unfolding.

In order to quantify the shape of the unfolded distribution,  the mean
($\langle x \rangle  =  \int_0^1 x f(x) dx$) and
variance ($\sigma^2(x) = \int_0^1 (x-\langle x \rangle)^2 f(x) dx $)
have been calculated and the  results were:  $\langle \xb \rangle=0.7140 \pm 0.0007
({\rm stat.})$~and $\sigma^2 ( \xb ) =0.0308 \pm 0.0003 ({\rm stat.})$.
The mean value quoted has been
corrected to account for the effect of Initial State Radiation (ISR)
which is necessary since $x$~is formed by scaling
$\Eb$~by the nominal beam energy of $45.6$~GeV. This is
only strictly correct in the case of no ISR and in  about $ 10\%$~of cases ISR 
reduces the energy available for the fragmenting b-quark system from the
nominal value.
The size of this effect on the analysis was evaluated from the simulation
and the resulting mean value 
for $\Eb$ was shifted by $+50$~MeV. The corresponding 
shift of $\avxb$ is $\delta \avxb = + 0.0011$.
The full bin-to-bin unfolding results including
covariance matrices, are listed in Appendix B.

\subsubsection{Systematic uncertainties}
\label{systematics}
Systematic uncertainties on the unfolded distribution of $\xb$ have been 
evaluated from a wide variety of sources, the effects of which on 
$\avxb$ are presented in Table~\ref{errors}.
In addition, statistical and systematic
uncertainties for each of the nine unfolded bins of Figure~\ref{ufoerg}a
are given in Appendix B together with the associated
covariance matrices.

{\bf \underline {Technical systematics:}}

Some crosschecks of the method were made on the simulation to ensure 
that the result of the unfolding was independent of the 
prior fragmentation function embedded in the simulation.
In addition, an investigation
was made of the sensitivity to the following technical aspects
of the {\sc Run} unfolding procedure:
\begin{itemize}
\item[(a)] The number of degrees of freedom, defined in 
Section~\ref{sec:unfold}, was increased from five (default value) to seven. 
The change in the results seen was 
then assigned as a systematic uncertainty to account for the 
degree of uncertainty present in determining at which point
to terminate the summation described in Equation~(\ref{weightsum}). 
\item[(b)] The number of knots in the basis spline 
representation of the
weight $W(\xb)$~(defined in Equation~(\ref{weightsum})) was varied
and found to have a negligible effect on the results.
\item[(c)] The binning of the reconstructed variable, 
$g(\xbrec)$ in Equation~(\ref{gAfb}),  
should be well matched to the resolution achieved in order to use
the information optimally. A wide range of different binnings
around the default choice was investigated and the results
found to be  consistent within the total 
systematic uncertainties quoted. Also, 
no improvement on the statistical precision was found.
\end{itemize}

{\bf \underline {Selection cuts and background dependence:}}

The hemisphere selection described in Section~\ref{hemsel}, includes 
selection cuts for
${\mathrm b\overline{b}}$~event enhancement and on the reconstructed scaled
hemisphere energy $E_{hem}/E_{beam}$, both of which could potentially have an
effect on the analysis if not accurately modelled in the simulation.
The DELPHI b-tagging is based on impact parameter
measurements which degrade at low momenta due to the increased
effects of multiple scattering. This effect correlates
the b-tagging information to the B-energy. Any  
variation in the
unfolding result was checked when scanned over a wide range of
b-tagging selection cuts i.e. different ${\mathrm b \overline{b}}$ purities.
The results were found to be stable around the working point
of ${\mathrm b \overline{b}}$ purity $\approx$ 96\%. In addition,
the effect of scanning around the nominal selection cut value of
$E_{hem}/E_{beam}=0.5$~was investigated and the results found to be stable.
No explicit systematic was assigned due to these two analysis selection cuts.

Uncertainties in the size  and composition of the background,
i.e.  $b(\xbrec)$ in Equation~(\ref{gAfb}), were also evaluated.
Approximately 75\% of the background was from non-${\mathrm b\overline{b}}$
events, primarily ${\rm c\overline{c}}$~events, which was accounted for as one of the b-physics
modelling weights described later.
The remainder was composed of cases where both b-quarks
were found in the same hemisphere which occasionally happens e.g. in three-jet
events or when a gluon splits into two b-quarks
leaving a topology with four b-quarks in the initial state.
In these cases, which occur in about $2\%$~of all hemispheres,
the connection between the generated
b-hadron energy and the reconstructed quantity becomes
confused and hence were assigned to the background.
It is assumed that the overall jet rate is well modelled in the 
simulation
but the gluon splitting rate to ${\mathrm b\overline{b}}$ is varied, from the default
value of $0.5\%$
by $\pm 50 \%$ \cite{ref:gluontobb}, 
and the change seen in the unfolding result is recorded
as a systematic uncertainty.

{\bf \underline{Reconstructed energy:}} 

\label{esys}
The relationship between the reconstructed variable distribution
in the simulation, $g_{sim}(\xbrec)$, and the underlying
physics p.d.f., $f_{sim}(\xb)$, is
\begin{equation}
  \label{response}
  g_{sim}(\xbrec)  =    \int R(\xbrec;\xb)f_{sim}(\xb)d\xb~,
\end{equation}
where $R(\xbrec;\xb)$~is the response function
defined in Equation~(\ref{gAfb}).
The unfolding
is, by construction, insensitive to details of the prior
fragmentation function $f_{sim}(\xb)$~but only under the assumption
that the  response function,  as derived from the 
simulation, is correct.
 It is therefore crucial that $R(\xbrec;\xb)$~be
as close to the situation in the data as possible. 
Separate uncertainty contributions
were assigned for each of the three
corrections, described in Section~\ref{EB}, that affect directly modelling of the
B-energy scale.
Half of the full change in the result was taken as an uncertainty 
when: (a) the  shifting/smearing procedure was turned off, (b) the spread
of the multiplicity weights of about $1.0$~was
changed by $\pm50\%$ and
(c) the  hemisphere energy weight was switched  off.

Since the multiplicity tuning was dependent on a specific selection cut 
on the TrackNet variable around the $0.5$~point, it was checked that 
the results were not a strong function of this choice. The 
multiplicity weights were recalculated based on considering three regions
in the TrackNet variable i.e. TrackNet$<0.2$,  $0.2<$TrackNet$<0.8$
and TrackNet$>0.8$~and the analysis repeated. The results were 
found to be consistent to well within the quoted systematic uncertainties 
and no
additional uncertainty was assigned.

A further crosscheck was made by using a
different choice for $\Eb$~other than the Bayesian neural network
variable described in Section~\ref{EB}. For this test, $\Eb$~was
estimated by applying a rapidity algorithm (described in Appendix A) and
corrected for missing neutral energy based on a parameterisation from 
the simulation.
A detailed description of this correction is given elsewhere~\cite{ref:delphiLifetimes}.
Repeating the analysis, the
change seen in the result for $\xb$~was -0.0011, well
contained within the assigned total systematic uncertainty.

\begin{table}[htb]
\begin{center}
\begin{tabular}{|c|c|c|} \hline
 uncertainty class& item &  $\delta \avxb $ \\
  \hline
technical &  number of degrees of freedom      &   $ +0.0025 $     \\
\hline
\multirow{2}{4cm}{selection cuts and backg. dependence}&   ${\rm g\to b\overline{b}}$  &   $ +0.0004 $     \\
  & & \\
\hline
\multirow{5}{4cm}{reconstructed energy}&  neutral energy smearing       &   $~~0.0023 $     \\
  &  fragmentation track multiplicity             &   $ +0.0030 $     \\
  & b-decay track multiplicity                   &   $ -0.0004 $     \\
 & neutral multiplicity                         &   $ +0.0010 $     \\
 &  hemisphere scaled energy $E_{hem}/E_{beam}$ &   $~~0.0003 $     \\
\hline
\multirow{10}{4cm}{b-physics modelling}&  b-hadron lifetimes                &   $ -0.0004 $     \\
 &  b-hadron production fractions     &   $~~0.0002 $     \\
 &  hemisphere quality                           &   $ -0.0018 $     \\
 & $\Bsstar$~rate                               &   $ -0.0018 $     \\
 & $\Bsstar$~$Q$-value dependence               &   $~~0.0003 $     \\
& $\rm K^0$ rate                               &   $ +0.0005 $     \\
&  $\Bstar$ rate                                &   $ -0.0001 $     \\
&  semileptonic decay rate                      &   $ -0.0001 $     \\
&  wrong sign charm rate                        &   $ +0.0001 $     \\
&  c- and b-quark efficiency                    &   $~~0.0001 $     \\
\hline
\multirow{2}{4cm}{calibration stability \& simulation statistics} &  calibration periods    &   $~~0.0025 $     \\  
 &  finite simulation statistics    &   $~~0.0005 $     \\
 
  \hline
Total &                  &   $~~0.0060 $     \\

  \hline
\end{tabular}

\caption
{\label{errors} Systematic uncertainty on the mean value of the  unfolded
 $\xb$~distribution. The total is the sum in
quadrature of all contributions. The sign indicates the correlation between
the change in an uncertainty  source and the shift in the final result.
Uncertainties assigned by turning a weight on/off have no sign. }
\end{center}
\end{table}

\clearpage
{\bf \underline{b-physics modelling:}} 

\label{mcweights}
The remaining systematic contributions
concern quantities for which the simulation was
weighted in order to account for known discrepancies
with the data. Weights were constructed to change the
lifetimes and production fractions of the b-hadron
species to more recent world average values~\cite{PDG}:
\begin{tabbing}
ttttt\=tttttttttttttttttttttttttttttttttttttttt\=tttttttttttttt \kill
\>$\tau(\BP)=1.638 \pm 0.011$ ps,      \> $f(\BP)=(39.9 \pm 1.1)$\% \\
\>$\tau(\BD)=1.530 \pm 0.009$ ps,    \> $f(\BD)=(39.9 \pm 1.1)$\% \\
\>$\tau(\BS)=1.470 \pm 0.027$ ps,    \> $f(\BS)=(11.0 \pm 1.2)$\% \\
\>$\tau({\mathrm b-baryon})=1.383 \pm 0.049$ ps, \> $f({\mathrm b-baryon})=(9.2 \pm 1.9)$\% .
\end{tabbing}
Systematic uncertainties from these sources were based on
varying them within the quoted one standard deviation uncertainties 
for the case of
the lifetimes and by switching the weights on/off for the case of
the production fractions.
The `hemisphere quality' was a quantity flagging
the presence of potentially badly reconstructed tracks
in the hemisphere. Improved agreement with the data
was achieved in many reconstructed quantities
by weighting the hemisphere quality distribution in
the simulation to agree with that seen in data. The
change induced by varying  the spread of the weight
around $1.0$~by $\pm 50\%$~from the  nominal value
was assigned as a systematic uncertainty.

By default the production rate of excited $\Bsstar$~states
was adjusted in the simulation to be $25\%$ per $\B$~meson hemispheres.
This rate was then
varied from $15\%$ to $35\%$~and half the total change seen in the results,
assigned as a systematic. 
In addition, sensitivity to the $\Bsstar$~$Q$-value\footnote{The $Q$-value is defined as:
$Q= m(\B^{**})-m(\B)-m({\rm T})$,~where e.g. for $\B^{**}\rightarrow \B^+ \K^-$,
$\B$ is the $\B^+$ and ${\rm T}$ is the $\K^-$. It is therefore the kinetic
energy available in the decay process for the decay products to take. }
was tested by applying a weight to force the simulated $Q$-value distribution to be that suggested
by a previous DELPHI analysis~\cite{delphi_newb2star},
and the change in the results was assigned as a systematic uncertainty.

Systematic uncertainties from the $\Bstar$~rate, ${\rm K^0_S}$~rate and the b-hadron
semi-leptonic branching fraction were accounted for by changing their
values in the 
simulation by the same relative uncertainty quoted on
current world averages~\cite{PDG}. In addition an uncertainty was assigned
due to changing the `wrong-sign' $\Ds$~production rate, i.e. $\Ds$~production
from ${\rm W}^- \rightarrow \overline{c}s$~decay, by $100\%$.

Finally a weight was applied to the 
simulation based on
the results of a double hemisphere tagging analysis in order to correct
 the efficiency to tag \Ztocc~events and
 \Ztobb~events to that measured from
the data.
At the analysis working
point of ${\mathrm b \overline{b}}$~purity of $96\%$, the correction to the
b~efficiency was about $-2\%$~and the correction
to the c~efficiency about $-12\%$. A systematic from
this source was assigned to be the full difference
in the results when this weight was removed.

{\bf \underline{Calibration stability and simulation statistics:}}

A spread is observed in the results as a function of 
time slices dividing up the data.
The likely source of this effect is the
division of the period into different calibration periods of the
vertex detector and half of the full spread in results has been
assigned as a systematic uncertainty. 
The effect of having finite simulation statistics for the
determination of the transfer matrix was small and was evaluated by
varying the elements of the matrix up and down by one statistical
standard deviation. 
\subsection{The weighted fitting analysis}
\label{sec:Orsay}

The procedures used for b-hadron energy reconstruction and measurement of 
the b-hadron fragmentation
distribution are different from those applied in the previous approach.
The B hadron energy is obtained by subtracting the energy taken by fragmentation 
tracks from the reconstructed energy of the jet containing the B candidate.
The b-hadron fragmentation distribution is determined by fitting a weight 
distribution on simulated events such that the corresponding reconstructed B energy 
distribution agrees with the one measured using real data events.

\subsubsection{Hadronic event selection}
\label{sec:Orsayhadsel}

Hadronic $\Zz$ decays were selected using the following requirements:
\begin{itemize}
\item $|\cos{(\theta_{thrust})}|<0.95$;
\item at least 15 particles, charged and neutrals, reconstructed.

\end{itemize}

Charged particles from b-hadron decays
can be identified from other charged hadrons using their positive impact
parameter measured relative to the event main vertex. 
For a hadronic event
resulting from the hadronisation of light quarks, charged particle impact 
parameters are expected to be compatible with the beam interaction position.
A variable, $P_{\rm btag}$, has been used, which has a flat distribution for such 
events 
and which is peaked at low values for events containing heavy quarks whose 
decay generates charged particles with offsets \cite{btag}.
In Table~\ref{tab:btag} are given the fraction of selected events in data and 
simulation,
the expected fraction of non-$\mathrm b \overline{b}$ events and the efficiency for $\mathrm b \overline{b}$ events.
According to these values, samples of hadronic
events containing about 10$\%$ contamination from non-$\mathrm b \overline{b}$ events
can be isolated with an efficiency higher than 60$\%$ for those 
originating from b-quarks.
Remaining differences between real and simulated events have been included
in the evaluation of systematics.

\begin{table}[htb]
\begin{center}
  \begin{tabular}{|c|c|c|c|c|c|c|}
    \hline
selection on $P_{\rm btag}$  &$<10^{-3}$ &$<10^{-4}$ &$<10^{-5}$ & $<10^{-6}$&  $<10^{-8}$ & $<10^{-10}$\\
\hline
Data: fraction of &    & & & & & \\
 selected events ($\%$) &  17.6 &14.3 &11.8 &9.9 &6.8 &4.5 \\
    \hline
MC: fraction of &   & & & & & \\
 selected events ($\%$)   &17.2 &14.0 &11.5 &9.5 &6.4 &4.2 \\
    \hline
MC: b-purity ($\%$)  &88.7  &93.5 &96.1 &97.6 &99.0 &99.9 \\
    \hline
MC: b-efficiency ($\%$)  &69.4  &59.3 &50.1 &41.9 &28.6 &19.1 \\
    \hline
  \end{tabular}
  \caption[Variation of the selected event sample composition
and efficiency for $\mathrm b \overline{b}$ events versus the selection cut on the b-tagging variable.]
{{Variation of the selected event sample composition
and efficiency for $\mathrm b \overline{b}$ events versus the selection cut on the $P_{\rm btag}$-variable.}
  \label{tab:btag}}
\end{center}
\end{table}
In the following,
samples of hadronic events depleted in b flavour have been selected by a 
selection cut on the 
b-tagging probability ($P_{\rm btag}\geq$ 10 $\%$) evaluated for the whole event, whereas
b-enriched samples have been retained using $P_{\rm btag}\leq 10^{-3}$.
 
\subsubsection {b-hadron energy reconstruction}
\label{sec:OrsayBener}

The b-hadron energy is determined in two steps. Jets are firstly reconstructed 
and their energies are obtained from a constrained fit requiring energy-momentum
conservation for the whole event. Then, considering only those jets for which the
axis is inside the VD acceptance ($\left | \cos{\theta_{jet}}\right |<0.75$), 
particles are classified as B decay products or
fragmentation particles. For charged particles, their offsets relative to the 
event main vertex, and their rapidity measured relative to the jet axis
are used in this classification, whereas for neutrals only the rapidity is 
used. 

Differences between real and simulated events can originate from a 
behaviour of the detector that differs from its expected performances or
from different  particle production characteristics in the events. 
As the reconstruction accuracy for charged particles depends on the type
of sub-detectors used and as differences remain between
the fractions of sub-detectors involved in the data and in the simulation, 
corrections have been applied. The procedure, equivalent to the removal
of a sub-detector, consists in 
rescaling the values of measurement uncertainties 
and in smearing the 
corresponding track parameter values. These corrections, which apply to about
$ 4 \%$ 
of all charged particles, depend on the type
of the removed sub-detector and were determined using the simulation,
by comparing uncertainty matrix elements for tracks with and 
without the corresponding sub-detector involved.
In addition, as the mass distribution of reconstructed weakly decaying
particles (such as those corresponding to the $\Do$ or $\Dp$ mesons) has 
a width which is larger in real data by about $20~\%$,
a smearing corresponding to the same fraction of their measurement 
uncertainty has 
been applied to simulated tracks.

After these corrections individual
particle momentum distributions have been compared in real and simulated events.
These distributions considered separately for b-depleted and
b-enriched samples have been normalised using the respective number of 
selected hadronic events in each category.
To match corresponding data/simulation distributions a momentum dependent correction 
is then applied, which
consists in removing tracks 
alternatively in data or in the 
simulation depending if the measured ratio is larger or lower than unity.
This correction has been determined separately for b-depleted
and b-enriched samples and also, independently, for charged and neutral
particles.

To avoid a possible bias induced by a correlation between the assumed shape
of the fragmentation function and the applied correction, the latter
has been evaluated iteratively using as input
in its determination the fragmentation distribution
measured at the previous step. In practice one iteration was used, as the
observed absolute variation between the second and first step on the resulting
$\avxb$ value was of the order of $10^{-3}$.

In a given event, jets are reconstructed using the Lund LUCLUS algorithm 
\cite{ref:luclus} with the $d_{\rm join}$ parameter (${\rm PARU}(44)$)
value set to 5.0 GeV/$c$. 
A first evaluation of the jet energies is obtained using the jets directions,
energies, masses and imposing total energy-momentum conservation for
the whole event. 
If the missing energy in a jet is larger than 1 GeV, a 4-vector
is added to the jet. Its direction
is taken to be the same as the jet direction and the missing momentum 
is evaluated
assuming that the missing particle mass is zero. 
Analysing simulated events, the relative uncertainty on the missing energy is
measured to be 20$\%$,
and uncertainties on angles of the missing particle
are 50 mrad. 
Energy momentum conservation is then applied again
to the whole event,
and particle parameters (for charged, neutral and possibly missing) are fitted.
After this procedure, 4-vectors of charged and neutral particles have been fitted,
and possibly new 4-vectors corresponding to missing energy in each jet have been 
obtained. 
Jets are 
reevaluated ($p_{\mathrm jet}$)
using this set of tracks and applying the same LUCLUS algorithm.
Fractions of the fitted charged, neutral and missing energy are compared 
in Table \ref{tab:evtener}.
Relative differences are at the level of a few $10^{-3}$. 
A comparison between data and the simulation has been also made for
the averages and variances of charged and neutral particle multiplicities.
The results are given in Table \ref{tab:multiplicities}.

\begin{table}[p]
\begin{center}
  \begin{tabular}{|c|c|c|c|c|}
    \hline
$\mathrm b \overline{b}$-depleted events & & & & \\
    \hline
Sample &$E_{ch.}~+~ E_{neu.}$& $E_{ch.}$ & $E_{neu.}$ & $E_{miss.}$\\
    \hline
Data &$0.8644$ &$0.5759$ & $0.2879$ & $0.1363$ \\
MC &$0.8676$ &$0.5778$ & $0.2893$ & $0.1323$ \\
\hline
(Data-MC)/MC&$-0.0037$ &$-0.0033$ & $-0.0048$ & $+0.030$ \\ 
    \hline
$\mathrm b \overline{b}$-enriched events & & & & \\
    \hline
Sample &$E_{ch.}~ +~ E_{neu.}$& $E_{ch.}$ & $E_{neu.}$ & $E_{miss.}$\\
    \hline
Data &$0.8423$ &$0.5891$ & $0.2528$ & $0.1589$ \\
MC &$0.8423$ &$0.5885$ & $0.2535$ & $0.1579$ \\
\hline
(Data-MC)/MC&$0.0000$ &$+0.0010$ & $-0.0027$ & $+0.0063$ \\ 
\hline
  \end{tabular}
  \caption[Fitted fractions of charged and neutral energies reconstructed in $\mathrm b \overline{b}$-depleted and $\mathrm b \overline{b}$-enriched event samples.]
{{Fitted fractions of charged energy $(E_{ch.})$, neutral 
energy $(E_{neu.})$ and their sum 
reconstructed in $\mathrm b\overline{b}$-depleted and $\mathrm b \overline{b}$-enriched
event samples in data and simulation. The missing energy $(E_{miss.})$
fitted fraction is also given.}
  \label{tab:evtener}}
\end{center}
\end{table}

\begin{table}[p]
\begin{center}
  \begin{tabular}{|c|c|c|}
    \hline
$\mathrm b \overline{b}$-depleted events & &  \\
    \hline
Sample & charged & neutrals \\
    \hline
Data &$22.93~(7.94)$ &$10.47~(3.84)$ \\
MC &$22.96~(7.62)$ &$10.56~(3.82)$ \\
\hline
$\mathrm b \overline{b}$-enriched events & & \\
    \hline
Sample & charged & neutrals \\
    \hline
Data &$25.34~(7.60)$ &$10.95~(3.80)$ \\
MC default&$24.74~(7.45)$ &$10.88~(3.82)$ \\
MC fitted&$25.16~(7.48)$ &$10.96~(3.83)$ \\
\hline
  \end{tabular}
  \caption[Charged and neutral particle multiplicities measured in data and simulation after corrections.]{{Charged and neutral particle multiplicities (variances)
measured in data and simulation.}
  \label{tab:multiplicities}}
\end{center}
\end{table}

\begin{figure}[htb!]
  \centering
  \subfloat[][]{\includegraphics[width=0.49\textwidth]{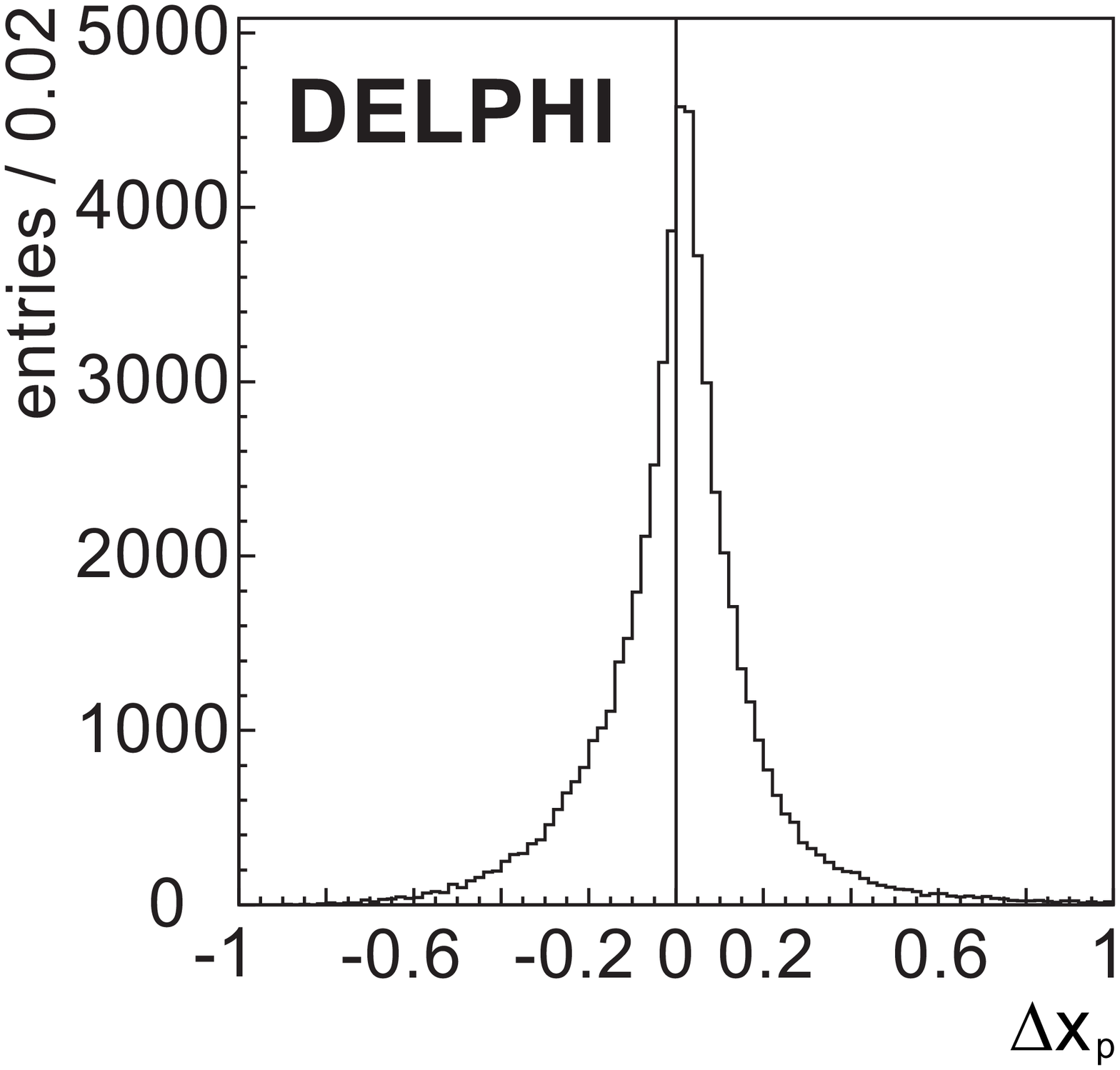}}\ 
  \subfloat[][]{\includegraphics[width=0.49\textwidth]{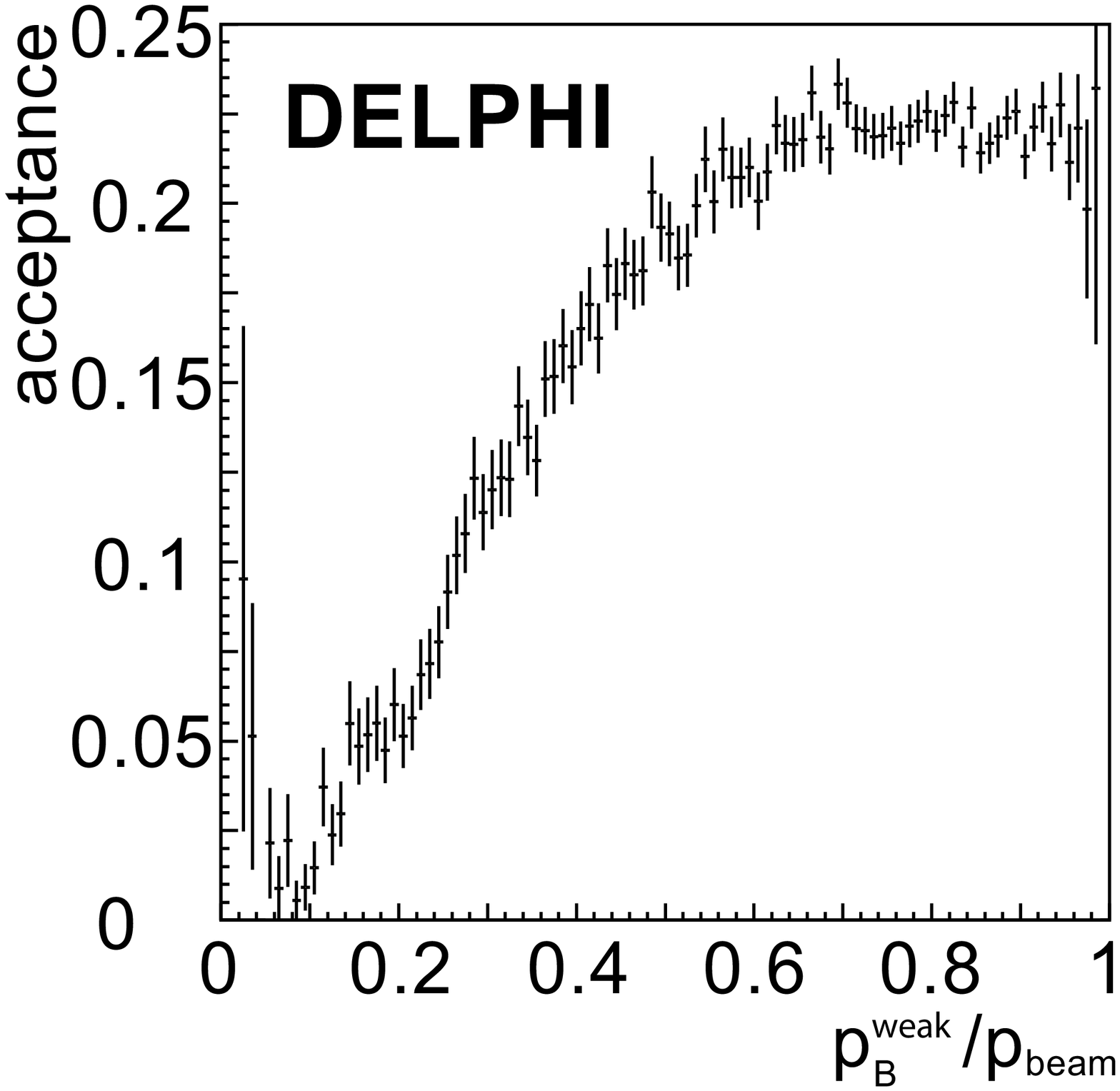}}
  \caption{{\bf a} distribution of $\Delta x_p = (p^{rec.}_{\mathrm B}-p^{gen.}_{\mathrm B})/p^{gen.}_{\mathrm B}$
   for weakly decaying b-hadrons. The full width at half maximum is equal to 16~$\%$.
   {\bf b} acceptance for signal events versus $p_{\mathrm B}^{\mathrm weak}/p_{\mathrm beam}$.}
  \label{fig:resol}
\end{figure}

Each jet pointing through the detector barrel region
defined by $\left | \cos{\theta_{jet}} \right |<0.75$ is considered
in turn, and charged particles belonging to the jet are used to
reconstruct a B decay vertex candidate.
It is then required that these tracks have at least two VD hits 
associated in $R\phi$ and a minimum
positive impact parameter with significance larger than $\sqrt{3} \sigma$
relative to the main vertex of the event. A secondary vertex is then
reconstructed. Tracks with a too large contribution to the $\chi^2$ are
removed from the fit in an iterative way. 
For a candidate to be accepted, it is required that at least three tracks
with the $Z$-coordinate measured in the VD remain,
and that the distance between the secondary and the primary vertex
projected along the jet direction is larger than 500 $\mu m$.
The reconstructed mass must not exceed the B mass 
(all particles are assumed to be pions). If not, particles ordered
by increasing values of their rapidity measured relative to the jet axis,
are eliminated in turn.
If the reconstructed mass is smaller than the B mass, particles belonging to 
the same jet ordered by decreasing rapidity values, are added in turn. 
For charged particles,
offsets relative to the primary and secondary vertices are also examined.
To possibly include a track,
it is required that its offset relative to the secondary vertex
is smaller than its offset relative to the primary vertex.
The procedure is stopped when the mass of selected particles is closest 
to the B mass.

The B momentum is obtained by subtracting from the fitted jet momentum the
momentum of the tracks from the jet, 
which have not been assigned to the B candidate. For the candidate to be accepted,
the sum of the jet neutral energy and of the charged energy
for tracks that are simultaneously compatible
with the primary and secondary vertices has to be smaller than 20 GeV.
Figure \ref{fig:resol}a shows the difference between the
reconstructed and the simulated B momentum, divided by the simulated value.

According to the simulation the applied algorithm has an average efficiency 
for the signal of $19\%$ (see Figure \ref{fig:resol}b) and a contamination 
of $5\%$ from non-b jets. The efficiency is rather flat 
for $p_{\mathrm B}^{\mathrm weak}/p_{\mathrm beam}>0.5$ and is still 
50$\%$ of
its maximum value around $p_{\mathrm B}^{\mathrm weak}/p_{\mathrm beam}=0.3$. 
There are 134282 candidates selected in the data sample.
The quoted efficiency for the signal differs from values given in Table~\ref{tab:btag},
because the latter refers to the whole event whereas the former is for
b-jets after applying the additional cuts used in the analysis.

\begin{figure}[htb!]
  \begin{center}
    \mbox{
\epsfig{file=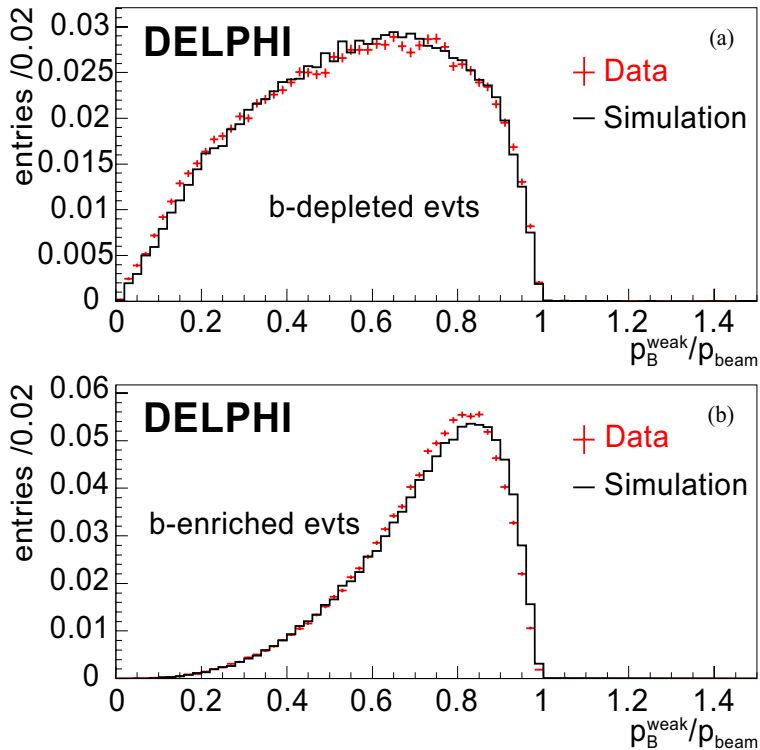,width=10cm,height=10cm}}
  \end{center}
  \caption[Comparison between the measured distributions
of the beam momentum fraction taken by a b-hadron, 
obtained in data and in the  MC ${\mathrm q\overline{q}}$
simulation.]{ {Comparison between the measured distributions
of the beam momentum fraction taken by a b-hadron, 
obtained in data (points with error bars) and in the ${\mathrm q\overline{q}}$
simulation (histogram). {\bf a} Depleted b-sample.
{\bf b} Enriched b-sample. The distributions have been normalised to unity.}
   \label{fig:xprec}}
\end{figure}

Measured $p_{\mathrm B}^{\mathrm weak}/p_{\mathrm beam}$ distributions 
are compared in Figure \ref{fig:xprec} with expectations from
the simulation. The two distributions agree for $\mathrm b \overline{b}$-depleted events and show a
marked difference for events in the b-enriched sample. In what follows,
the  transformation of the non-perturbative QCD distribution used
in the simulation, required to make the weighted distribution
of simulated events agree with the data, has been determined. 

\subsubsection{Determination of the b-hadron fragmentation distribution}
\label{sec:Orsayfragdist}

The binned 
distribution of the reconstructed $p_{\mathrm B}^{\mathrm weak}/p_{\mathrm jet}$ variable 
has been fitted
by minimising a $\chi^2$, which includes effects from
the data and simulation statistics and from the weighting procedure.

In each bin the
number of measured events is compared with an estimated number obtained in the
following way:
\begin{itemize}  
\item contributions from background events are taken from the ${\mathrm q}\overline{\mathrm q}$
simulation. They comprise three components: non-b jets in non-$\mathrm b\overline{b}$
events, non-b jets in $\mathrm b\overline{b}$ events and b jets from gluon splitting.
In simulated
events, the fractions of these components are respectively equal to
$5.2\%,~0.45\%~{\rm and}~0.24\%$ of the analysed events.
The number of gluon splitting candidates has been multiplied by 1.5 
to account for
its measured rate at LEP \cite{ref:gluontobb}.

\begin{figure}[htb!]
  \begin{center}
    \mbox{\epsfig{file=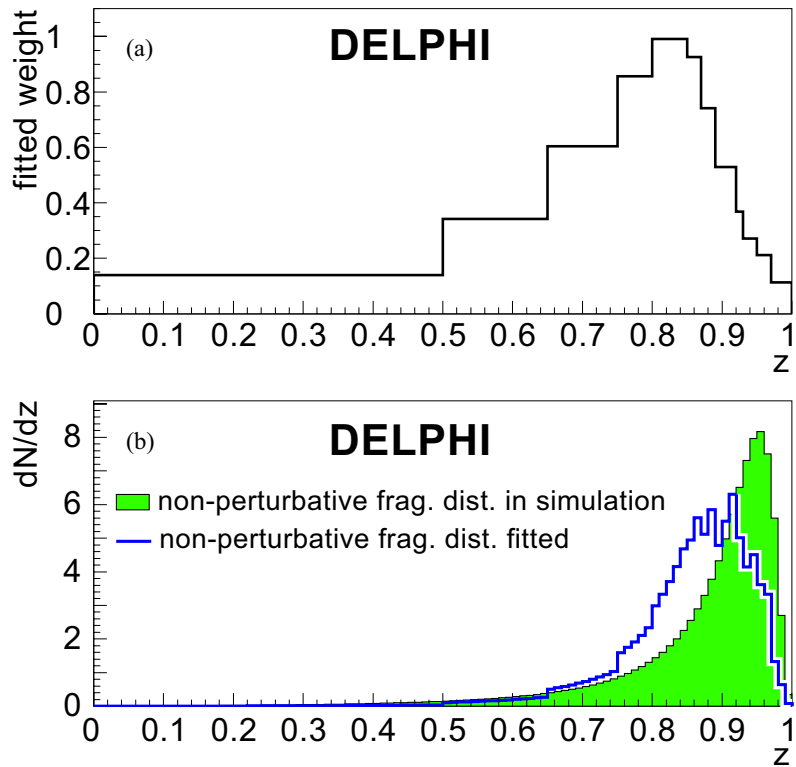,width=10.5cm}}
  \end{center}
  \caption[Fitted $z$ distribution on 1994 selected events.]{{ {Fitted $z$ distribution on data selected events.
{\bf a} Distribution of the fitted weights.
{\bf b} Comparison between the initial Peterson distribution used in the
simulation generator (shaded), and the corresponding distribution favoured by data events (solid line). The visible steps on this last distribution correspond 
to the applied weights, which have constant values over each bin as illustrated 
in {\bf a}.}}
   \label{fig:fitdat_z_94}}
\end{figure}

\item The distribution of signal events is obtained by weighting  $\mathrm b\overline{b}$ 
simulated events. This weight contains several components, which have been
determined to correct the values of 
parameters used in the simulation so that they agree with 
corresponding measured quantities as:
$\B$ lifetimes, $\B$ charged particle multiplicity and $\B^{**}$ fraction in jets.
The used values of these measured quantities are the same as those used in the regularized
unfolding analysis, as detailed in Section~\ref{systematics}.
\item A weight,
whose parameters are fitted, is also applied for each value of the simulated 
$z$ variable (see Section \ref{FRAGFUN}). The weights are constant over intervals in $z$
(the weight function is a histogram with a non-uniform binning).
\item The normalisation of $\mathrm b \overline{b}$ events is taken as a free parameter.
\end{itemize}  

To prevent oscillations between the contents of nearby bins of the
weight histogram, a regularisation term is included in the $\chi^2$:
\begin{equation}
\chi^2_{\rm reg} = C \times \left [ 2 \times n(i) - n(i-1) - n(i+1) \right ]^2~,
\label{eq:chi2}
\end{equation}
where $C$ is a parameter whose value ($C=1$)
has been determined empirically
using simulated events; and $n(i)$ is the content of bin $i$.

Distributions corrected for all effects are then obtained using corresponding 
generated distributions from simulated events before any selection criteria,
and by applying the weight 
distribution fitted on real events,
which depends on the $z$ variable generated value for each simulated b-hadron.
Statistical uncertainties in each bin, of these distributions, have
been obtained using the full covariance matrix of the fitted parameters
and generating toy experiments.

The fitted weight distribution obtained with the data sample is given
in Figure \ref{fig:fitdat_z_94}. This figure shows also the $z$ distribution 
as favoured by the data. It is rather different from the Peterson distribution
which was used in the simulation, and shown on the same figure.

As in the companion analysis described in Section \ref{sect:unfold},
the differential b-quark fragmentation distribution is evaluated in nine 
intervals of the $\xb$ variable whose averaged value is equal to
$\avxb=0.6978 \pm 0.0010$. It is displayed in Fig. \ref{fig:Orsay_Karl}.
Measured values of the distribution in each bin and the corresponding
statistical error matrix are given in Appendix C. Its integral
has been normalised to unity.

\subsubsection{Systematic uncertainties}
\label{sec:Orsaysystematics}

Systematic uncertainties have been evaluated for each value
of the fragmentation distribution obtained in the nine 
$\xb$ intervals. In Table \ref{tab:Orsayerrors}, systematics
on $\avxb$ have been reported.
Sources of systematic uncertainties have been ordered as in the previous
analysis (Section \ref{systematics}).

{\bf \underline {Technical systematics:}}

The  weight function consists of twelve bins in $z$
whose content is fitted\footnote{The content of one of these bins
is fixed to one as the normalisation of the signal $\mathrm b \overline{b}$ events
is also fitted}. Choices for the bin definition
and bin number can induce a systematic uncertainty on the extracted
distribution. This has been studied by comparing the generated and fitted
distributions in simulated events.
Generated events correspond to the average value $\avxb_{gen.}^{sim.}=0.7057$ and have been
reconstructed at $\avxb_{rec.}^{sim.}=0.7060$. The quoted values 
have been corrected for
the effect of the beam radiation which corresponds to an increase of $0.0015$.
The observed difference on $\avxb$ is 
equal to $+0.0003$.

These results depend also on the choice for the value of the curvature
parameter $C$ introduced in the $\chi^2$ expression
(see Equation (\ref{eq:chi2})). Changing the value of this parameter
between $0.02$ and $5.0$ gives variations on $\avxb$
at the level of $\pm 0.002$ on simulated events and even
smaller values on real data events.
In the following analysis the value $C=1$ has been used and effects of the variation
of this parameter between $0.02$ and $5.0$ are included in the evaluation of systematic 
uncertainties.

{\bf \underline {Selection cuts and background dependence:}}

In the analysed sample with  the selection $P_{\rm btag} \leq 10^{-3}$, the estimated 
fraction of non-b candidates amounts to $5.2\%$. 
In Table \ref{tab:btag} it was observed that the fraction of selected events
is a few $\%$ (relative) higher in real data. 
As this effect remains in samples of high purity
in $\mathrm b \overline{b}$ events, its main origin comes most probably from a difference in 
efficiency
between real and simulated $\mathrm b \overline{b}$ events. A possible underestimate of the 
selection efficiency to 
non-$\mathrm b \overline{b}$ events in the simulation amounts then to $10\%$ (relative) 
at maximum. 
The effect of a $\pm 20\%$ variation on the non-b background
level has been evaluated; it gives 
$\delta \avxb = \pm 0.0012$. 

The stability of the measured $\xb$ distribution has been
studied for different selections on the value of the $P_{\rm btag}$ variable.
The resulting $\avxb$ is stable within $\pm 0.001$. For the corresponding systematic
evaluation, half the difference obtained using selection cuts
at $10^{-4}$ and $10^{-10}$ has been used.

Hadronic jets have been reconstructed using the LUCLUS algorithm with the
value of the parameter defining the jets, $d_{\rm join}$=5 GeV/$c$. Sensitivity of present 
results on the value of this parameter has been studied by redoing the
measurements using $d_{\rm join}$=10 GeV/$c$. The  variation on $\avxb$ is 
equal to $+0.0002$.

In a jet, there are charged particles which can be compatible simultaneously
with the primary and the secondary vertex. Concerning neutral particles, 
the angular resolution
does not allow them to be attached with confidence to one of the two vertices.
The energy taken by these two classes of tracks is denoted 
``ambiguous'' energy.  In the analysis, events have
been selected requiring that the ``ambiguous'' energy is lower than $20$~GeV. 
The stability of the results has been studied
by changing the value for this selection criterion. A change from $20$~GeV 
to 15 GeV results in a $25\%$ decrease in the number of selected events, and no variation is measured for $\avxb$.
A change from $15$~GeV to $10$~GeV keeps $50\%$ of the initial statistics. The corresponding variation is taken as a systematic uncertainty, which corresponds to a variation of $\avxb$ by $-0.0034$. 

Events have been selected requiring at least three charged particles
at the candidate B decay vertex. 
Taking the difference observed for selections with at least 
three and five charged particles as an
evaluation for the corresponding systematic,
the  variation on $\avxb$ is equal to $+0.0012$.

The rate for b-hadron production originating from gluon coupling
to $\mathrm b \overline{b}$ pairs has been measured by LEP experiments
and found to be larger than the rate used in the simulation 
by a factor $1.5$.
The corresponding systematic uncertainty has been evaluated,
considering the uncertainty, of $30\%$, obtained by DELPHI on this quantity
\cite{ref:delphigbb}.
The variation on $\avxb$ is equal to $-0.0001$.

{\bf \underline{Reconstructed energy:}} 

The analysis uses the beam energy as a constraint in a global fit
of 4-momenta of charged and neutral particles, such that the total energy
and momentum of the event is conserved.

Corrections applied on charged and neutral energy distributions
have been described in Section \ref{sec:OrsayBener}. They induce
a  variation on $\avxb$ of $+0.0017$. The corresponding
systematic uncertainty has been evaluated taking the effect of this correction. 

Measured jet multiplicities are not identical in data and in simulated events.
Taking as reference the fraction of two-jet events, fractions of
three- and four-jet events have to be corrected respectively by $-5\%$ 
and $+13\%$ in the simulation. 
Simulated events have been weighted accordingly so that the two distributions
agree. From the statistical
accuracy of this correction, the systematic uncertainty has been evaluated 
to be
one third of the correction.
This corresponds to 
$\delta \avxb=  -0.0001$.

\begin{table}[htb]
\begin{center}
\begin{tabular}{|c|c|c|} \hline

uncertainty class & item &  $\delta \avxb$ \\
  \hline
%
\multirow{2}{4cm}{technical} &  fitted function shape             &   $ ~0.0015 $    \\
 & curvature parameter in $\chi^2$             &   $ ~0.0020 $    \\
\hline
%
%
\multirow{6}{4cm}{selection cuts and backg. dependence}&  b-tagging selection cut  &   $ ~0.0010 $    \\
  &  non-b background level     &   $ ~0.0012 $    \\
  & jet clustering parameter value               &   $ ~0.0002 $    \\
  & ambiguous energy level                      &   $ ~0.0034 $    \\
  & secondary vertex multiplicity                &   $ ~0.0012 $    \\
  & $\mathrm g\to b\overline{b}$                 &   $ -0.0001 $    \\
\hline
%
%
\multirow{2}{4cm}{reconstructed energy}&  corrections on tracks          &   $ ~0.0017 $    \\
   &  jet multiplicity                     &   $ -0.0001 $    \\
\hline
%
%
\multirow{3}{4cm}{b-physics modelling}&  b-hadron lifetimes                &   $ -0.0005 $    \\
 &  $\mathrm B^{**}$~rate             &   $ -0.0008 $    \\
 & b-decay track multiplicity                   &   $ -0.0008 $    \\
  \hline
calibration stability &  calibration periods   &   $ ~0.0038 $    \\
\hline
 Total &                  &   $ ~0.0064 $    \\
  \hline
\end{tabular}

\caption
{\label{tab:Orsayerrors} {Systematic uncertainty on the mean value of the 
 $\xb$~distribution in the weighted fitting analysis. The total is the sum in
quadrature of all contributions. The sign indicates the correlation between
the change in an uncertainty  source and the shift in the final result.
Uncertainties assigned by turning a weight on/off have no sign. }}
\end{center}
\end{table}

{\bf \underline{b-physics modelling:}} 

Variations of the values of parameters that govern decay
properties or production characteristics of b-hadrons have been 
also considered.

Simulated events have been generated using the same lifetime value
of $\tau_{\mathrm B}=1.6~ ps$. Events have been weighted such that 
each type of b-hadron
is distributed according to its corresponding lifetime, as given
in reference \cite{PDG}.
Taking, as systematics, the total variation induced by this correction,
the variation on $\avxb$ is equal to $-0.0005$.

In the simulation, the $\Bsstar$ production rate in a b-quark
jet amounts to $32\%$.
A weight is applied on b-hadrons which
originate from $\Bsstar$ decays to lower the effective $\Bsstar$ rate to $25\%$. 
The corresponding systematic has been taken as the variation on $\avxb$, namely $-0.0008$.
 
The difference between simulated, $n_{ch}^{sim.}({\mathrm B})$, and measured,
$n_{ch}^{meas.}({\mathrm B})$, average charged
multiplicities in b-hadron decays amounts to $0.06$:
\begin{equation}
n_{ch}^{meas.}({\mathrm B})=4.97 \pm 0.03 \pm 0.06, ~n_{ch}^{sim.}({\mathrm B})=4.91.
\end{equation}
This difference has been corrected by weighting events using a weight that 
has a linear variation with the actual b-hadron charged  
multiplicity
in a given event.
The simulated multiplicity distribution has been fitted with  a Gaussian
of standard deviation $(\sigma_{nch})$ equal to $2.03$ charged particles. Probability values, 
for a given charged 
multiplicity $i,~P_i,$ have been transformed into:
\begin{equation}
P_i^T= P_i \left [ 1+\beta (n_{ch}^{sim.} -i)\right ]~.
\end{equation}
The value of $\beta$ is obtained by requiring that the new average 
multiplicity computed using $P_i^T$ is equal to $n_{ch}^{meas.}(B)$.
Then:
\begin{equation}
\beta=\frac{n_{ch}^{sim.}-n_{ch}^{meas.}}{\sigma_{nch}^2}~.
\end{equation}
 The corresponding systematic uncertainty has been evaluated
by considering an uncertainty of $\pm0.1$ charged particles on $n_{ch}^{meas.}$.
The variation on $\avxb$ is equal to $\mp 0.0008$.

{\bf \underline{Calibration stability and simulation statistics:}} 

The stability of the energy calibration
has been studied dividing the analysed data samples in five 
time ordered subsamples of similar statistics. The statistical accuracy
of each $\avxb$ measurement is of about $0.002$. The 
systematic uncertainty attached to the energy reconstruction has
been evaluated by taking half the difference between the two extremes
of the five measurements of $\avxb$: $\pm 0.0038$.

Uncertainties corresponding to the finite statistics of simulated events
have been included in the statistical uncertainty of the measurements.

%
%
\subsection{Combination of the $\xb$ distributions}
\label{SECT-COMBINE}

The results of the two $\xb$ measurements obtained in Sections \ref{sect:unfold} and 
\ref{sec:Orsay} have been averaged. In this combination a complete correlation 
has been assumed between statistical uncertainties, due to the common data used by the two analyses. 
The following sources of systematic uncertainties have been considered also as fully correlated:
\begin{itemize}
\item neutral energy smearing in the regularised unfolding analysis with
ambiguous energy level in the weighted fitting analysis; 
\item $g \rightarrow b\overline{b}$ branching fraction; 
\item $\Bsstar$ production rate; 
\item b-hadron lifetimes;
\item b-decay track multiplicity;
\item b-hadron production fractions;
\item wrong sign charm rate.
\end{itemize}
\begin{table}[htb!]
\begin{center}
  \begin{tabular}{|c|c|c|c|c|}
 \hline
 bin           & value    & statistical & systematic
 & $ \sqrt{\sigma_{stat}^2+\sigma_{syst}^2} $ \\
 borders          &     & uncertainty & uncertainty & \\
 \hline
 \hline
  0.10  --  0.30      &   0.194  &   0.004      &0.020   & 0.020  \\
  0.30  --  0.42      &   0.474  &   0.008      &0.031   & 0.032  \\
  0.42  --  0.54      &   0.734  &   0.009      &0.037   & 0.038  \\
  0.54  --  0.64      &   1.112  &   0.013      &0.048   & 0.050  \\
  0.64  --  0.73      &   1.753  &   0.021      &0.057   & 0.060  \\
  0.73  --  0.80      &   2.641  &   0.029      &0.064   & 0.070  \\
  0.80  --  0.88      &   3.013  &   0.029      &0.119   & 0.122  \\
  0.88  --  0.94      &   1.787  &   0.028      &0.119   & 0.122  \\
  0.94  --  1.00      &   0.227  &   0.015      &0.046   & 0.049  \\
\hline
 \end{tabular}

\caption
{\label{tab:comb_xbweak} {The combined unfolded and weighted results, 
per bin, for $f(\xb)$. Quoted uncertainties have been scaled by $1.31$.}}
\end{center}
\end{table}
\begin{figure}[htb!]
	\centering
		\includegraphics[width=0.85\textwidth]{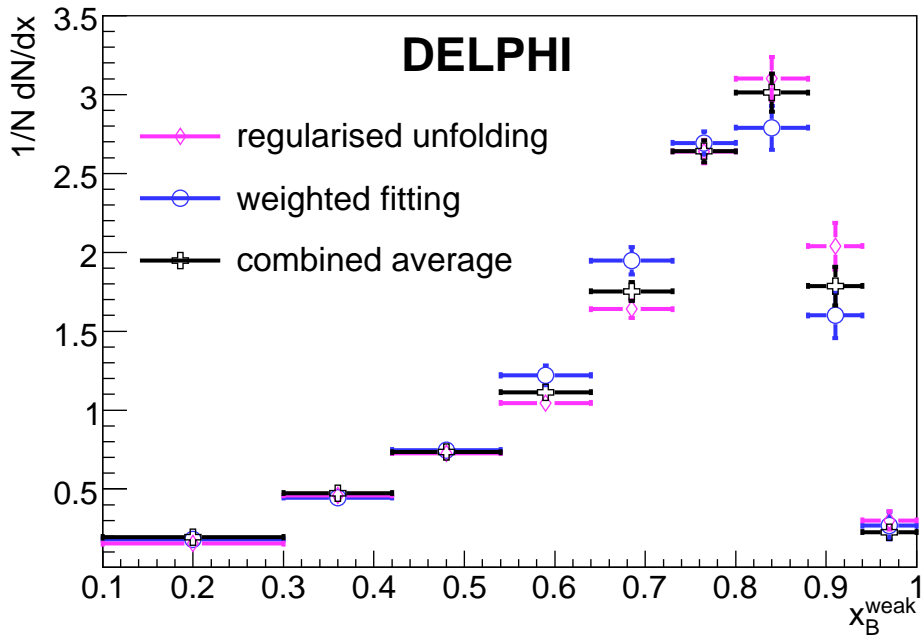}	  
	\caption{ Measured fragmentation distributions in the two analyses and their combined average. Uncertainties on the combined average are scaled by $1.31$.}
	\label{fig:Orsay_Karl}
\end{figure}
Other systematic uncertainties, some of them large, 
have been taken as uncorrelated as the two
analyses are using different techniques. 
No significant correlation was observed between the two measurements when considering event samples recorded
during the same time periods.

The combined $\xb$ distribution has been obtained by a fit using the full 
error matrix of the two analyses.
This matrix has two insignificant eigenvalues which have been removed.
The fit has therefore $7$ degrees of freedom and the $\chi^2$ value is $11.96$ 
(probability of $10.2\%$).

The combined value of the $f(\xb)$ distribution 
in each bin is given in Table 
\ref{tab:comb_xbweak}
and in Figure \ref{fig:Orsay_Karl}. 
The corresponding statistical and total error matrices are given in Appendix D. In the following, all the quoted uncertainties on the $f(\xb)$ distribution bins are scaled by $1.31$. This corresponds to $\chi^2/\mbox{NDF}=1$.
By rescaling the uncertainties it is ensured 
that possible poor fit probabilities of models
with the combined measurement 
do not originate from an underestimate of quoted measurement uncertainties.
The average value of this distribution is equal to:
\begin{equation}
\langle \xb \rangle =  0.699 \pm 0.011~.
\end{equation}
This value is largely influenced by correlations between the $\xb$ distributions from the two analyses.

The combined result is compared with model predictions 
in Section~\ref{modelfits} and with
other experimental results from ALEPH \cite{ref:aleph1}, 
OPAL \cite{ref:opal1} and SLD \cite{ref:sld1} at the $\Zz$ pole
in Section~\ref{sec:world_average}.

\subsection{Fits to hadronisation models}
\label{modelfits}

The measured $f(\xb)$ distribution has been compared to functional forms that are in
common use inside event generators. 
Since the Lund \cite{ref:lund1}, Lund-Bowler \cite{ref:bowler1}
and Peterson \cite{peterson}
models are functions of $z$ and, in the case of Lund and Lund-Bowler, of a
transverse mass variable $m_{{\mathrm b} \bot}^2$ 
that varies event-to-event\footnote{The transverse mass squared $m_{{\mathrm b} \bot}^2= m^2+p_{\bot}^2$
is defined within the Lund generator in terms of the mass ($m$) of the primary b-hadron, and its transverse momentum
($p_{\bot}$) relative to the string axis.},
these functions cannot simply be fitted to the unfolded distributions.
Instead, parameters
of these models have been fitted to data using a high
statistics Monte-Carlo sample at the generator level by applying weights.
The configuration of the event generator used for these studies is as given in Table~\ref{tab:orsmc}. For further details see reference~\cite{tuningdelphi}.

\begin{table}[ht!]
\begin{center}
\begin{tabular}{|l|l|} \hline
  Event Generator         & JETSET/PYTHIA 6.156 \\
  Perturbative ansatz     & Parton shower ($\Lambda_{QCD}=0.297$~GeV,~$Q_0=1.56$~GeV) \\
  Non-perturbative ansatz & String fragmentation\\
  Fragmentation function  & Peterson with $\epsilon_{\mathrm b}=0.004$\\ 
  Bose-Einstein correlations & Disabled \\ \hline
\end{tabular}
\caption
{\label{tab:orsmc} Details of the
event generator used together
with some of the more relevant parameter values that have been
tuned to the DELPHI data.}
\end{center}
\end{table}

\begin{figure}[htb!]
\begin{center}
\leavevmode
   \includegraphics[width=0.85\textwidth]{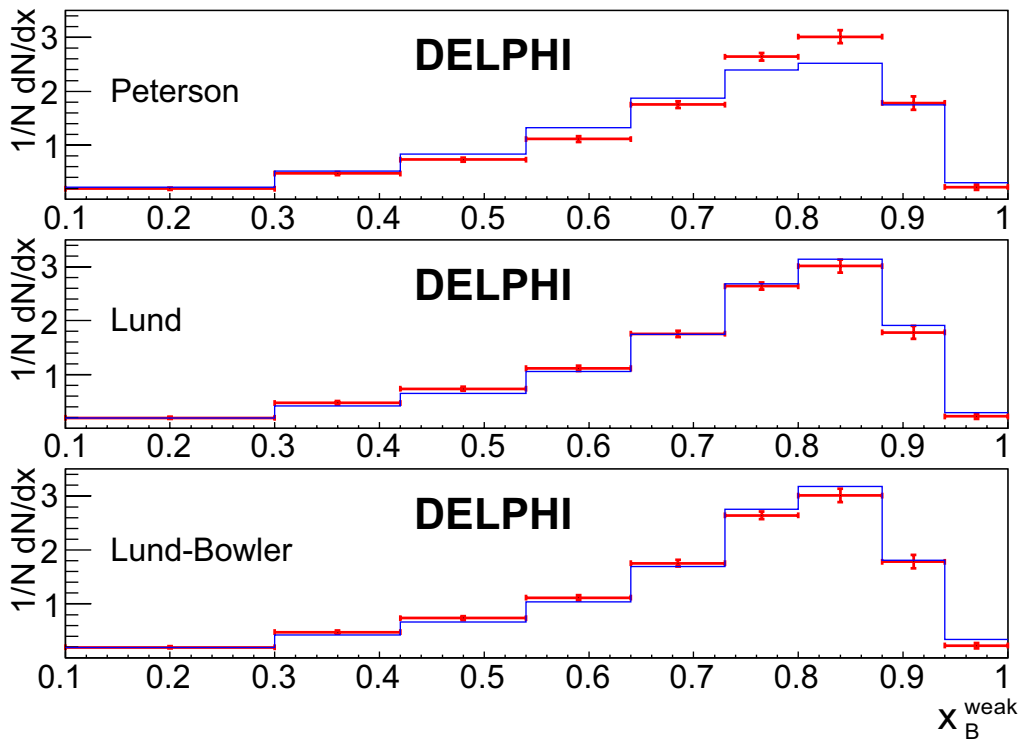}
\caption {\label{fig:modelfits_reweight} { The result of fitting various hadronisation model
functions to the measured $\xb$~distributions.
Points with error bars represent the data, and histograms represent the reweighted Monte-Carlo simulation with the best fit result.}}
\end{center}
\end{figure}

\begin{figure}[htb!]
	\centering
		\includegraphics[width=0.85\textwidth]{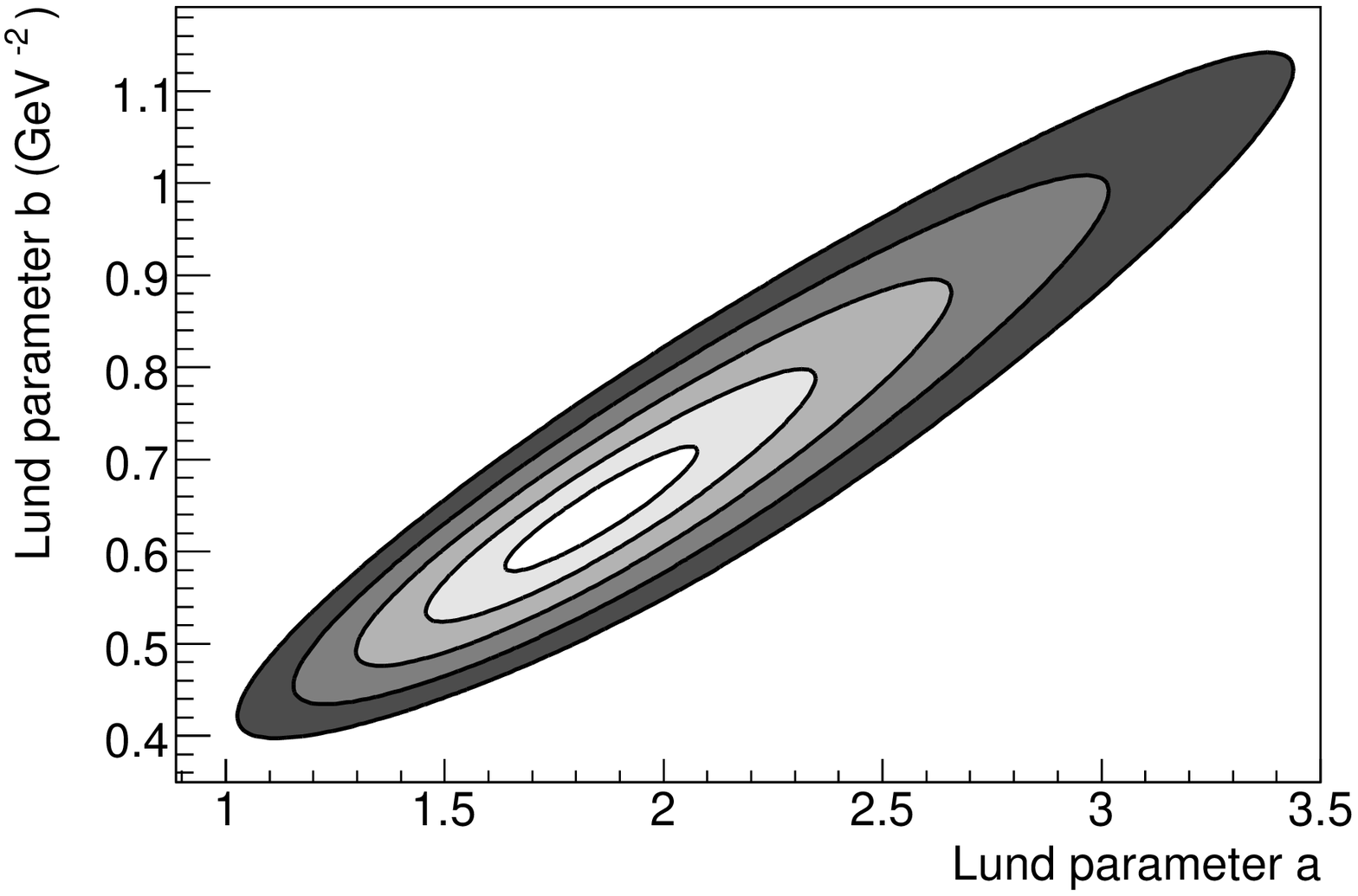}
	\caption{ The one standard deviation (lightest grey) to five standard deviations
	(darkest grey) contours of the Lund parameters $a$ and $b$. These contours
	correspond to coverage probabilities of $39.3\%$, $63.2\%$, $77.7\%$, $86.5\%$ and
	$91.8\%$. The $\chi^2$ has been obtained comparing the measured $f(\xb)$
	distribution in data to the generated model prediction.}
	\label{fig:DelphiLundContours}
\end{figure}

For each event in the generated sample, the values of the internal variables
$z$ and $m_{\mathrm b\perp}^2$ are used to define a weight
$w=f_{\rm fit}(z,m_{\mathrm b\perp}^2;\vec{X}_{\rm fit})/f_{\rm Peterson}(z;\epsilon_{\mathrm b})$,
where $f_{\rm fit}(z,m_{\mathrm b\perp}^2;\vec{X}_{\rm fit})$ stands for the Lund, 
Lund-Bowler\footnote{The predicted value $r_Q = 1$~\cite{ref:bowler1} has been used.}
or Peterson\footnote{Note that when $f_{\rm fit}$ represents the Peterson fragmentation function it does not depend on $m_{\mathrm b\perp}^2$.} fitted distributions,
$\vec{X}_{\rm fit}$ to their corresponding parameters and
$f_{\rm Peterson}(z;\epsilon_{\mathrm b})$ to the Peterson distribution used
in the generated sample.
The choice of using the Peterson fragmentation function is motivated by the fact
that, unlike Lund and Lund-Bowler, this model has a tail at small $z$ values, which ensures a
non-vanishing probability over all the $z$ spectrum.
Values of the model parameters $\vec{X}_{\rm fit}$ have been fitted by requiring that the
weighted generated distribution of $\xb$ agrees with
the measured one within uncertainties. As explained in Section \ref{SECT-COMBINE}, the measured distribution has $7$ degrees of freedom, and therefore $2$ eigenvalues have been
cut away in the present fit. The Lund model results in the best fit to data, followed by the Lund-Bowler model. 
Fit results are detailed in Table~\ref{tab:modelfits_reweight}, and the corresponding
$\xb$ distributions are shown in Figure~\ref{fig:modelfits_reweight} in comparison with the
measured distribution. The one to five standard deviation contours of the  Lund parameters $a$ and $b$ are presented in Figure~\ref{fig:DelphiLundContours}. Clearly, the data suggest that the Lund and Lund-Bowler functions yield better fits than
those explicitly constructed to describe the fragmentation of heavy quarks
e.g. the Peterson function.

It must be noted that the fitted values for the parameters of the ``universal'' Lund fragmentation distribution are rather different from
those determined using hadronic events at LEP which are dominated by
light flavours $(a=0.35,~b=0.52\ \bunit)$.

\begin{table}[htb!]
\begin{center}
{\small
\begin{tabular}{|c|c|c|c|}
  \hline
  Model & Parameters & $\chi^2/NDF$ & Correlation \\ \hline
  Peterson$
\left[\frac{1}{x}\left ( 1 - \frac{1}{x}-\frac{\epsilon_{\mathrm b}}{1-x} \right )^{-2}\right]$  & $\epsilon_{\mathrm b} = (4.06^{+0.46}_{-0.41})\times 10^{-3}$
             & 55.8/6 &  ---\\ \hline

  Lund  $\left[\frac{1}{x}(1-x)^a \exp{\left ( -\frac{bm_{\mathrm b\bot}^2}{x}\right )} \right]$     & \renewcommand{\arraystretch}{1.}
               $\begin{array}{l} a=1.84^{+0.23}_{-0.21} \\ b=0.642^{+0.073}_{-0.063}\ \bunit \end{array}$
              & 9.8/5  & $92.2\%$ \\ \hline
 Lund-Bowler $\left[ \frac{1}{x^{1+r_Qbm_{\mathrm b\bot}^2}}(1-x)^a \exp{\left (
     -\frac{bm_{\mathrm b\bot}^2}{x}\right )} \right]$& \renewcommand{\arraystretch}{1.}
               $\begin{array}{l} a=1.04^{+0.14}_{-0.12} \\ b=3.08^{+0.45}_{-0.39}\ \bunit \end{array}$
             & 20.7/5 & $85.6\%$ \\
             ($r_Q = 1$) & & & \\
\hline
\end{tabular}
}
\caption {{Results of the $f(\xb)$~hadronisation model fits.
For the Lund and Lund-Bowler models, also the correlation between the $a$ and $b$ parameters
is given.}
\label{tab:modelfits_reweight}}
\end{center}

\end{table}

\renewcommand{\arraystretch}{1.}

\section{Analytic extraction of the non-perturbative QCD fragmentation function}
\label{SECT-NONPERTB}

The combined DELPHI measurement of $f(\xb)$ is used to
extract the b-quark fragmentation function. For this study, 
the variable $\xb$ is transformed to $\xp$, which is preferred because it 
varies exactly between $0$ and $1$.
As explained in Section~\ref{SECT-INTRO}, $f(\xp)$~as measured
in the experiment, can be viewed as the result of perturbative and 
non-perturbative QCD processes: 
\begin{equation}
f(\xp) = \int_0^{\infty}{\frac{dx}{x} ~f_{pert.}(x)~
f_{non-pert.}\left ( \frac{\xp}{x} \right )}~.
\label{eq:start}
\end{equation}
In order to separate out the
non-perturbative contribution, a choice for the perturbative part must
be made. This problem is addressed in two ways:
\begin{itemize}
	\item the perturbative contribution is taken from a parton
shower Monte-Carlo generator. In this case parameters of the (non-perturbative)
fragmentation function $f(x)$~are
 also fitted within the context of commonly-used hadronisation models;
  \item the perturbative contribution is taken to be a NLL QCD calculation
and the corresponding non-perturbative component is computed to reproduce the measurements.
\end{itemize}

\begin{figure}[htb!]
  \begin{center}
  		\includegraphics[width=1.00\textwidth]{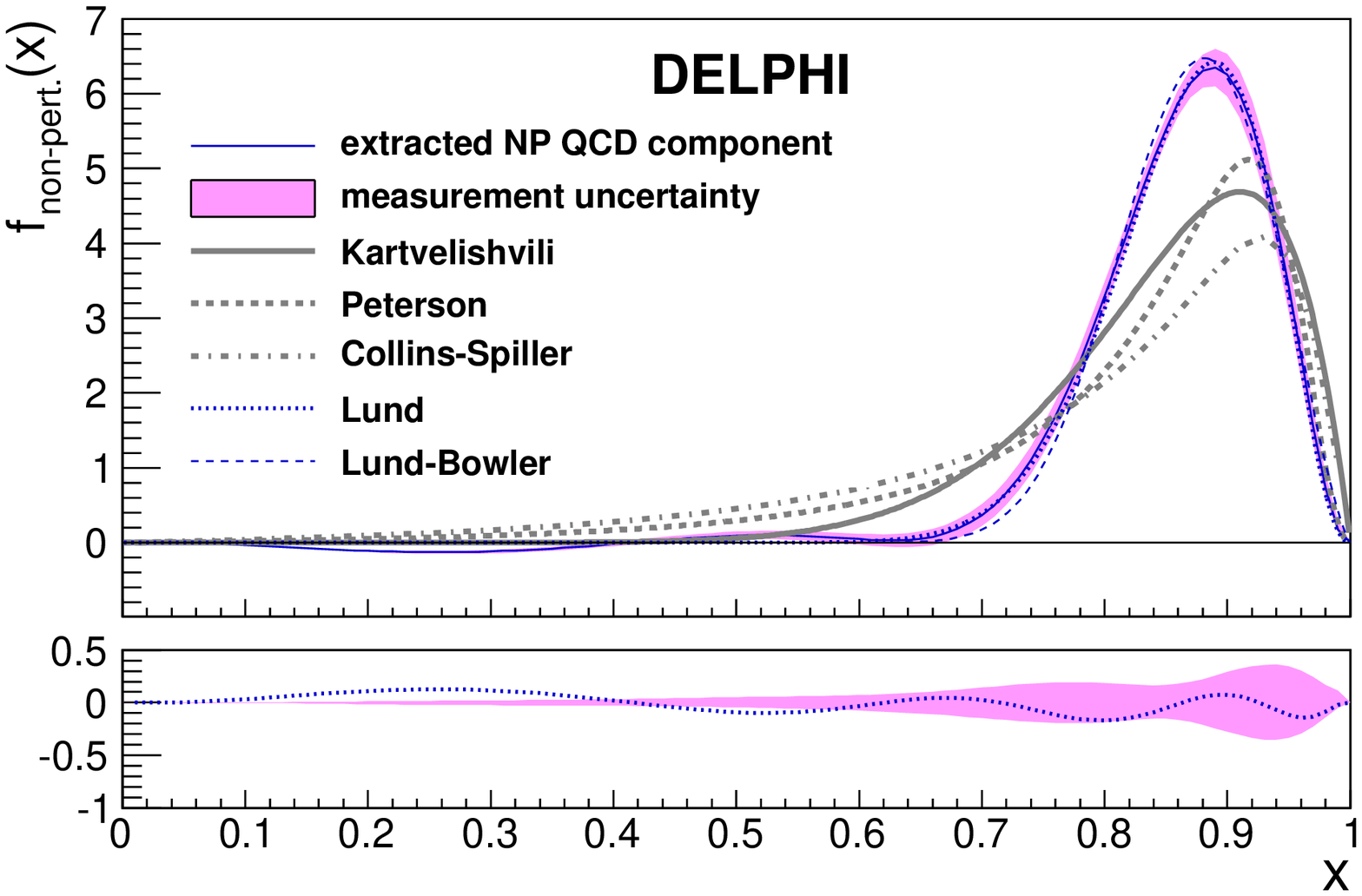}
  \end{center}
  \caption[
  The extracted non-perturbative QCD component when the perturbative 
one is taken from JETSET~7.3.] {{Dependence of the non-perturbative (NP)
QCD component (full line), entering in Equation (\ref{eq:start}), 
when the perturbative component is taken
from JETSET~7.3. The shaded area corresponds
to measurement uncertainties.
These uncertainties are correlated for different $x$-values. 
The other curves correspond to different models whose parameters have been
obtained after a fit on present measurements of the fragmentation distribution.
The lower plot shows the difference between the extracted non-perturbative
QCD component and the fitted Lund model.
Note that the variable $x$ used to display the variation of the different distributions
is not $x_p^{\mathrm weak}$, but the integration variable from Equation~(\ref{eq:start}).
} 
   \label{fig:jetset}}
\end{figure}

\begin{table}[htb!]
\begin{center}
\scriptsize{
  \begin{tabular}{|l|c|c||c|c|}
    \hline
~ \hfill Model \hfill ~ & \multicolumn{2}{c||}{JETSET~7.3} & \multicolumn{2}{c|}{PYTHIA~6.156}\\
    \cline{2-5}
& Fitted Parameters       & $\chi^2/NDF$ & Fitted Parameters       & $\chi^2/NDF$ \\ 
& and Correlation ($\rho$)&              & and Correlation ($\rho$)&              \\ 
    \hline
Kartvelishvili \cite{ref:kart1} &$\epsilon_{\mathrm b}=10.17 \pm 0.37 $ & $126/6$  & $\epsilon_{\mathrm b}=13.50 \pm 0.56$ & 57/6\\
 $x^{\epsilon_{\mathrm b}}(1-x)$         &  & & & \\
\hline
Peterson \cite{peterson}   &$\epsilon_{\mathrm b}=(7.4 \pm 0.6) \times 10^{-3}$  & $83/6$  &$\epsilon_{\mathrm b}=(4.4 \pm 0.4) \times 10^{-3}$  & $153/6$  \\
 $\frac{1}{x}\left ( 1 - \frac{1}{x}-\frac{\epsilon_{\mathrm b}}{1-x} \right )^{-2}$         &  & & & \\
\hline
Collins-Spiller\cite{ref:cs1}              &$\epsilon_{\mathrm b}=(6.74^{+0.72}_{-0.64}) \times 10^{-3}$  & $358/6$    & $\epsilon_{\mathrm b}=(2.56^{+0.38}_{-0.32}) \times 10^{-3}$  & $322/6$ \\
 $\left (\frac{1-x}{x}+\frac{\epsilon_{\mathrm b}(2-x)}{1-x} \right )\left ( 1 +x^2 \right ) \times$         &  & & &\\
 $\left ( 1 - \frac{1}{x}-\frac{\epsilon_{\mathrm b}}{1-x} \right )^{-2}$         &  & & &\\
\hline
Lund \cite{ref:lund1}           &$a = 2.35^{+0.25}_{-0.22}$ & $2.3/5$    & $a = 2.06^{+0.28}_{-0.24}$ & $3.8/5$ \\
$\frac{1}{x}(1-x)^a \exp{\left ( -\frac{bm_{\mathrm b\bot}^2}{x}\right )}$ &$bm_{\mathrm b\bot}^2=17.7^{+1.9}_{-1.6}$&         &$bm_{\mathrm b\bot}^2=19.8^{+2.4}_{-2.0}$& \\
&$\left(\rho = 89.1\%\right)$                     &         & $\left(\rho = 89.7\%\right)$ & \\
\hline
Lund-Bowler \cite{ref:bowler1}       &$a= 1.32\pm 0.14 $& $12.2/5$ & $a= 1.02^{+0.15}_{-0.13} $& $5.1/5$ \\
$\frac{1}{x^{1+bm_{\mathrm b\bot}^2}}(1-x)^a \exp{\left ( -\frac{bm_{\mathrm b\bot}^2}{x}\right )}$ &$bm_{\mathrm b\bot}^2=80.^{+12.}_{-11.}$       &  & $bm_{\mathrm b\bot}^2=95.^{+13.}_{-12.}$ & \\
&$\left(\rho = 75.0\%\right)$       &         & $\left(\rho = 76.5\%\right)$& \\
    \hline
  \end{tabular}
}
  \caption{{Values of the parameters and of the $\chi^2/NDF$ obtained 
when fitting results from Equation \eqref{eq:start}, obtained for different 
models of the non-perturbative QCD component, to the measured 
b-fragmentation distribution. 
Results are shown for perturbative QCD components taken
from JETSET~7.3 and PYTHIA~6.156. The Lund and Lund-Bowler models have been simplified by assuming 
that the transverse mass of the b-quark, $m_{\mathrm b\bot}$, is  a constant. 
}
  \label{tab:fittedmodels}}
\end{center}
\end{table}

The method is based on the use of the Mellin transformation which is 
appropriate when dealing with integral equations as given in  (\ref{eq:start}).
The Mellin transformation of $f(x)$ is:
\begin{equation}
\tilde{f}(N) = \int_0^{\infty}{dx ~x^{N-1}~f(x)}~,
\label{eq:mellint}
\end{equation}
where $N$ is a complex variable. For real integer values of $ N \ge 2$, the values 
of $\tilde{f}(N)$ correspond to the moments of the initial $x$ distribution\footnote{By definition  
$\tilde{f}(1)~(= 1)$ corresponds to the normalisation of $f(x)$.}. For 
physical processes, $x$ is restricted to be within the $[0,1]$ interval. 
The interest in using Mellin transformed expressions is that 
Equation (\ref{eq:start}) becomes a simple product:
\begin{equation}
\tilde{f}(N) = \tilde{f}_{pert.}(N) \times  
\tilde{f}_{non-pert.}(N)~.
\label{eq:mellintb}
\end{equation}

Having computed  distributions of the measured and 
perturbative QCD components in the $N$-space, the non-perturbative distribution,
$\tilde{f}_{non-pert.}(N)$, is obtained from Equation~(\ref{eq:mellintb}).
Applying the inverse Mellin transformation on this distribution 
gives $f_{non-pert.}(x)$  without any need for a model input:
\begin{equation}
f_{non-pert.}(x) =
\frac{1}{2\pi i}\oint{dN~\frac{\tilde{f}_{meas.}(N)}{\tilde{f}_{pert.}(N)}~x^{-N}}~,
\label{eq:inv_mellin}
\end{equation}
in which the integral runs over a contour in the complex $N$-plane. 
More details on this approach can be found in \cite{ref:benhaim,thesis_eli}.

In practice, the Mellin transformed distribution $\tilde{f}_{meas.}(N)$ of the measured distribution 
has been obtained after having adjusted an analytic 
expression to the measured distribution in $\xp$, and by applying the Mellin 
transformation on this fitted function. The following expression, which 
depends on five parameters, has been used:
\begin{equation}
f(x)=p_0 \times \left [ p_1 x^{p_2}(1-x)^{p_3}+(1-p_1)x^{p_4}(1-x)^{p_5}\right ]~,
\label{eq:eqdata}
\end{equation}
where $p_0$ is a normalisation coefficient. Values of the parameters have 
been obtained by comparing, in each bin, the measured bin content with the 
integral of $f(x)$ over the bin. In order to check the effect of a 
given choice of parameterisation, the whole procedure has been repeated,
replacing
the expression of Equation \eqref{eq:eqdata} by another function: a cubic 
spline, with five intervals between $0<x<1$, continuous up to the second 
derivative, normalised to $1$, and forced to be $0$ at $x=0$ and $x=1$. 
This function also depends on five parameters. The results obtained with 
the two parameterisations have been found to be similar~\cite{thesis_eli}. 
The $N$ representation of the fitted function given in
Equation (\ref{eq:eqdata}) is:
\begin{equation}
\tilde{f}(N)=p_0 \left [
p_1 \frac{\Gamma (p_2+N)}{\Gamma (p_2+p_3+N+1)}
+(1-p_1)\frac{\Gamma (p_4+N)}{\Gamma (p_4+p_5+N+1)}
\right ]~.
\label{eq:momdata}
\end{equation}

The Mellin transformed distribution of the perturbative QCD component in a
parton shower Monte-Carlo generator 
has been obtained from the  b-quark $\xp$ distribution generated after
gluon radiation\footnote{In practice, this distribution has been fitted using an expression similar to the one of Equation~\eqref{eq:eqdata}, with three $x^{p_i}(1-x)^{p_j}$ terms, which provided a good description.}.
The NLL QCD perturbative component
has been computed, directly as a function of $N$, in \cite{ref:catani1}. 

The $x$ distribution of the  non-perturbative QCD component extracted in 
the present approach is independent of any hadronic modelling, but it depends
on the procedures adopted to compute the perturbative QCD component. 

\subsection{Results obtained using a generated perturbative QCD component}
\label{sec:gen}

The JETSET~7.3 and PYTHIA~6.156 event generators, with values of the parameters tuned on 
DELPHI data at the $\Zz$ pole, have been both used for this study. Events have been 
produced using the parton shower option of the generator. The corresponding 
non-perturbative QCD component has been extracted, and is displayed in 
Figure~\ref{fig:jetset} for the case of JETSET~7.3. 
The experimental uncertainty on the extracted non-perturbative QCD component is shown as a band.
To estimate this uncertainty, a large number of sets of the parameters $p_{1,..,5}$
has been generated, according to their measured error matrix. This matrix has been 
obtained by propagating the uncertainties of the measured distribution to the 
fitted parameters. The extraction has been performed for each set of parameters. The root mean square of the resulting distributions for a given value of $x$ has been taken as the uncertainty.

\begin{figure}[htbp!]
	\centering
		\includegraphics[width=0.85\textwidth]{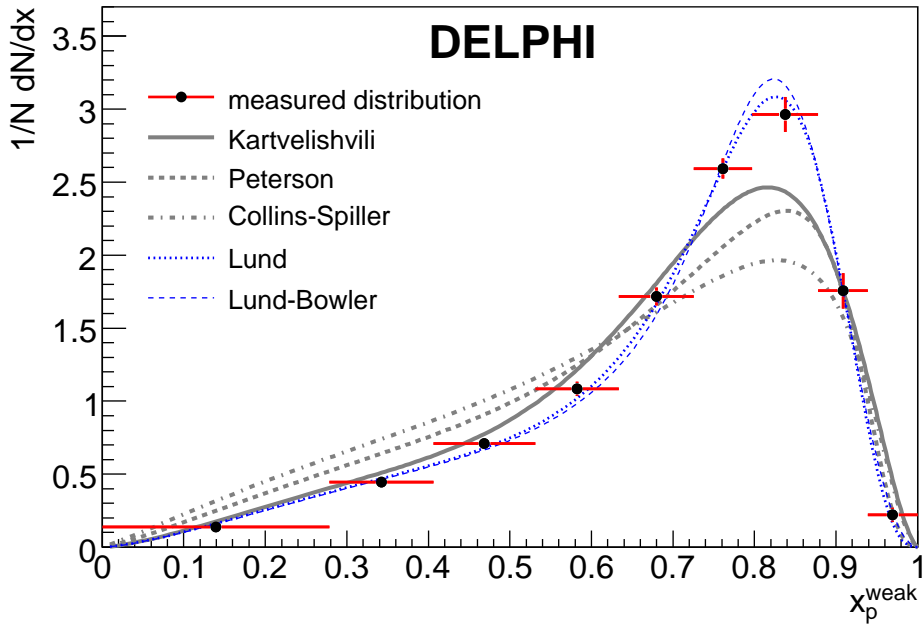}
	\caption{ {The measured $\xp$ distribution (data points), compared to the folding products of fitted hadronisation models with the perturbative QCD component from JETSET~7.3 (curves). The fits have been performed by comparing the integral of the resulting folding product in each
bin to the measured fragmentation distribution bin content.}}
	\label{fig:ModelFitsConv_MC}
\end{figure}

Parameters of several commonly used hadronisation models have been
fitted.
In this case, the same perturbative component 
(as extracted from JETSET~7.3 or PYTHIA~6.156) is used whereas the 
non-perturbative components are taken from models.
These two components are folded according to Equation (\ref{eq:start}). 
The integrals of the resulting folding product in each bin are compared to the measurements,
and values of the model parameters are fitted.
They are given in
Table~\ref{tab:fittedmodels} for both event generators,
for which the fitted parameters differ, in some cases significantly
(illustrating that the non-perturbative component of the fragmentation distribution
depends on the algorithm employed to generate the perturbative component).
The corresponding distributions, obtained
for the different models from the fits with JETSET~7.3, are compared in Figure~\ref{fig:jetset}
with the distribution extracted directly from data, using
the same perturbative QCD input from JETSET.
Figure~\ref{fig:ModelFitsConv_MC} shows the fragmentation distributions that have been compared in the fit: the measured $\xp$ data points and the folding products resulting from fitted hadronisation models.

Data favour the Lund and Lund-Bowler models whereas other parameterisations
are excluded.

It has to be noted that values obtained in this approach for model 
parameters, are compatible with those listed in Table~\ref{tab:modelfits_reweight},
when the same generator is used. The conversion between the Lund and Lund-Bowler $bm_{\mathrm b\bot}^2$
parameters, as fitted here, and the $b$ fitted in Section~\ref{modelfits} is done using
$m_{\mathrm b\bot}^2 = 30.1\ {\rm GeV}^2$, which corresponds to the mean value of $m_{\mathrm b\bot}^2$ in the
generated events. The approximation of a constant $m_{\mathrm b\bot}^2$ is possible due to the small
dispersion of this variable in generated events.

\subsection{Results using a perturbative QCD component obtained by an analytic computation based on QCD}
\label{sec:subsecth}

\begin{figure}[htbp!]
  \begin{center}
  		\includegraphics[width=1.00\textwidth]{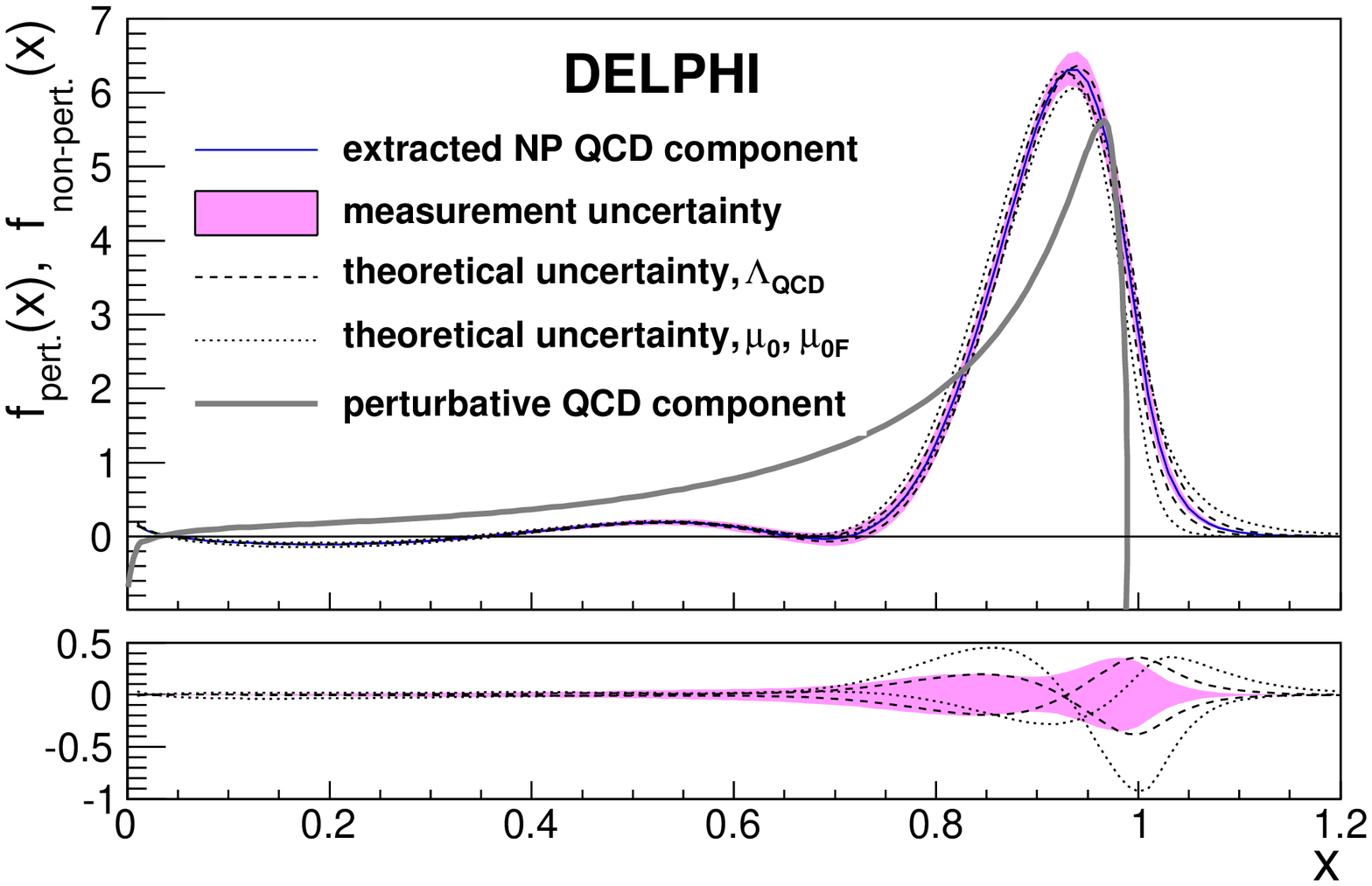}
  \end{center}
  \caption[The extracted non-perturbative QCD component when the perturbative 
one is taken from a theoretical NLL QCD calculation.]{{The $x$-dependence of 
the non-perturbative QCD component of the measured 
b-fragmentation distribution (thin full line).
This curve is obtained by interpolating corresponding values determined at
numerous x values.
The shaded area corresponds to measurement uncertainties.
These uncertainties are correlated for different $x$-values. 
The perturbative QCD component (thick full line) is given by the analytic
computation of \cite{ref:catani1}.
It has to be complemented by a $\delta$-function containing 5$\%$ of the events,
located at $x=1$.
The thin lines on both sides of the non-perturbative 
distribution correspond to $\mu_0=\mu_{0F}=\{m_b/2,2m_b\}$ (dotted lines) 
and $\Lambda^{(5)}_{{\rm QCD}}=(0.226\pm0.025) ~\GeV$ (dashed lines). Variations induced by the other parameters,
$\mu=\mu_F=\{Q/2,2Q\}$ and $m_b=(4.75\pm0.25)~\GeV/c^2$
are smaller. The lower plot shows the variation of the different uncertainties.
Note that the variable $x$ used to display the variation of the different distributions
is not $x_p^{\mathrm weak}$, but the integration variable from Equation~(\ref{eq:start}).
}
   \label{fig:npx}}
\end{figure}

\begin{figure}[htb!]
	\centering
		\includegraphics[width=0.85\textwidth]{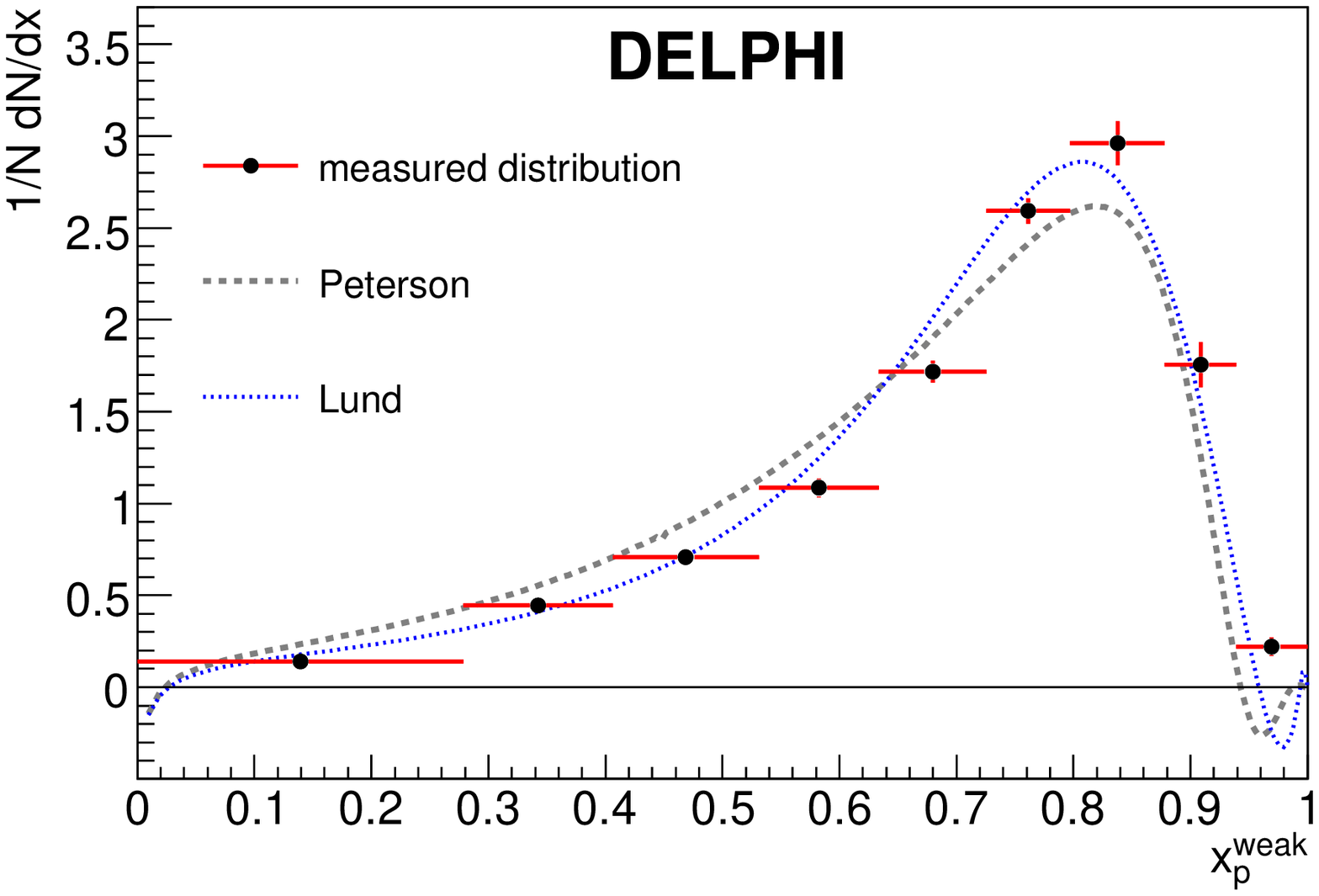}
  \caption{ {
  Fit of the $\xp$-dependence of the fragmentation distribution 
using an analytic evaluation of the perturbative component from theory, and
a model for the non-perturbative component.
The histogram with error bars gives
the measured $x_P$ distribution. The curves correspond to the folding products of the Lund and Peterson
hadronisation models with the perturbative QCD component. The fitted parameters 
for the Lund model are $a=0.551$ and $b=10.27~\bunit$. For the Peterson model, 
$\epsilon_{\mathrm b} = 0.0038$ has been obtained.
The negative parts of the folding products are due to the unphysical region of the
perturbative component. 
}
   \label{fig:ModelFitsConv_theo}}
\end{figure}

The perturbative QCD fragmentation function is evaluated according to the approach 
presented in  \cite{ref:catani1}. 
Computations are done directly in the $N$-space and are 
expected to be reliable when $|N|$ is not too large (typically less than 20).
This function has singularities at large values of $Re(N)$; in particular, a
zero is present at $Re(N)=N_0\simeq 41.7$ (this number depends on the exact
values assumed for the other parameters entering into the computation).
 To obtain distributions for the variable $x$ from results in moment space
the inverse Mellin transformation is applied, that consists in
integrating over a contour in $N$ (see Equation \eqref{eq:inv_mellin}).
The choice of the contour has to take into account the presence of
singularities at large $Re(N)$.
When $x$ gets close to 1, large values of $|N|$ contribute and thus the 
perturbative fragmentation distribution is not reliable in these regions. 
This behaviour affects also values of the distribution at lower $x$, 
which can be understood by the fact that 
moments of the total distribution are fixed.
Unlike the perturbative QCD component which is  defined in \cite{ref:catani1} 
within the $[0,1]$ interval, the non-perturbative component has to be 
extended in the region $x>1$.
This ``non-physical'' behaviour
comes from the zero of $\tilde{f}_{pert.}(N)$ for $N=N_0$ which gives a 
pole in the expression to be integrated in Equation \eqref{eq:inv_mellin}. 
In other words, this behaviour is directly related to the break-down of the theory
for large values of $|N|$ (i.e. for $x$ close to $1$).
Using properties of integrals in the complex plane, it can be shown that, 
for $x>1$, the non-perturbative QCD distribution can be well approximated 
by $x^{-N_0}$. Uncertainties attached to the determination of the theoretical perturbative 
QCD component are related to the definition of the scales entering into the 
computation \cite{ref:benhaim,ref:catani1}.

The extracted non-perturbative component is given in Figure \ref{fig:npx}. Its shape
depends on the same quantities as those used to evaluate the perturbative distribution, and thus similar variations appear, as drawn also in the Figure. The measurement uncertainty band has been obtained using the 
same procedure as explained in Section \ref{sec:gen}.

It has to be noted that the data description in terms of a product of two 
QCD components, perturbative and non-perturbative, is not directly affected 
by uncertainties attached to the determination of the perturbative component. 
This is because the non-perturbative component, as determined in the present 
approach, compensates for a given choice of method or of parameter values.

This study indicates that a perturbative QCD component
obtained analytically from theory must not be folded with a 
non-perturbative QCD component
taken from a model.
All model distributions are physical and cannot
compensate for unphysical behaviour of the perturbative computation.
This is illustrated in Figure \ref{fig:ModelFitsConv_theo},
where the measured distribution is shown together with the folding product of the NLL QCD perturbative component and hadronisation models. Model parameters have been obtained from fits to data using the folding product of Equation (\ref{eq:start}), in the same way as in Section~\ref{sec:gen}.
The negative parts of the folding products are due to the unphysical behaviour of the perturbative component.

 \section {Combined fit of results from all experiments}
\label{sec:world_average}

\begin{figure}[htb]
  \begin{center}
    \mbox{\epsfig{file=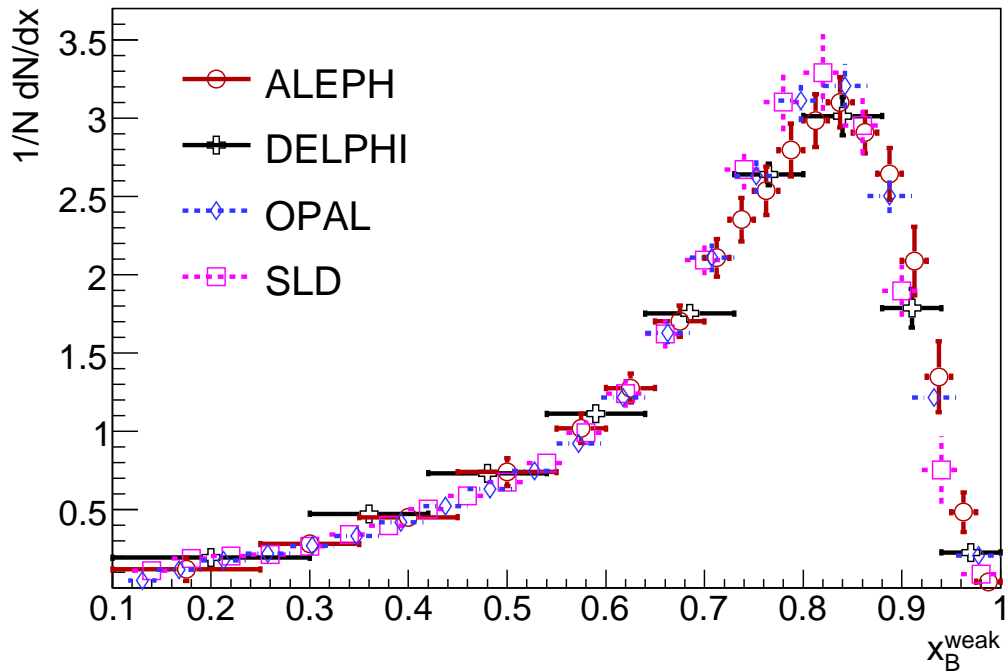,height=10cm}}
  \end{center}
  \caption{{{Comparison between the various measurements of the b-quark
 fragmentation distribution versus $\xb$.  
}}
   \label{fig:Orsay_Karl_all}}
\end{figure}

In addition to the DELPHI results presented in this paper, the b-quark fragmentation distribution has been measured by ALEPH, OPAL and SLD \cite{ref:aleph1, ref:opal1, ref:sld1}. 
These measurements are displayed in Figure \ref{fig:Orsay_Karl_all}.
Each experiment used a different technique. These four results have been fitted to give a world average b-quark
 fragmentation distribution and to determine the corresponding non-perturbative QCD component. Global fits of the Lund  and Lund-Bowler parameterisations have been also obtained.

\subsection{Combined fit to fragmentation distributions}
\label{sec:world_average_3gamma}

Each of the four measurements of the b-quark fragmentation distribution is given with a different choice of binning and has a different number of effective degrees of freedom.
In order to obtain a combined distribution,
a global fit has been done, using the smooth parameterisation of Equation~\eqref{eq:eqdata}. The $\chi^2$ minimised in the fit is the sum of $\chi^2$ for the different experiments, computed by comparing, in each bin, the integral of the parameterisation to the measured bin content.
The number of degrees of freedom for ALEPH, DELPHI and OPAL are $7$, $7$ and $5$, respectively.
When one more degree of freedom is used for one of the three experiments, fits show a large increase in the $\chi^2$ value.
For SLD, the diagonal error matrix of the $22$ bin values has been used, as the full error matrix was not detailed in \cite{ref:sld1}. 
A comparison of the sources of systematic uncertainties between the 
different analyses shows that, due to the various methods which have been used \cite{ref:aleph1, ref:opal1, ref:sld1}, their origins are very different\footnote{The correlated and the total systematic uncertainties are 
 $\pm 0.0012$ and 
$(+0.0038,~-0.0033)$ for OPAL;  and $\pm 0.0009$ and $\pm 0.0027$ respectively 
for SLD.
The ALEPH measurement uses B-meson semileptonic decays and there is
almost no correlation with the other three results.}.
As a result, systematic uncertainties from the different experiments have been supposed to be uncorrelated. 
The fit has been done using both the $\xb$ and $\xp$ distributions. 
For $36$ degrees of freedom, the fit of $\xb$ yields a $\chi^2$ of $55.8$ (probability of $2\%$), and the one for $\xp$ yields a $\chi^2$ of $67.7$ (probability of $0.1\%$).
The large $\chi^2$ values are not likely to originate from the smooth function itself,
as this function gives a good fit quality to all the individual distributions.
The minimum $\chi^2$ probability obtained in these fits is $31\%$.
This marginal compatibility comes rather from the dispersion of the results
mainly between ALEPH and SLD measurements which are respectively peaked
on the high and low sides of the distribution. Quoted uncertainties for $\xb$ and $\xp$ in the following are therefore rescaled by factors $1.24$ and $1.37$, respectively.

\begin{table}[htb!]
\begin{center}
\begin{tabular}{|c|c|c|}
 \hline
          & $\xb$                 & $\xp$                  \\
 \hline
    $p_1$ & $12.97^{+0.77 }_{-0.71 }$ & $12.50^{+0.82 }_{-0.76 }$   \\
    $p_2$ & $2.67 ^{+0.15 }_{-0.14 }$ & $2.63 ^{+0.17 }_{-0.15 }$   \\
    $p_3$ & $2.29 ^{+0.19 }_{-0.17 }$ & $2.05 ^{+0.19 }_{-0.18 }$   \\
    $p_4$ & $1.45 ^{+0.28 }_{-0.22 }$ & $1.31 ^{+0.24 }_{-0.20 }$   \\
    $p_5$ & $0.663^{+0.035}_{-0.036}$ & $0.664 \pm 0.036        $   \\ \hline
\end{tabular}  \caption[Parameters for the combined world average fragmentation distribution.]{{Parameters for the combined world average fragmentation distribution. The quoted uncertainties have been rescaled by factors $1.24$ and  $1.37$ for $\xb$ and $\xp$, respectively, as explained in the text.}}
  \label{tab:param_world_average}
\end{center}
\end{table}

\begin{table}[htb]
\begin{center}
\begin{tabular}{|c|cccccc|}
 \hline
parameter & $p_1$ & $p_2$ & $p_3$ & $p_4$ & $p_5$ & {      }\\
 \hline
    $p_1$ & $0.546693$  &              &               &               &              & \\
    $p_2$ & $0.102357$  & $ 0.021477$  &               &               &              & \\
    $p_3$ & $0.008932$  & $-0.002099$  & $ 0.032990$   &               &              & \\
    $p_4$ & $-0.088755$ & $-0.018877$  & $ 0.031327$   &  $ 0.059791$  &              & \\
    $p_5$ & $0.020556$  & $ 0.003772$  & $-0.001624$   &  $-0.007269$  &   $0.001277$ & \\ \hline
\end{tabular}  \caption[Error matrix on the fitted parameters of the combined world average fragmentation distribution for $\xb$.]{{Error matrix on the fitted parameters of the combined world average fragmentation distribution for the variable $\xb$. Quoted uncertainties are obtained after applying the scaling factor of $1.24$, as explained in the text.}}
  \label{tab:errmat_world_average_xe}
\end{center}
\end{table}

\begin{table}[htb]
\begin{center}
\begin{tabular}{|c|cccccc|}
 \hline
parameter & $p_1$ & $p_2$ & $p_3$ & $p_4$ & $p_5$ & {      }\\
 \hline
    $p_1$ & $0.624398$  &              &               &              &             & \\
    $p_2$ & $0.118819$  & $ 0.025406$  &               &              &             & \\
    $p_3$ & $0.035693$  & $ 0.002269$  & $ 0.032581$   &              &             & \\
    $p_4$ & $-0.070514$ & $-0.016499$  & $ 0.024564$   &  $ 0.047581$ &             & \\
    $p_5$ & $0.022370$  & $ 0.004205$  & $-0.000181$   &  $-0.006008$ &   $0.001316$& \\ \hline
\end{tabular}  \caption[Error matrix on the fitted parameters of the combined world average fragmentation distribution for $\xp$.]{{The error matrix on the fitted parameters of the combined world average fragmentation distribution for the variable $\xp$. Quoted uncertainties are obtained after applying the scaling factor of $1.37$, as explained in the text.}}
  \label{tab:errmat_world_average_xp}
\end{center}
\end{table}

The fitted parameters $p_1,..,p_5$ of Equation~\ref{eq:eqdata}, that represent the world average fragmentation distribution for $\xb$ and $\xp$ are given in Table~\ref{tab:param_world_average}. The full error matrices on $p_1,..,p_5$ for the two distributions are given in Tables \ref{tab:errmat_world_average_xe} and \ref{tab:errmat_world_average_xp}.

The moments of the combined $\xb$ distribution are given in Table~\ref{tab:mom_world_average}. They have been calculated using the analytic expression of Equation \eqref{eq:momdata} with the fitted values of the parameters $p_1,..,p_5$. 
The uncertainties have been propagated from the error matrix on the parameters. 
\begin{table}[htb]
	\centering
		\begin{tabular}{|c|c|c|c|c|c|c|}
			\hline
N    & moment  & uncertainty  & &   N    & moment  & uncertainty \\
\hline
\hline
    &           &            & & 21  & $0.0333$  &  $0.0011$ \\
2   & $0.7092$  &  $0.0025$  & & 22  & $0.0301$  &  $0.0010$ \\
3   & $0.5334$  &  $0.0031$  & & 23  & $0.0272$  &  $0.0009$ \\
4   & $0.4158$  &  $0.0032$  & & 24  & $0.0247$  &  $0.0009$ \\
5   & $0.3322$  &  $0.0031$  & & 25  & $0.0225$  &  $0.0008$ \\
6   & $0.2703$  &  $0.0029$  & & 26  & $0.0205$  &  $0.0008$ \\
7   & $0.2231$  &  $0.0027$  & & 27  & $0.0188$  &  $0.0008$ \\
8   & $0.1864$  &  $0.0026$  & & 28  & $0.0172$  &  $0.0007$ \\
9   & $0.1573$  &  $0.0024$  & & 29  & $0.0158$  &  $0.0007$ \\
10  & $0.1339$  &  $0.0022$  & & 30  & $0.0146$  &  $0.0006$ \\
11  & $0.1148$  &  $0.0021$  & & 31  & $0.0134$  &  $0.0006$ \\
12  & $0.0992$  &  $0.0019$  & & 32  & $0.0124$  &  $0.0006$ \\
13  & $0.0862$  &  $0.0018$  & & 33  & $0.0115$  &  $0.0006$ \\
14  & $0.0754$  &  $0.0017$  & & 34  & $0.0107$  &  $0.0005$ \\
15  & $0.0662$  &  $0.0016$  & & 35  & $0.0099$  &  $0.0005$ \\
16  & $0.0585$  &  $0.0015$  & & 36  & $0.0092$  &  $0.0005$ \\
17  & $0.0518$  &  $0.0014$  & & 37  & $0.0086$  &  $0.0005$ \\
18  & $0.0462$  &  $0.0013$  & & 38  & $0.0080$  &  $0.0004$ \\
19  & $0.0413$  &  $0.0012$  & & 39  & $0.0075$  &  $0.0004$ \\
20  & $0.0370$  &  $0.0011$  & & 40  & $0.0070$  &  $0.0004$ \\ \hline
		\end{tabular}
	\caption[Moments of the world average fragmentation distribution for $\xb$.]{{Moments of the world average fragmentation distribution for $\xb$. $N=2$ 
corresponds to the mean value of the distribution. The moments have been calculated using the analytic expression of Equation \eqref{eq:momdata}. Uncertainties have been propagated using the error matrix of Table \ref{tab:errmat_world_average_xe}. }}
	\label{tab:mom_world_average}
\end{table}
The $\xb$ distribution vanishes in the interval $\xb\in[0.,~0.116=2m_B/\sqrt{s}\:]$ whereas the parameterisation of Equation~\eqref{eq:eqdata} has non zero values in this interval. Nevertheless, the effect of this caveat has been found to be negligible.
The integral of the fitted parameterisation in the region $[0.,~0.116]$ is $0.0019$. The contribution of this region to the average value of the distribution is of ${\cal O}(10^{-4})$, which is $\sim\!20$ times smaller than the uncertainty. The effect decreases rapidly for higher order moments (${\cal O}(10^{-12})$ for the $10^{th}$ moment).

%
\subsection{Combined fit of the Lund and Lund-Bowler models}
\label{sec:world_average_lund}

\begin{figure}[htb!]
  \centering
  \subfloat[][]{\includegraphics[width=0.85\textwidth]{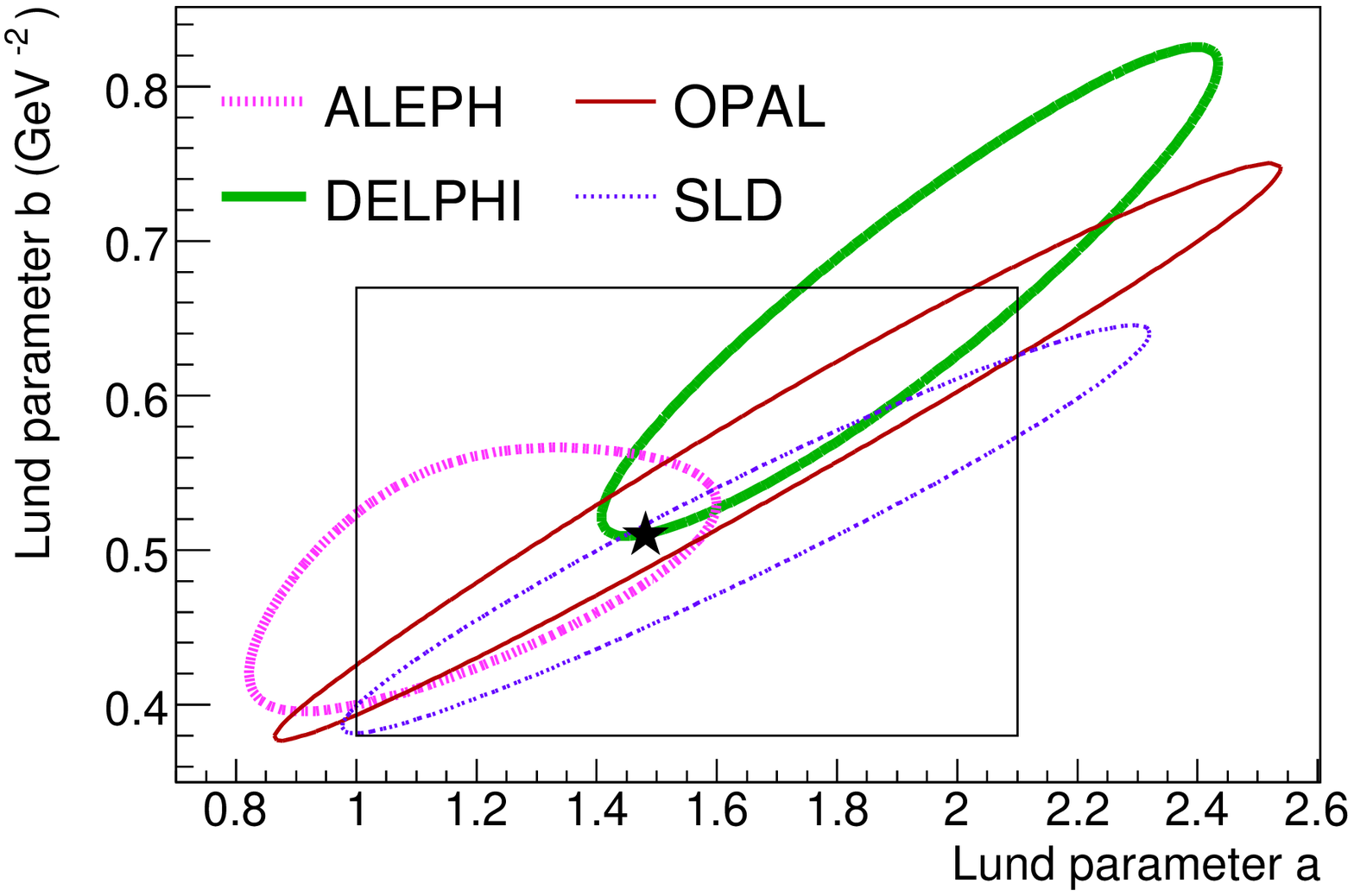}}\\
  \subfloat[][]{\includegraphics[width=0.85\textwidth]{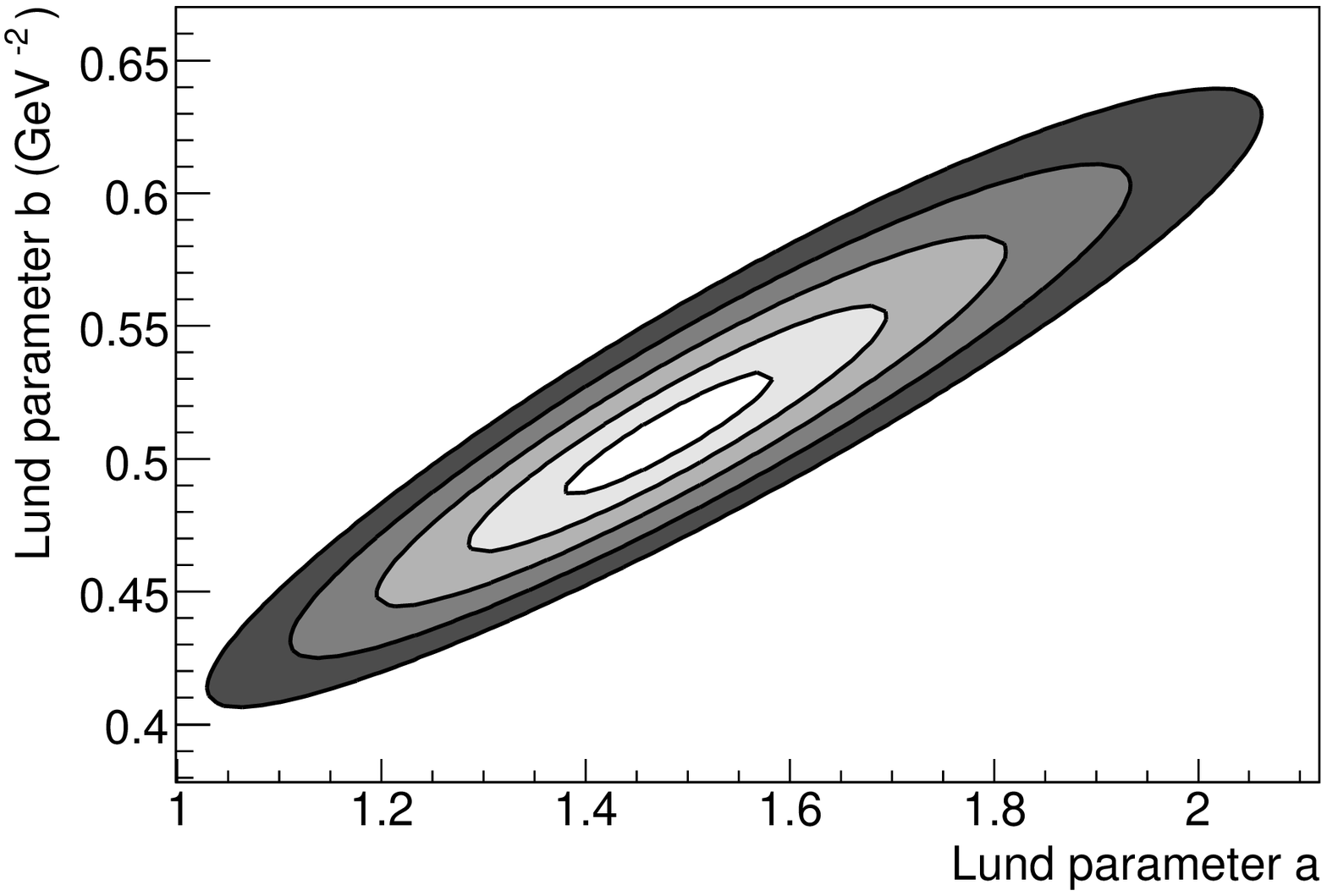}}
  \caption{{\bf a} Contours of $68.3\%$ coverage probability for the $a$ and $b$
	Lund parameters corresponding to a separate fit to each experiment 
	and the result obtained in the combined fit marked by $\star$.
	{\bf b} Contours varying from $1$ standard deviation (lightest grey)
	to $5$ standard deviations
	(darkest grey) for the $a$ and $b$ Lund parameters obtained in the combined fit.
	These contours
	correspond to coverage probabilities of $39.3\%$, $63.2\%$, $77.7\%$, $86.5\%$ and
	$91.8\%$. The box drawn in {\bf a} corresponds to the area presented 
	in {\bf b}.}
  \label{fig:WorldAverageLundContours}
\end{figure}
The $a$ and $b$ parameters of the Lund and Lund-Bowler fragmentation distributions,
in the framework of PYTHIA~6.156, have been fitted using
the weighting approach described in Section~\ref{modelfits} and minimising the sum of $\chi^2$ for the four experiments. For the Lund-Bowler model, the predicted value $r_Q = 1$~\cite{ref:bowler1} has been used.
The fit yields the Lund parameters $a= 1.48^{+0.11}_{-0.10}$; $b=0.509^{+0.024}_{-0.023}\ \bunit$
with a statistical correlation of $\rho = 92.6\%$.
For the Lund-Bowler parameters the result is $a= 0.795^{+0.062}_{-0.059}$; $b=2.28^{+0.13}_{-0.12}\ \bunit$
with a statistical correlation of $\rho = 84.2\%$. The minimum $\chi^2$ is $39.5$ with $39$ degrees of freedom (probability of $44.6\%$) for the Lund model and $92.4$ for the Lund-Bowler model.
Experimental data clearly favours the Lund model to the Lund-Bowler (with $r_Q = 1$) one.

Figure~\ref{fig:WorldAverageLundContours} shows the $68.3\%$ coverage probability contours for the Lund parameters $a$ and $b$ , as obtained individually for each experiment, compared to the combined fit result. The one to five standard deviations contours of the fitted parameters are also presented.

Using the parametric form fitted on all
measured $x_p^{\mathrm weak}$ distributions, given in 
Table \ref{tab:param_world_average},
and the direct folding approach described in Section \ref{sec:gen}, the
fitted values of the parameters for the Lund fragmentation distribution
are: $a= 1.48 \pm 0.10$; $bm_{{\mathrm b}\perp}^2=16.62 \pm 0.71$
with a statistical correlation of $\rho = 90.7\%$. These values are very
similar to those given at the beginning of this section,
which were obtained by reweighting 
the simulation event-by-event.

\subsection {Extraction of the non-perturbative QCD component from the combined distribution}
\label{sec:np_world_average}

\begin{figure}[htbp!]
	\centering
		\subfloat[][]{\includegraphics[width=0.78\textwidth]{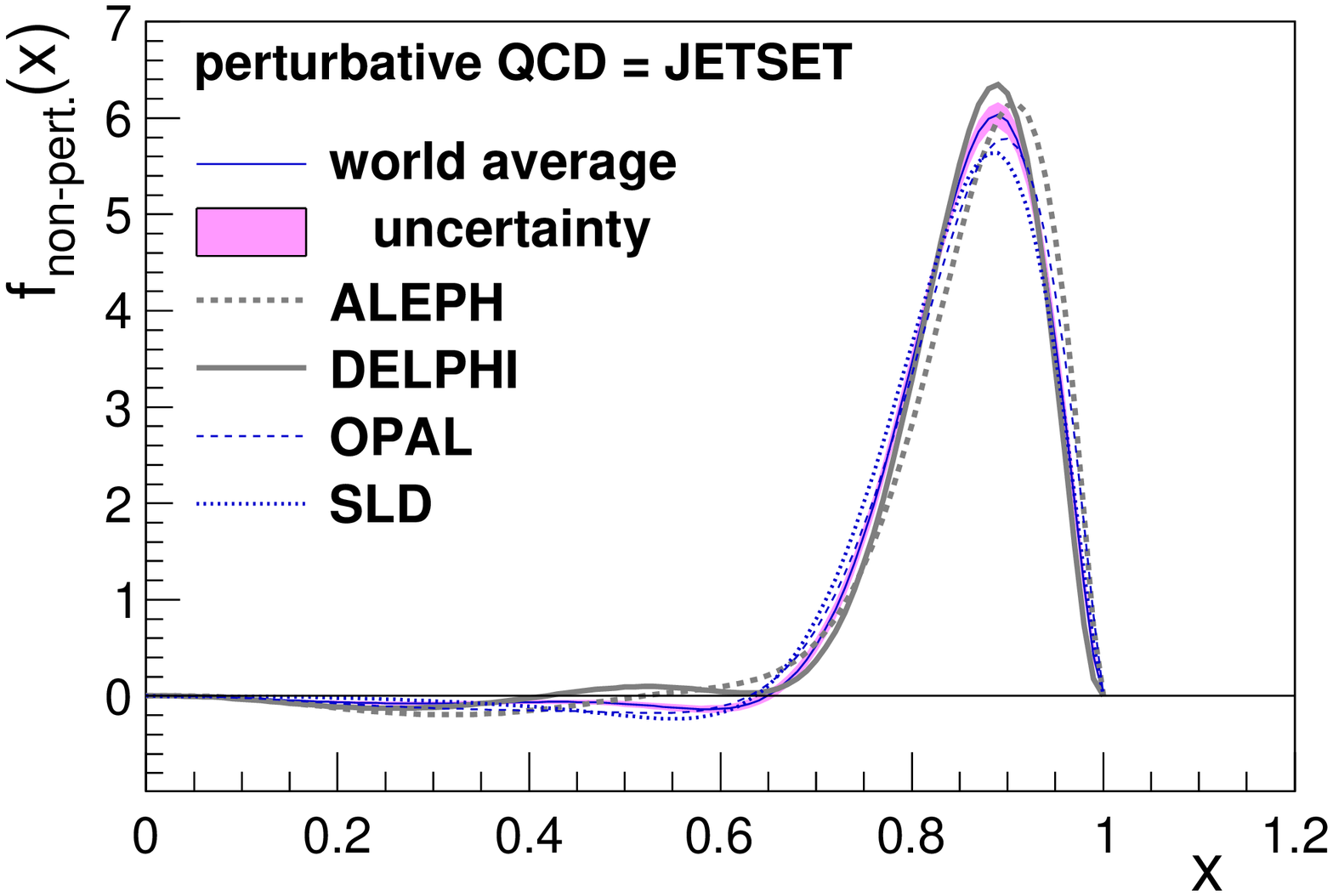}}\\
		\subfloat[][]{\includegraphics[width=0.78\textwidth]{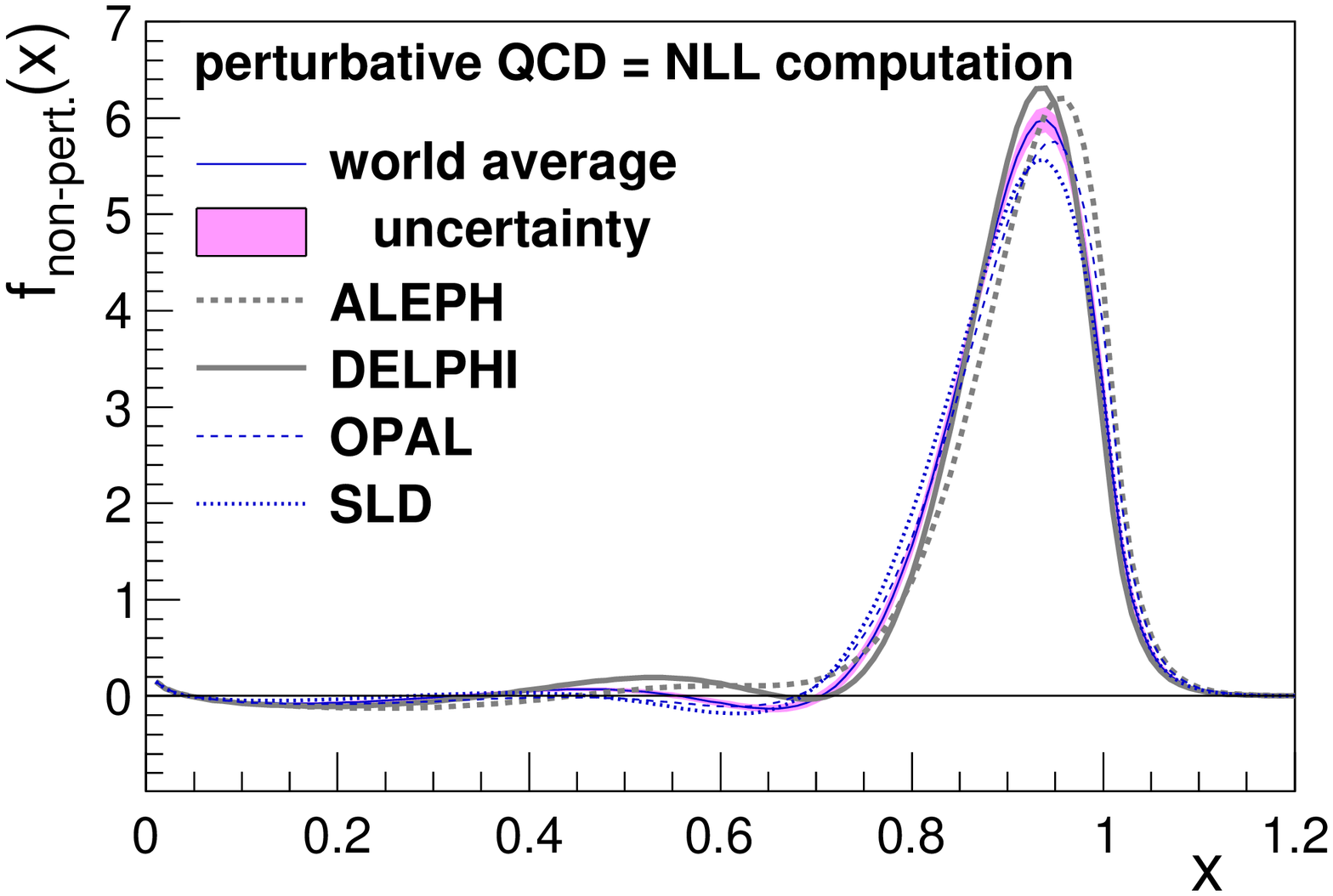}}
\caption[Comparison of the extracted non-perturbative QCD component for different experiments.]{{Comparison of the extracted non-perturbative QCD component of the b-quark fragmentation function for the result from the present analysis, ALEPH \cite{ref:aleph1}, OPAL \cite{ref:opal1}, SLD \cite{ref:sld1} and the combined $\xp$ distribution. {\bf a} The perturbative QCD component has been taken from JETSET~7.3. {\bf b} The perturbative QCD component has been taken from NLL QCD \cite{ref:catani1}. The shaded error bands represent the experimental uncertainty of the combined distributions.}}
	\label{fig:comp_manip_np}
\end{figure}

\begin{figure}[htbp!]
	\centering
		\includegraphics[width=0.78\textwidth]{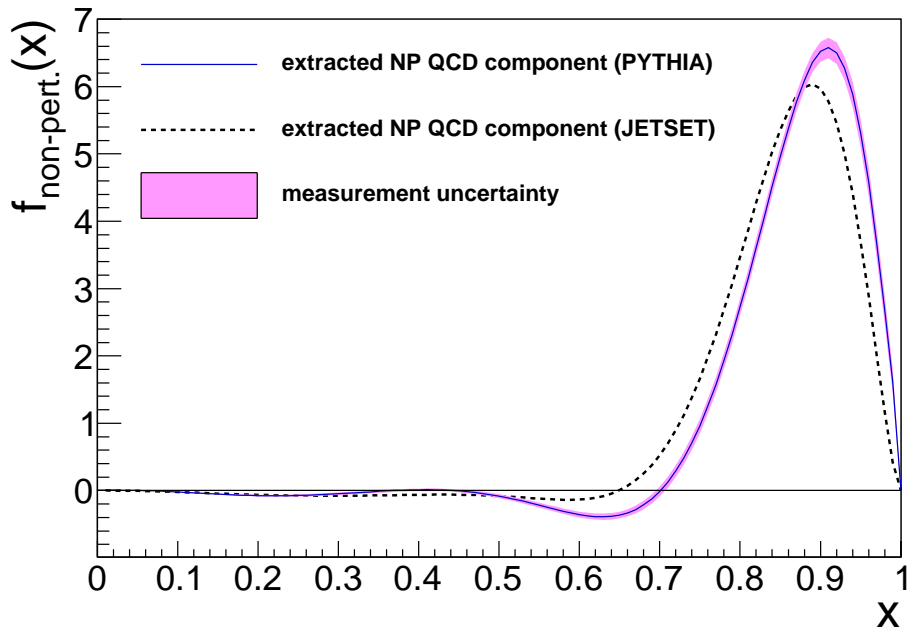}
	\caption{{ Comparison of the extracted non-perturbative QCD components from the combined $\xp$ distribution, corresponding to the perturbative QCD components from JETSET~7.3 and PYTHIA~6.156 parton shower Monte-Carlo. The latter is given with its corresponding error band.}}
	\label{fig:WorldAvgExtNp_PythiaAndJetset}
\end{figure}

\begin{figure}[htbp]
	\centering
		\includegraphics[width=0.78\textwidth]{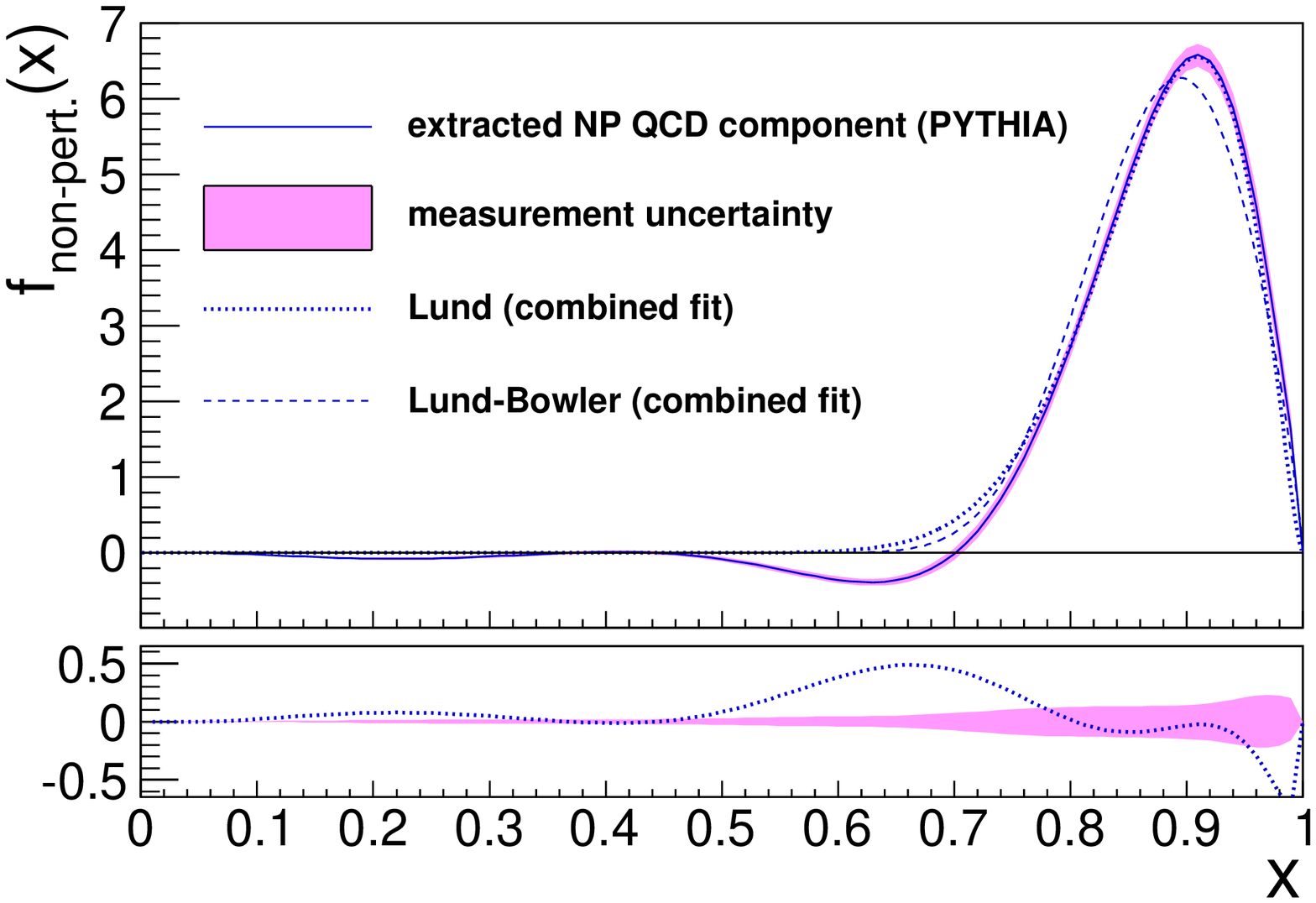}
	\caption{{The extracted non-perturbative QCD component corresponding to PYTHIA 6.156 parton shower Monte-Carlo (presented with error band), compared with the Lund and Lund-Bowler models, obtained from the combined fit. The Lund model is scaled by $1.068$, which corresponds to the integral of the extracted function in the region $x\in[0.701, 1.0]$, where it has positive values. 
For this comparison, $m_{b \bot}^2$ has been taken as $30.1\ {\rm GeV}^2$, which corresponds to its average value in generated events used for the fit.
The lower plot shows the difference between the extracted non-perturbative QCD component corresponding to the PYTHIA 6.156 parton shower Monte-Carlo and the fitted Lund model.
}}
	\label{fig:WorldAvg_extNPandLund}
\end{figure}

The non-perturbative QCD component has been extracted from the combined fragmentation function obtained in Section \ref{sec:world_average_3gamma} and from 
the b-quark fragmentation functions measured separately by
each experiment at  $e^+e^-$ colliders \cite{ref:aleph1, ref:opal1, ref:sld1}, including the result from the present analysis. The extraction has been performed using the $\xp$ distributions,  following the prescription given in Sections \ref{sec:gen} and \ref{sec:subsecth} using the two perturbative QCD approaches considered in this paper: parton shower Monte-Carlo (both JETSET~7.3 and PYTHIA~6.156) and the theoretical NLL QCD computation from \cite{ref:catani1}. Comparisons of the results obtained for the combined distribution and for the different experiments are shown in Figure \ref{fig:comp_manip_np}.
The extracted distributions corresponding to JETSET~7.3 and PYTHIA~6.156 parton shower Monte-Carlo are compared in Figure~\ref{fig:WorldAvgExtNp_PythiaAndJetset}, and show significant differences. The former is clearly softer than the latter. Finally, the extracted distribution corresponding to PYTHIA~6.156 is compared
in Figure~\ref{fig:WorldAvg_extNPandLund}
with the Lund and Lund-Bowler models, obtained in Section~\ref{sec:world_average_lund} from the combined fit. There is a good agreement between the Lund model and the extracted non-perturbative QCD component in the peak region. However, from the above figures, together with Figure \ref{fig:jetset} it is quite
apparent that, in spite of some dispersion, all measurements favour a 
model similar to the Lund or Lund-Bowler shapes. All other parameterisations
are excluded.

 \section{Conclusion}

The fragmentation distribution of the b-quark has been measured
using two analyses based on very different approaches. 
The average fractions of the beam energy taken by weakly decaying 
b-hadrons are:
\begin{equation}
\langle \xb \rangle = 0.7140 \pm 0.0007{\rm (stat.)} \pm 0.0060{\rm (syst.)}
\end{equation}
and
\begin{equation}
\langle \xb \rangle = 0.6978 \pm 0.0010{\rm (stat.)} \pm 0.0064{\rm (syst.)}~,
\end{equation}
in the regularised unfolding and weighted fitting analyses, respectively.

The combined $\xb$ distribution has been obtained by a fit using the full error
matrix of the two analyses. The average value of this distribution is equal to:
\begin{equation}
\langle \xb \rangle =  0.699 \pm 0.011~.
\end{equation}

The non-perturbative QCD b-quark fragmentation 
distribution is obtained from
the combined DELPHI measurement in a way that does not depend on any
non-perturbative hadronisation model. In order to describe the measured
fragmentation function, the non-perturbative distribution has to be folded
with the adequate perturbative component, which can be given by a
generator,
as provided for example by the JETSET/PYTHIA parton shower
or by an analytic
perturbative QCD computation. It has been demonstrated that for the latter,
it is not legitimate to use a physical distribution, as given by 
commonly used hadronisation models,
for the non-perturbative QCD component. Parameters obtained for the non-perturbative component depend on the choice for the perturbative evaluation.
This has been illustrated by comparing results obtained with JETSET~7.3, 
PYTHIA~6.156 and analytic QCD computation.

The distributions obtained by folding analytically the perturbative and the non-perturbative
QCD components are found to be similar to the ones obtained by a Monte-Carlo generator.
We stress that the non-perturbative component depends on the exact procedure used to obtain the perturbative one.

The combined measurement from DELPHI has been compared with expectations from
different non-perturbative hadronisation models of the b-quark fragmentation
distribution within a Monte Carlo simulation. Only the Lund and Lund-Bowler
models give reasonable descriptions of the data, the Lund ansatz being
favoured. The parameters of the Lund fragmentation 
that fit best the data are obtained 
within the framework
of PYTHIA~6.156 to be:
\begin{equation}
a= 1.84^{+0.23}_{-0.21}~{\rm and} ~b=0.642^{+0.073}_{-0.063}\ \bunit
\end{equation}
with a  correlation factor $\rho = 92.2\%$.

The present measurement is combined with previous results 
from the ALEPH, OPAL and SLD experiments and 
a world averaged b-quark fragmentation distribution is obtained
giving:
\begin{equation}
\langle \xb \rangle =  0.7092\pm0.0025.
\end{equation}

The 
corresponding non-perturbative QCD component is also determined.
Particularly, a global fit to all the available fragmentation distributions
is performed to obtain the parameters of the Lund fragmentation model.
\begin{equation}
a= 1.48^{+0.11}_{-0.10}~{\rm and} ~b=0.509^{+0.024}_{-0.023}\ \bunit
\end{equation}
with a correlation factor $\rho = 92.6\%$.

\subsection*{Acknowledgements}
\vskip 3 mm
We are greatly indebted to our technical collaborators, to the members of the
CERN-SL Division for the excellent performance of the LEP collider, and to the funding agencies for their support in building and operating the DELPHI detector.\\
We acknowledge in particular the support of \\
Austrian Federal Ministry of Education, Science and Culture,
GZ 616.364/2-III/2a/98, \\
FNRS--FWO, Flanders Institute to encourage scientific and technological 
research in the industry (IWT) and Belgian Federal Office for Scientific, 
Technical and Cultural affairs (OSTC), Belgium, \\
FINEP, CNPq, CAPES, FUJB and FAPERJ, Brazil, \\
Ministry of Education of the Czech Republic, project LC527, \\
Academy of Sciences of the Czech Republic, project AV0Z10100502, \\
Commission of the European Communities (DG XII), \\
Direction des Sciences de la Mati$\grave{\mbox{\rm e}}$re, CEA, France, \\
Bundesministerium f$\ddot{\mbox{\rm u}}$r Bildung, Wissenschaft, Forschung 
und Technologie, Germany,\\
General Secretariat for Research and Technology, Greece, \\
National Science Foundation (NWO) and Foundation for Research on Matter (FOM),
The Netherlands, \\
Norwegian Research Council,  \\
State Committee for Scientific Research, Poland, SPUB-M/CERN/PO3/DZ296/2000,
SPUB-M/CERN/PO3/DZ297/2000, 2P03B 104 19 and 2P03B 69 23(2002-2004),\\
FCT - Funda\c{c}\~ao para a Ci\^encia e Tecnologia, Portugal, \\
Vedecka grantova agentura MS SR, Slovakia, Nr. 95/5195/134, \\
Ministry of Science and Technology of the Republic of Slovenia, \\
CICYT, Spain, AEN99-0950 and AEN99-0761,  \\
The Swedish Research Council,      \\
The Science and Technology Facilities Council, UK, \\
Department of Energy, USA, DE-FG02-01ER41155, \\
EEC RTN contract HPRN-CT-00292-2002. \\


\clearpage
\begin{appendix}
\section*{Appendix A: Input variables of the $E_B^{\rm{weak}}$~reconstruction neural network}
\label{appena}

Below is a listing of the main input variables to the  neural network 
used to reconstruct $E_B^{\rm{weak}}$ in the regularized unfolding analysis. This variable is an output of the
DELPHI inclusive b-physics package BSAURUS and further details of the general approach
and specifics of the  reconstructed quantities used can be found in~\cite{bsaurus}.
Central to this approach are: (a) the accurate reconstruction of a secondary
vertex in each hemisphere as a candidate b-hadron decay vertex and (b)
the reconstruction of the quantity: TrackNet. The TrackNet is itself a neural network
variable providing, for each charged particle in an event hemisphere, a number
ranging between zero and one, related to the probability that the particle
originates from the decay of a b-hadron.

\begin{itemize}
\item{Rapidities\footnote{ Defined as
$y=\frac{1}{2} \cdot \log\left((E+p_{\|})/(E-p_{\|})\right)$,
where $p_{\|}$~is the momentum component of the particle in the direction of
the b-quark. The direction is estimated as the axis of the jet associated
with the b-hadron.}
of the particles with the highest, and next highest, values in the event hemisphere.}

\item{In $2$-jet events, the weakly decaying b-hadron energy as estimated by a
TrackNet-weighted sum of charged particle
energies in the event hemisphere added to a rapidity-weighted sum for neutral 
particles. For events with more than $2$ jets, the energy as reconstructed by 
the {\it rapidity algorithm} is taken, which sums over all particles in a
hemisphere that pass a selection cut of $y>1.6$. This value provides an 
efficient separation of particles likely to have  originated in b-hadron decays,
from those produced in the fragmentation process. Note that jets were
reconstructed using the LUCLUS algorithm based on a transverse momentum cutoff
parameter value of $d_{\rm join}={\rm PARU}(44)=5.0$~GeV/$c$.}

\item{Mass of the weakly decaying b-hadron based on all particles with $y>1.6$. }

\item{Energy of the jet with the highest $b-$tag value calculated by
    applying the standard DELPHI b-tagging procedure~\cite{btag} to the jet.}

\item{
The summed charged and neutral energy reconstructed in the event hemisphere and in the event as a whole.
}

\item{An estimate of the missing $p_T$~between the b-hadron candidate direction
 and  the thrust axis, calculated using only fragmentation particles (identified
 by demanding ${\rm TrackNet}<0.5$) in the same hemisphere as the b-hadron candidate and all particles
in the  opposite hemisphere. }

\item{The mass of the reconstructed secondary vertex.}

\item{The polar angle of the b-hadron candidate momentum vector.}

\item{The thrust value of the event.}

\item{The total number of all charged and neutral particles.}

\item{The number of particles passing a TrackNet cut of $>0.5$.}

\item{For the best electron or muon candidate in the
hemisphere, with the correct charge correlation,  
a number related to the probability that it originates from the
b-hadron candidate. Assuming that the track
comes from the primary vertex,
this number is the probability of getting an impact parameter 
significance at least as large as that observed.}

\item{The gap in rapidity between the particle of highest rapidity in the
    hemisphere (with
  TrackNet value less than 0.5) and the particle of lowest rapidity 
(with TrackNet value greater than 0.5).}

\end{itemize}
Note that some variables used as inputs are not directly
correlated with the  b-hadron energy but instead provided the network
with information about how {\it reliable} the other input variables might be,
e.g.  measures of the total hemisphere energy and the number of particles in the hemisphere that  fail selection cuts. 
The network was thus able to learn
during the training phase, e.g. to give extra weight
to the variables of event hemispheres when there is a good chance of a
hemisphere containing a well reconstructed b-hadron.  

\begin{figure}[htb]
	\centering
		\includegraphics[width=0.8\textwidth]{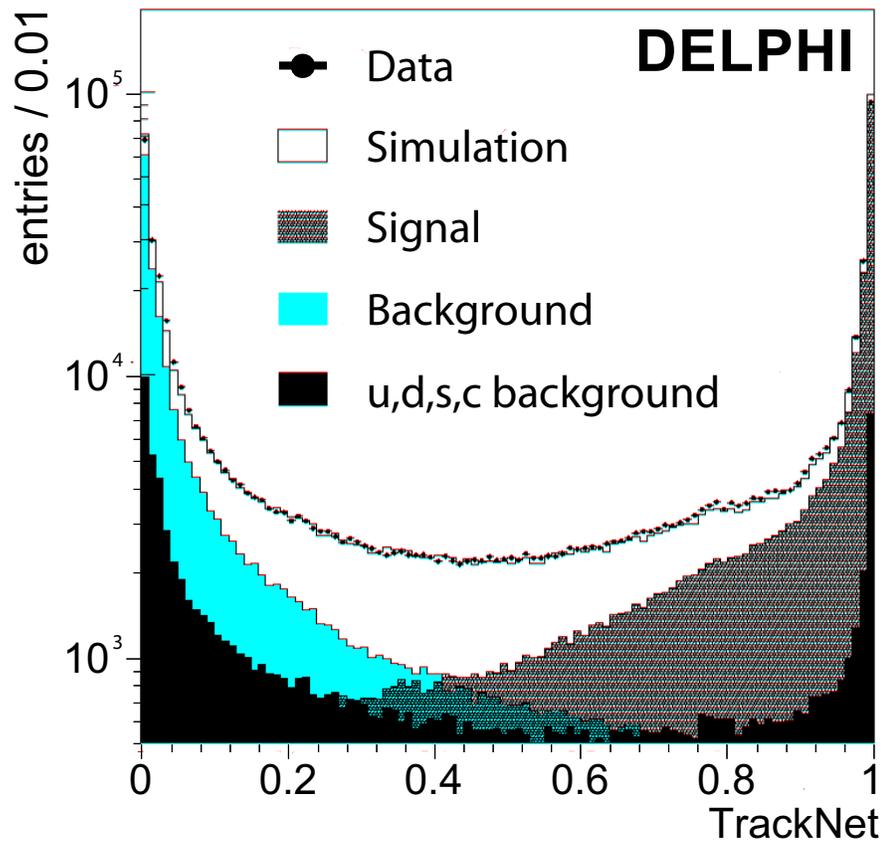}
	\caption{
The TrackNet distribution for different track species.
`Signal' refers to tracks originating from the B-hadron decay chain, 
`Background' are tracks from fragmentation or excited b-hadron decay and 
`u,d,s,c background' are any tracks in non-b decays of the $\Zz$.}
	\label{fig:bfrag_trknet}
\end{figure}

The TrackNet distribution for different track species is shown in Figure~\ref{fig:bfrag_trknet}.
Tracks from the b-decay chain are identified with high purity and efficiency via a cut at $\rm{TrackNet}>0.5$.

\end{appendix}

\clearpage
\begin{appendix}
\section*{Appendix B: Regularised unfolding analysis: central values and  covariance matrices}
\label{appenb}

The full bin-to-bin results of the $\xb$ distribution from the regularised unfolding analysis are listed in Table~\ref{tab:resweak}.
In each bin the quoted value corresponds to the average of the fitted
distribution over this bin:
\begin{equation}
\frac{\Delta f}{\Delta \xb} = \frac{1}{x_{\mathrm B}^{\mathrm weak,~max}-x_{\mathrm B}^{\mathrm weak,~min}}\int_{x_{\mathrm B}^{\mathrm weak,~min}}^{x_{\mathrm B}^{\mathrm weak,~max}}
{\frac{df}{d\xb}~d\xb}~.\nonumber
\end{equation}
Values $x_{\mathrm B}^{\mathrm weak,~max}$ and
$x_{\mathrm B}^{\mathrm weak,~min}$ are the bin limits.
The corresponding statistical and total covariance matrices are given in Tables~\ref{tab:covweak1}
and~\ref{tab:covweak2}, respectively.
 \begin{table}[htb!]
 \begin{center}
{\footnotesize
  \begin{tabular}{|c|c|c|c|c|}
 \hline
 bin borders & value & stat. error & syst. error
 & $ \sqrt{\sigma_{stat}^2+\sigma_{syst}^2} $ \\
 \hline
 \hline
  0.100     --  0.299      &  0.154    & 0.005    & 0.044    & 0.044    \\
  0.299     --  0.419      &  0.458    & 0.007    & 0.035    & 0.036    \\
  0.419     --  0.535      &  0.724    & 0.008    & 0.039    & 0.039    \\
  0.535     --  0.637      &  1.041    & 0.010    & 0.056    & 0.057    \\
  0.637     --  0.726      &  1.635    & 0.012    & 0.063    & 0.064    \\
  0.726     --  0.803      &  2.632    & 0.014    & 0.073    & 0.075    \\
  0.803     --  0.877      &  3.091    & 0.016    & 0.140    & 0.141    \\
  0.877     --  0.939      &  2.033    & 0.013    & 0.159    & 0.159    \\
  0.939     --  1.000      &  0.300    & 0.009    & 0.067    & 0.067    \\
 \hline
 \end{tabular}

}
 \caption[]{The unfolding result, per bin, for $f(\xb)$.}
  \label{tab:resweak}
 \end{center}
\end{table}
\begin{table}[htb!]
 \begin{center}
 {\footnotesize
  \begin{tabular}{|c|ccccccccc|}
 \hline
 bin& 1& 2& 3& 4& 5& 6& 7& 8& 9\\
 \hline
 1
 &    0.021
 &
 &
 &
 &
 &
 &
 &
 &
 \\
 2
 &    0.022
 &    0.047
 &
 &
 &
 &
 &
 &
 &
 \\
 3
 &   -0.019
 &    0.005
 &    0.069
 &
 &
 &
 &
 &
 &
 \\
 4
 &   -0.018
 &   -0.038
 &    0.026
 &    0.103
 &
 &
 &
 &
 &
 \\
 5
 &    0.011
 &   -0.011
 &   -0.055
 &    0.026
 &    0.151
 &
 &
 &
 &
 \\
 6
 &    0.008
 &    0.019
 &   -0.025
 &   -0.073
 &    0.057
 &    0.194
 &
 &
 &
 \\
 7
 &   -0.006
 &    0.003
 &    0.029
 &   -0.022
 &   -0.099
 &    0.057
 &    0.245
 &
 &
 \\
 8
 &   -0.001
 &   -0.006
 &    0.003
 &    0.024
 &   -0.018
 &   -0.079
 &    0.025
 &    0.163
 &
 \\
 9
 &    0.002
 &    0.001
 &   -0.008
 &   -0.001
 &    0.031
 &   -0.011
 &   -0.085
 &    0.052
 &    0.088
 \\
 \hline
 \end{tabular}

 }
\caption[]{The statistical covariance matrix, in units of $10^{-3}$, for the unfolded bins in $f(\xb)$.}
 \label{tab:covweak1}
  \end{center}
\end{table}
 \begin{table}[htb!]
\begin{center}
{\footnotesize
 \begin{tabular}{|c|ccccccccc|}
 \hline
 bin& 1& 2& 3& 4& 5& 6& 7& 8& 9\\
 \hline
 1
 &    1.927
 &
 &
 &
 &
 &
 &
 &
 &
 \\
 2
 &    0.201
 &    1.296
 &
 &
 &
 &
 &
 &
 &
 \\
 3
 &    0.127
 &    0.814
 &    1.552
 &
 &
 &
 &
 &
 &
 \\
 4
 &    0.252
 &    0.819
 &    1.252
 &    3.271
 &
 &
 &
 &
 &
 \\
 5
 &    0.247
 &    0.717
 &    0.996
 &    2.587
 &    4.120
 &
 &
 &
 &
 \\
 6
 &   -0.191
 &    0.167
 &    0.308
 &    1.200
 &    2.222
 &    5.578
 &
 &
 &
 \\
 7
 &   -0.928
 &   -1.163
 &   -1.682
 &   -2.615
 &   -2.980
 &   -1.146
 &   19.847
 &
 &
 \\
 8
 &   -1.676
 &   -2.818
 &   -3.920
 &   -6.877
 &   -8.562
 &   -4.806
 &    8.827
 &   25.342
 &
 \\
 9
 &   -0.669
 &   -1.120
 &   -1.557
 &   -2.725
 &   -3.197
 &   -1.535
 &    3.025
 &    9.007
 &    4.538
 \\
 \hline
 \end{tabular}

 }
\caption[]{The total (i.e. including statistical and systematic
uncertainties) covariance matrix, in units of $10^{-3}$, for the unfolded
bins in $f(\xb)$.}
 \label{tab:covweak2}
 \end{center}
 \end{table}
\clearpage

\end{appendix}

\begin{appendix}
\section*{Appendix C: Weighted fitting analysis: central values and  covariance matrices}
\label{appenc}

The full bin-to-bin results of the $\xb$ distribution from the weighted fitting analysis are listed in Table~\ref{tab:resweakorsay}. For explanation of the quoted bin values see Appendix A. The corresponding statistical and total covariance matrices are given in Tables~\ref{tab:covweak1orsay} and~\ref{tab:covweak2orsay}, respectively.

\begin{table}[htb!]
 \begin{center}
{\footnotesize
  \begin{tabular}{|c|c|c|c|c|}
 \hline
 bin borders & value & stat. error & syst. error
 & $ \sqrt{\sigma_{stat}^2+\sigma_{syst}^2} $ \\
 \hline
 \hline
  0.100     --  0.299      &  0.180    & 0.003    & 0.010    & 0.010    \\
  0.299     --  0.419      &  0.445    & 0.008    & 0.028    & 0.029    \\
  0.419     --  0.535      &  0.746    & 0.012    & 0.034    & 0.036    \\
  0.535     --  0.637      &  1.220    & 0.017    & 0.060    & 0.062    \\
  0.637     --  0.726      &  1.947    & 0.025    & 0.082    & 0.086    \\
  0.726     --  0.803      &  2.694    & 0.040    & 0.062    & 0.074    \\
  0.803     --  0.877      &  2.790    & 0.040    & 0.133    & 0.139    \\
  0.877     --  0.939      &  1.600    & 0.043    & 0.137    & 0.144    \\
  0.939     --  1.000      &  0.268    & 0.020    & 0.086    & 0.088    \\
 \hline
 \end{tabular}

}
 \caption[]{The unfolding result, per bin, for $f(\xb)$.}
 \label{tab:resweakorsay}
 \end{center}
\end{table}
\begin{table}[htb!]
 \begin{center}
 {\footnotesize
  \begin{tabular}{|c|ccccccccc|}
 \hline
 bin& 1& 2& 3& 4& 5& 6& 7& 8& 9\\
 \hline
 1
 &    0.007
 &
 &
 &
 &
 &
 &
 &
 &
 \\
 2
 &    0.021
 &    0.058
 &
 &
 &
 &
 &
 &
 &
 \\
 3
 &    0.026
 &    0.077
 &    0.138
 &
 &
 &
 &
 &
 &
 \\
 4
 &    0.021
 &    0.063
 &    0.166
 &    0.292
 &
 &
 &
 &
 &
 \\
 5
 &   -0.003
 &   -0.015
 &    0.024
 &    0.210
 &    0.613
 &
 &
 &
 &
 \\
 6
 &   -0.015
 &   -0.061
 &   -0.140
 &   -0.124
 &    0.526
 &    1.587
 &
 &
 &
 \\
 7
 &    0.024
 &    0.055
 &    0.017
 &   -0.098
 &   -0.172
 &    0.415
 &    1.603
 &
 &
 \\
 8
 &    0.033
 &    0.082
 &    0.158
 &    0.216
 &   -0.040
 &   -0.707
 &   -0.183
 &    1.838
 &
 \\
 9
 &   -0.002
 &   -0.005
 &   -0.008
 &    0.001
 &    0.079
 &    0.190
 &   -0.067
 &   -0.352
 &    0.416
 \\
 \hline
 \end{tabular}

 }
\caption[]{The statistical covariance matrix,
in units of $10^{-3}$, for the unfolded
bins in $f(\xb)$.}
 \label{tab:covweak1orsay}
 \end{center}
\end{table}
 \begin{table}[htb!]
 \begin{center}
  {\footnotesize
 \begin{tabular}{|c|ccccccccc|}
 \hline
 bin& 1& 2& 3& 4& 5& 6& 7& 8& 9\\
 \hline
 1
 &    0.106
 &
 &
 &
 &
 &
 &
 &
 &
 \\
 2
 &    0.292
 &    0.841
 &
 &
 &
 &
 &
 &
 &
 \\
 3
 &    0.248
 &    0.818
 &    1.323
 &
 &
 &
 &
 &
 &
 \\
 4
 &   -0.025
 &    0.202
 &    1.602
 &    3.849
 &
 &
 &
 &
 &
 \\
 5
 &   -0.397
 &   -0.806
 &    0.855
 &    4.463
 &    7.415
 &
 &
 &
 &
 \\
 6
 &   -0.337
 &   -1.027
 &   -1.057
 &   -0.042
 &    2.877
 &    5.440
 &
 &
 &
 \\
 7
 &   -0.048
 &   -0.656
 &   -3.257
 &   -7.428
 &   -9.097
 &   -0.679
 &   19.353
 &
 &
 \\
 8
 &   -0.432
 &   -1.745
 &   -3.326
 &   -5.212
 &   -6.299
 &   -1.804
 &   13.991
 &   20.555
 &
 \\
 9
 &    0.286
 &    0.902
 &    0.677
 &   -0.202
 &   -0.564
 &   -0.573
 &   -2.817
 &   -6.084
 &    7.864
 \\
 \hline
 \end{tabular}

 }
\caption[]{The total (i.e. including statistical and systematic
uncertainties) covariance matrix, in units of $10^{-3}$, for the unfolded
bins in $f(\xb)$.}
 \label{tab:covweak2orsay}
 \end{center}
 \end{table}

\end{appendix}

\clearpage
\begin{appendix}
\section*{Appendix D: Combined DELPHI result: covariance matrices}
\label{append}

The statistical and total covariance matrices for the combined DELPHI result in bins of the $f(\xb)$ distribution are given in Tables~\ref{tab:covweak1comb} and~\ref{tab:covweak2comb}, respectively.

 \begin{table}[htb!]
 \begin{center}
 {\footnotesize
  \begin{tabular}{|c|ccccccccc|}
 \hline
 bin& 1& 2& 3& 4& 5& 6& 7& 8& 9\\
 \hline
1 & 0.019 &       &       &       &       &       &       &       &       \\
2 & 0.034 & 0.069 &       &       &       &       &       &       &       \\
3 & 0.022 & 0.049 & 0.072 &       &       &       &       &       &       \\
4 & 0.025 & 0.050 & 0.086 & 0.181 &       &       &       &       &       \\
5 & 0.044 & 0.078 & 0.061 & 0.194 & 0.422 &       &       &       &       \\
6 & 0.076 & 0.133 & 0.098 & 0.177 & 0.468 & 0.819 &       &       &       \\
7 & 0.077 & 0.144 & 0.137 & 0.181 & 0.257 & 0.629 & 0.833 &       &       \\
8 & 0.078 & 0.146 & 0.137 & 0.243 & 0.337 & 0.449 & 0.568 & 0.755 &       \\
9 & 0.037 & 0.063 & 0.054 & 0.110 & 0.207 & 0.240 & 0.140 & 0.308 & 0.227 \\
 \hline
 \end{tabular}

 }
\caption[]{The statistical covariance matrix,
in units of $10^{-3}$, for the unfolded
bins in $f(\xb)$. Elements are scaled by a factor 1.71.}
 \label{tab:covweak1comb}
 \end{center}
 \end{table}

\vspace{0.7cm}

 \begin{table}[htb!]
 \begin{center}
  {\footnotesize
 \begin{tabular}{|c|ccccccccc|}
 \hline
 bin& 1& 2& 3& 4& 5& 6& 7& 8& 9\\
 \hline
1 &  0.408 &        &        &        &        &        &         &         &        \\
2 & -0.139 &  1.013 &        &        &        &        &         &         &        \\
3 &  0.554 &  0.346 &  1.403 &        &        &        &         &         &        \\
4 & -0.134 &  0.730 &  1.013 &  2.514 &        &        &         &         &        \\
5 & -0.109 &  0.073 &  0.723 &  2.393 &  3.729 &        &         &         &        \\
6 & -0.122 &  0.044 & -0.216 &  0.398 &  2.056 &  4.850 &         &         &        \\
7 & -0.111 & -0.319 & -1.721 & -3.815 & -4.596 & -0.505 &  14.838 &         &        \\
8 & -0.372 & -1.123 & -2.413 & -3.922 & -4.605 & -2.410 &  10.305 &  14.939 &        \\
9 &  0.075 & -0.210 & -0.441 & -1.102 & -1.149 &  0.123 & \,1.550 & \,3.507 &  2.373 \\
 \hline
 \end{tabular}
 }
\caption[]{The total (i.e. including statistical and systematic
uncertainties) covariance matrix, in units of $10^{-3}$, for the unfolded
bins in $f(\xb)$. Elements are scaled by a factor 1.71.}
 \label{tab:covweak2comb}
 \end{center}
 \end{table}

\end{appendix}

\clearpage


\end{document}